\documentclass[ejs]{imsart}

\RequirePackage[OT1]{fontenc}
\RequirePackage{amsthm,amsmath}
\RequirePackage[colorlinks,citecolor=blue,urlcolor=blue]{hyperref}

\doi{10.1214/154957804100000000}
\pubyear{0000}
\volume{0}
\firstpage{1}
\lastpage{1}

\usepackage{amsmath,amsthm,amssymb,amsfonts,verbatim,bm,graphicx,enumerate,epstopdf,lscape,subfigure,multirow} 
\usepackage{algorithm,algorithmicx}
\usepackage{algpseudocode}
\usepackage{boxedminipage,longtable}
\usepackage[dvipsnames,usenames]{color}
\usepackage{tikz,verbatim,bibentry}
\usepackage{bm,mathrsfs,cite}
\usepackage[comma,authoryear]{natbib}

\newtheorem{lemma}{Lemma}
\newtheorem{theorem}{Theorem}
\newtheorem{proposition}{Proposition}

\theoremstyle{definition}
\newtheorem{rmk}{Remark}

\newtheorem{assumption}{Assumption}

\newcommand{\utwi}[1]{\mbox{\boldmath $ #1$}}

\newcommand{\bO}{{\utwi{O}}}

\newcommand{\cA}{{\cal A}}
\newcommand{\cB}{{\cal B}}

\newcommand{\cE}{{\cal E}}
\newcommand{\cF}{{\cal F}}

\newcommand{\cM}{{\cal M}}

\newcommand{\cW}{{\cal W}}
\newcommand{\cX}{{\cal X}}

\newcommand{\cZ}{{\cal Z}}

\newcommand{\bel}{\begin{eqnarray}\label}
\newcommand{\eel}{\end{eqnarray}}

\def\I{\mathrm{I}}
\def\II{\mathrm{II}}
\def\III{\mathrm{III}}
\def\IV{\mathrm{IV}}

\def\Var{\textsf{Var}}
\def\E{\mathbb{E}}
\def\P{\mathbb{P}}
\def\RR{\mathbb{R}}
\def\R{\mathbb{R}}

\def\tr{\mathrm{tr}}

\DeclareMathOperator*{\argmin}{\arg\min}

\DeclareMathOperator*{\mat1}{\text{mat}_1}

\DeclareMathOperator*{\vec1}{\text{vec}}
\def\cano{(\text{\footnotesize cano})}

\begin{document}

\begin{frontmatter}
\title{Rank Determination in Tensor Factor Model\thanksref{t1}}
\runtitle{Determining the Number of Factors}

\thankstext{t1}{
Cun-Hui Zhang is the corresponding author. Han's research is supported in part by National Science Foundation grant IIS-1741390. Zhang's research is supported in part by NSF grants DMS-1721495, IIS-1741390, CCF-1934924 and DMS-2052949 Chen's research is supported in part by National Science Foundation grants DMS-1503409, DMS-1737857, IIS-1741390 and DMS-2052949}

\begin{aug}
\author{\fnms{Yuefeng} \snm{Han}
\ead[label=e1,mark]{yuefeng.han@rutgers.edu}},
\author{\fnms{Rong} \snm{Chen}
\ead[label=e2,mark]{rongchen@stat.rutgers.edu}}
\and
\author{\fnms{Cun-Hui} \snm{Zhang}
\corref{}\ead[label=e4,mark]{czhang@stat.rutgers.edu}}

\address{Department of Statistics,
Rutgers University,
Piscataway, NJ 08854,
USA,
\printead{e1,e2,e4}}

\runauthor{Y. Han, C. Zhang and R. Chen}
\affiliation{Rutgers University}

\end{aug}

\begin{abstract}
Factor model is an appealing and effective analytic tool for high-dimensional time series, with a wide range of applications in economics, finance and statistics. 
This paper develops two criteria for the determination of the number of factors for tensor factor models where the signal part of an observed tensor time series assumes a Tucker decomposition with the core tensor as the factor tensor. The task is to determine the dimensions of the core tensor. 
One of the proposed criteria is similar to information based criteria of model selection, and the other is an extension of the approaches based on the ratios of consecutive eigenvalues often used in factor analysis for panel time series. 
Theoretically results, including sufficient conditions and convergence rates, are established.
The results include the vector factor models as special cases, with an additional convergence rates.
Simulation studies provide promising finite sample performance for the two criteria.

\end{abstract}



\end{frontmatter}

\section{Introduction} \label{section:introduction}

Factor models have become a popular dimensional reduction tool in economics and statistics, especially for analyzing high dimensional time series. In practice a few common factors can often capture a large amount of variations and dynamics among a large pool of variables and time series. In the finance literature, \citet{chamberlain1983} exploited factor analysis to extend classical arbitrage pricing theory. In macroeconomics, \citet{bai2002,bai2003,stock2002} considered static factor models for modeling macroeconomic time series. \citet{forni2005} studied the identification of economy-wide and global shocks using generalized dynamic factor models. \citet{fan2011, fan2013, fan2019} established large covariance matrix estimation based on the static factor model. Factor models are also used to evaluate the impacts of various policies; see, e.g., \citet{bai2014}, \citet{ouyang2015} and \citet{li2017estimation}. Recently, large matrix or tensor (multi-dimensional array) data has become ubiquitous. \citet{wang2019} proposed a matrix factor model and applied it to matrix-valued financial data. \citet{chen2021factor} analyzed the multi-category import-export network data via tensor factor model.

A critical step in building a factor model is to correctly specify the number of factors used in the model. Estimation and forecasting procedures are all depended on the number of factors. Moreover, in some cases the number of factors may have some crucial economic interpretations and important theoretical consequences. For example, in finance and macroeconomics, it provides the number of sources of nondiversifiable risk or the fundamental shocks driving the macroeconomic dynamics. See, \cite{forni1998let}, \citet{stock2016dynamic}, \citet{giannone2006}, \citet{forni2009}, among others. 

Over the past decades, many methods have been developed
to determine the number of common factors needed for modelling
high dimensional vector time series. 
The most widely studied approach is to utilize the behavior of the eigenvalues of
the covariance matrix (see, e.g., \citet{bai2002}), or the singular values 
of the autocovariance matrix (see, e.g., \citet{lam2012}). 
By the definition of factor models, the eigenvalues or the singular values 
corresponding to the systematic components must increase with the number of cross-sectional
units. The rest of
the eigenvalues, which represents idiosyncratic components, stay bounded or
remain to be zero. In the static factor model, \citet{bai2002} proposed to estimate the number of factors by separating diverging eigenvalues from the rest using threshold functions,
in the form of an information criterion. Alternative criteria based on random matrix theory have been studied in \citet{kapetanios2010} and \citet{onatski2010} for the static factor model. Specifically
\citet{kapetanios2010} developed sequential tests and employed a subsampling method to obtain an approximation of the asymptotic distribution of the estimated eigenvalues. \citet{onatski2010} constructed tests based on the empirical distribution of the eigenvalues. In addition, \citet{onatski2012} proposed an alternative estimator using {the difference of consecutive eigenvalues}.
\citet{bai2007} and \citet{amengual2007} extended the work of \citet{bai2002} to the restricted dynamic factor model. \citet{hallin2007} further extended the framework to the generalized dynamic factor model through thresholding eigenvalues of the spectral density matrix. They also proposed a data-dependent method to adjust the multiplicative constant of the penalty function.
\cite{alessi2010improved} introduced a tuning multiplicative constant in the penalty for dealing with approximate factor models.
\citet{kong2017} employed similar ideas to study continuous time factor model with high frequency data. \citet{li2017} modified \citet{bai2002}'s procedure to the case that
the number of factors is allowed to increase with the sample size.
\cite{trapani2018randomized} proposed a randomized sequential test procedure to determine the number of factors.

An alternative approach is to study the ratio of each pair of adjacent eigenvalues, with the
insight that ratio of the smallest eigenvalue among these corresponding to the system component and the
largest eigenvalue among these corresponding to the idiosyncratic component goes to infinity. Under stationary conditions, \citet{ahn2013} developed such an estimator based on the sample covariance matrix. \citet{lam2012} used such a ratio based estimator based on singular values of the aucovariance matrix, under an alternative definition of factor models proposed in \citet{pan2008} and \citet{lam2011}.

Other than the eigenvalue-based methods, \citet{ye2003} developed an eigenvector based order determination procedure. \citet{luo2016} proposed a new estimator that combines both the eigenvalues and the bootstrap eigenvector variability. \citet{jung2018} suggested to sequentially test skewness of the squared lengths of residual scores that are obtained by removing leading principal components. However, these works assumed that the data are temporally independent, which are unlikely to hold for economic data.

These studies all focus on panel (vector) time series. Recently there is a growing
interest in analyzing matrix- or tensor-valued time series, as such time series is
encountered more and more frequently in applications, including Fama-French 10 by 10 series \citep{wang2019}, a set of economic indicator series among a set of countries \citep{chen2021autoregressive}, multi-category international trading volume series \citep{Hoff2011,chen2019matrix}, multi-type international action counts among a group of countries \citep{Hoff2015}, sequence of realized covariance matrices \citep{lunde2016econometric, kim2019factor}, sequence of gray-scale face recognition images \citep{chen2021statistical}, dynamic networks \citep{barabasi1999emergence,jiang2020autoregressive}, dynamic human brain transcriptome data \citep{liu2022characterizing}, multivariate spatial-temporal climate series \citep{chen2020semiparametric}, neuroimaging data \citep{zhang2019cross,zhou2013}.
Factor model is again developed as an effective dimension reduction tool \citep{wang2019,chen2021factor,han2020}. Same as for the vector
factor models, it is important to determine the number of factors in these models. 


In this paper, we consider the determination of the dimension of the core tensor factor in the tensor factor model in \citet{chen2021factor} and \citet{han2020}, which assumes the form 
\[ 
\cX_t=\cM_t+\cE_t=\cF_t\times_1 A_1\times_2\ldots\times_K A_K+\cE_t.
\]
Similar to \citet{lam2012}, the noise tensor $\cE_t$ is 
assumed to be a white tensor process with potentially strong contemporary correlations among the elements of the noise tensor, and all common dynamics is absorbed in the signal process $\cM_t$. 
This model setting is different from the approximate factor model in \cite{bai2002} and the dynamic factor model in \cite{hallin2007}, in which the noise process 
is allowed to have weak auto-correlations, but with strong restriction on the contemporary correlation.

\citet{chen2021factor} and \citet{han2020} studied the estimation procedures of the tensor factor model, assuming the ranks of the core tensor $\cF_t$ is given, with some ad hoc rank determination suggestions. In this paper we formally
propose two criteria for specifying the ranks of the core factor process, which we name ``the information criterion'' (IC) and ``the eigenvalue ratio'' (ER). They are all based on examining the eigenvalues of the sample cross-auto-moment of the observed tensor time series, utilizing the whiteness property of the noise process.
The IC estimators aim at truncating eigenvalues, which is similar to the
information criteria in vector factor models (e.g., \citet{bai2002} and \citet{hallin2007}). The ER estimators are obtained by minimizing the ratio of two adjacent eigenvalues arranged in ascending order, extending the standard ER estimator in \citet{lam2012} and \citet{wang2019} with an added small penalty term in both the numerator and denominator of the ratio. The penalty term behaves like a
lower bound correction to the true zero eigenvalues. 
We adopt similar ideas of the TOPUP and TIPUP procedures of \citet{chen2021factor}, and their corresponding
iterative versions, iTOPUP and iTIPUP, of \citet{han2020}, to construct sample auto-cross-moments. 
Our theoretical and empirical investigations show that estimators based on the iterative algorithms are much better than that based on the non-iterative ones, as the iterative algorithms significantly improve the estimation accuracy of the eigenvalues. The finite sample properties of the IC and ER criteria are also good. The empirical evidences show that the best estimators in tensor factor model are the IC and ER estimators based on iTIPUP, under some mild conditions on the level of signal cancellation typically associated with the TIPUP based procedures. 



This paper is organized as follows. Section \ref{section:model} briefly describes the tensor factor model and the corresponding estimation procedures proposed in \citet{chen2021factor} and \citet{han2020}. Section \ref{section:method} introduces the criteria for determining the ranks of the core tensor factor process,
and their iterative versions.
Section \ref{section:theories} investigates theoretical properties of the proposed estimators.
Section \ref{section:simulation} presents simulation studies of the finite sample properties of the proposed methods. Real data analysis is given in Section \ref{section:data}. Discussions 
are provided in Section \ref{section:conclusion}. All technical details are relegated to the Supplementary Material.

\section{General order determination criteria of semipositive definite matrices}\label{section:order}

In this section, we first propose two general order determination criteria based on
the properties of the estimated eigenvalues of a semipositive matrix. 

Let $\widehat W$ be 
a $p\times p$ symmetric and non-negative definite matrix, which is a sample version
of a true $p\times p$ symmetric and non-negative definite matrix. We assume
$W=\E \widehat W$. Also let $\hat \lambda_j$ be the eigenvalues of $\widehat W$ such that $\hat\lambda_1\ge \hat\lambda_2\ge\ldots \ge \hat\lambda_{p}$.
Let $\lambda_1\ge \ldots\ge \lambda_r> \lambda_{r+1}=\ldots =\lambda_{p}=0$ be the eigenvalues of $W$.
Note that the rank of $W$ is $r$.

Let $m^*<p $ be a predefined upper bound and functions $G(\widehat{W})$ and $H(\widehat{W})$ be
some appropriate positive penalty functions. We propose the following two quantities
\begin{eqnarray} 
  {\rm IC}(\widehat{W}) & = &  \argmin_{0\le m\le m^*}
  \left\{ \sum_{j=m+1}^{p} \hat\lambda_{j}+mG(\widehat W)\right\} \label{estimate:IC}\\
  {\rm ER}(\widehat W) & = &  \argmin_{1\le m\le m^*}
  \frac{\hat\lambda_{m+1}+H(\widehat W)}{\hat\lambda_{m}+H(\widehat W)}. \label{estimate:ER}
\end{eqnarray}

The first criterion in (\ref{estimate:IC})
is similar to an information criterion as its first term
mimics the residual sum of squares of using a rank $m$ matrix to approximate the
matrix $\widehat W$ while the second term $mG(\widehat W)$ penalizes the model complexity $m$.
We will call it {\it the information criterion (IC)}.
The second criterion in (\ref{estimate:ER}) uses
the ratio of two adjacent eigenvalues of $\widehat W$, with a small
penalty term $H(\widehat W)$ added to both the numerator and denominator.
We will call it {\it the eigen-ratio criterion (ER)}.

\begin{rmk}[\it The information criterion]
Note that, for a given $m$, the principle components can be viewed as
solutions of an optimization problem in which the ``sum of squared residuals'' is minimized,
\begin{align}\label{estimate:procedure}
  \widehat U_{m}=\argmin_{U_{m}}\tr\left\{ \left(I- U_{m} U_{m}^\top \right) \widehat W \right\}.
\end{align}
Note that $ \tr\left\{\left(I- \widehat U_{m}\widehat U_{m}^\top \right) \widehat W\right\}
  =\sum_{j=m+1}^p \hat\lambda_k$.
It plays the role of residual sum of squares classically appearing
in information criterion methods.
Criterion (\ref{estimate:IC}) has a structure comparable to that of \citet{bai2002} and \citet{hallin2007}. For vector factor models, the method proposed in \citet{bai2002} is the same of the IC criterion with $\widehat W$ being the sample covariance matrix, while that in \citet{hallin2007} used spectral density matrix estimation. 
The penalty $mG(\widehat W)$ is intimately related to the rate of convergence of the non-divergent eigenvalues, when $\widehat W$ is estimated from a set of data with diverging
dimensions, and balances between overestimation and underestimation.
\end{rmk}

\begin{rmk}[\it Eigen-ratio criterion]
Different from the standard ER estimator in \citet{lam2012}, we add
a penalty term $H(\widehat W)$ to both the numerator and denominator.
The intuition behind $H(\widehat W)$ is as follows. Since $\widehat W$ is a noisy version (an estimator) of $W$ of rank $r$, all estimated eigenvalues 
$\hat\lambda_{j}$ ($r+1\le j\le p$) correspond
to the zero eigenvalues of $W$.
Hence the ratio $\hat\lambda_{j+1}/\hat\lambda_{j}$ ($j> r$) theoretically can be arbitrary small.
The penalty $H(\widehat W)$ provides a lower bound correction to $\hat \lambda_{j}$
($r+1\le j\le p$). When it is of a proper order,
we can ensure that the ratio $(\hat\lambda_{m+1}+H(\widehat W))/(\hat\lambda_{m}+H(\widehat W))$ goes
to zero when $m=r$ (the true rank), 
while all other such ratios are asymptotically bounded from below. In vector factor models, \citet{ahn2013} exploited the ratio of eigenvalues of sample covariance matrix to determine the number of factors. Non-divergent eigenvalues therein are bounded below by a positive number asymptotically, as long as the eigenvalues of covariance matrix of idiosyncratic noises are bounded away from zero. Our criterion (\ref{estimate:ER}) has a similar flavor.
\end{rmk}


Here we show a consistency result for the general estimator. To be more precise, let $\widehat W^{(n)}$ and $W^{(n)}$ be two sequences of semi-positive symmetric matrices, with
$W^{(n)}=\E \widehat W^{(n)}$ and $n$ be an index associated with the sample size and dimension. Also let $\hat \lambda_j^{(n)}$ be the eigenvalues of
$\widehat W^{(n)}$ such that $\hat\lambda_1^{(n)}\ge \hat\lambda_2^{(n)}\ge\ldots \ge
\hat\lambda_{p}^{(n)}$. Let $\lambda_1^{(n)}\ge \ldots\ge \lambda_r^{(n)}> \lambda_{r+1}^{(n)}=\ldots =\lambda_{p}^{(n)}=0$ be the eigenvalues of $W$.
We assume $\lambda_r^{(n)}\to\infty$ as
$n\to\infty$. The following proposition provides the sufficient conditions for the consistency of the IC and ER estimators in \eqref{estimate:IC} and \eqref{estimate:ER}, respectively. It provides a guideline of choosing proper penalty functions $G(\cdot)$ and $H(\cdot)$ in order determination for any generic $\widehat W^{(n)}$.

\begin{proposition}\label{prop:generic}
Assume $|\hat \lambda_1^{(n)}-\lambda_1^{(n)}|= o_{\P}(\lambda_1^{(n)})+O_{\P}(\gamma_n)$ and $|\hat \lambda_r^{(n)}-\lambda_r^{(n)}| = o_{\P}(\lambda_r^{(n)})+O_{\P}(\gamma_n)$,
    and $|\hat \lambda_j^{(n)}-\lambda_j^{(n)}|=O_{\P}(\beta_n)$ for all $j> r$. Then, \\
  (i) $\P(IC(\widehat W^{(n)})=r)\to 1$, provided that
  $(G(\widehat W^{(n)})+\gamma _n)/\lambda_r^{(n)}\to 0$ and
  $G(\widehat W^{(n)})/\beta_n\to\infty$; \\
  (ii) $\P(ER(\widehat W^{(n)})=r)\to 1$, provided that
  $(H(\widehat W^{(n)})+\beta_n) / ((\lambda_r^{(n)})^2/\lambda_1^{(n)})\to 0$,
  $\gamma_n/\lambda_r^{(n)}\to 0$ and $H(\widehat W^{(n)})/ (\beta_n^2/\lambda_r^{(n)})\to \infty$. 
\end{proposition}



In the conditions of Proposition \ref{prop:generic}, $\gamma_n$ represents 
the convergence rate of the sample eigenvalues corresponding to the non-zero eigenvalues of $W^{(n)}$, and $\beta_n$ represents  
the rate of the sample eigenvalues corresponding to the zero eigenvalues of $W^{(n)}$. For example, under the strong factor model of \cite{lam2012}'s setting, $\gamma_n=p^2T^{-1/2}$ and $\beta_n=p^2T^{-1}$, where $T$ is the sample size.

\begin{rmk}\label{spike}
In our model setting, $\lambda_{r+1}^{(n)}=\ldots =\lambda_{p}^{(n)}=0$ and our objective is to separate the zero and non-zero eigenvalues. The proposition holds for general spiked eigenvalue detection as well. Specifically, let $\lambda_1^{(n)}\ge \ldots\ge \lambda_r^{(n)}> \gamma_n> \lambda_{r+1}^{(n)}\ge\ldots \ge\lambda_{p}^{(n)}\ge0$ be the eigenvalues of $W^{(n)}$, where $\lambda_{r+1}^{(n)}, \ldots, \lambda_p^{(n)}$ are called non-spiked eigenvalues \citep{cai2020limiting}. Again it is assumed that $\lambda_r^{(n)}\to\infty$ as $n\to\infty$. Then Proposition \ref{prop:generic} holds when
$\gamma_n$ and $\beta_n$ are the convergence rate of the sample eigenvalues corresponding to the spiked and non-spiked eigenvalues of $W^{(n)}$, respectively. Note that the approaches of \citet{bai2002}, \citet{amengual2007}, \citet{hallin2007}, \citet{lam2012} and \citet{ahn2013} all
fit in this generic setting or its variants,
with various forms of the penalty functions $G(\widehat W)$ and $H(\widehat W)$ 
to distinguish $\hat\lambda_r^{(n)}$ from $\hat\lambda_{r+1}^{(n)}$. 
For example, \citet{bai2002} suggest to use $G_1=p^{-1}T^{-1}(p+T)\log(p T(p+T)^{-1})$, $G_2=p^{-1}T^{-1}(p+T)\log(\min\{p,T\})$, or $G_3=\max\{p^{-1},T^{-1}\}\log(\min\{p,T\})$, where $T$ is the sample size and $p$ is the number of variables.

\end{rmk}

\begin{rmk}
When the dimensions $d_k$ are large, 
estimating eigenvalues of a matrix using its 
sample version 
is in general very difficult and potentially inaccurate. However, to determine the number of factors, only the leading eigenvalues need to be estimated relatively accurately to achieve the purpose, which requires relatively mild conditions on the sample version of the matrix.  
\end{rmk}

For the specific problems such as the tensor factor model problem we focus there, a detailed analysis of 
the rates $\gamma_n$ and $\beta_n$ is needed to construct the penalty functions $G(\cdot)$ and $H(\cdot)$ and to establish the consistency of the rank estimators. In fact one can establish the convergence rate with a more
detailed analysis beyond the simple consistency results in Proposition \ref{prop:generic}, as we will do for the tensor factor model.


\section{Order determination criteria for tensor factor models}\label{section:order2}

\subsection{The model} 
\label{section:model}

Here we briefly introduce the tensor factor model setup 
in \citet{chen2021factor} and \citet{han2020}. A tensor factor model can be written as
\begin{equation} \label{eq:tensorfactor}
\cX_t=\cM_t+\cE_t=\cF_t\times_1 A_1\times_2\ldots\times_K A_K+\cE_t,
\end{equation}
where $\cX_t\in\RR^{d_1\times\cdots\times d_K}$ is the observed tensor at time $t$, the core tensor $\cF_t$ is the unobserved latent tensor factor process of dimension $r_1\times\ldots \times r_K$, $A_k$ are the deterministic loading matrix of size $d_k\times r_k$ and $r_k\ll d_k$, and $\cE_t$ is the idiosyncratic noise components of $\cX_t$, which is assumed to be a white process. Here the $k$-mode product of $\cX\in\RR^{d_1\times d_2\times \cdots \times d_K}$ with a matrix $U\in\RR^{d_k'\times d_k}$, denoted as $\cX\times_k U$,
  is an order $K$-tensor of size
  $d_1\times \cdots \times d_{k-1} \times d_k'\times d_{k+1}\times \cdots \times d_K$ such that
$$ (\cX\times_k U)_{i_1,...,i_{k-1},j,i_{k+1},...,i_K}=\sum_{i_k=1}^{d_k} \cX_{i_1,i_2,...,i_K} U_{j,i_k}.$$
  The core tensor $\cF_t$ is usually much smaller than $\cX_t$ in dimension. We also assume that the rank of $A_k$ is $r_k$. Otherwise $\cX_t$ in \eqref{eq:tensorfactor} may be expressed equivalently with a lower-dimensional factor process. The parameters $r_1,...,r_K$ are assumed to be fixed but unknown. For more details of the tensor factor model
\eqref{eq:tensorfactor}, see \citet{chen2021factor} and \citet{han2020}.

It is obvious that the loading matrices $A_k$ are not identifiable in Model
\eqref{eq:tensorfactor}. Model \eqref{eq:tensorfactor} is unchanged if we replace $(A_1,...,A_K, \cF_t)$ by $(A_1H_1,...,A_KH_K, \cF_t\times_{k=1}^K H_k^{-1})$ for any invertible $r_k\times r_k$ matrix $H_k$.
However, the linear space spanned by the columns of $A_k$, called 
the factor loading space, is uniquely defined. Assume $A_k$ has a SVD representation $A_k=U_k\Lambda_k V_k^\top$. Then, the factor loading space of $A_k$ can be represented by the orthogonal projection $P_k$,
\begin{align}\label{eq:projection}
P_k=P_{A_k}:=A_k (A_k^\top A_k)^{-1} A_k^\top=U_k U_k^\top. 
\end{align}

\subsection{Rank selection criteria for tensor factor models} \label{section:method}

The two criteria introduced in Section~\ref{section:order} can be used to estimate the number of factors
in the tensor factor model (\ref{eq:tensorfactor}), using properly constructed matrices
$W$ and $\widehat W$. Particularly, we will study the following four constructions.

Since they have been proposed and used for loading space estimation in
\cite{chen2021factor} and \cite{han2020}, we adopt the same names to represent 
them.

\noindent
    {\bf (I) TOPUP:\ } 
    Let
\begin{equation*}
{\rm{TOPUP}}_k(\cX_{1:T})=\mat1\left(\sum_{t=h+1}^T \frac{{\rm mat}_k(\cX_{t-h}) \otimes {\rm mat}_k(\cX_t)} {T-h}, \ h=1,...,h_0 \right),
\end{equation*}
and
\begin{align*}
  &\widehat W_k=\widehat W_k(\cX_{1:T}) :={\text{TOPUP}}_k (\cX_{1:T})
  ({\text{TOPUP}}_k (\cX_{1:T}))^\top,
\end{align*}
where $\otimes$ is the tensor product such that, for any $\cA\in\RR^{m_1\times m_2\times \cdots \times m_K}$ and $\cB\in \RR^{r_1\times r_2\times \cdots \times r_N}$, 
$$(\cA\otimes\cB)_{i_1,...,i_K,j_1,...,j_N}=(\cA)_{i_1,...,i_K}(\cB)_{j_1,...,j_N} ,$$ and ${\rm mat}_k$ is
the tensor unfolding (into a matrix) operation along mode-$k$ of a tensor.
Here we emphasize that $\widehat W_k$ is constructed using
$\cX_{1:T}=(\cX_1,\ldots,\cX_T)$.
The constant $h_0$ is a (small) predetermined integer and the sum over $h$
in $\widehat W_k$ is to accumulate the information from different time lags $h$.
The rank of its population version can be shown to be $r_k$ under certain conditions, hence we can use
the IC and ER estimators presented in Section~\ref{section:order} to determine $r_k$.
\vspace{0.1in}

\noindent
{\bf (II) TIPUP: \ } Define a $d_k\times (d_k h_0)$ matrix as
\begin{equation*}
{\rm{TIPUP}}_k(\cX_{1:T})={\rm{mat}}_1\left(\sum_{t=h+1}^T \frac{{\rm{mat}}_k(\cX_{t-h}) {\rm{mat}}_k^\top(\cX_t)} {T-h}, \ h=1,...,h_0 \right),
\end{equation*}
which replaces the tensor product in $\mat1({\rm{TOPUP}}_k(\cX_{1:T})$
by the inner product. Let
\begin{align*}
&\widehat W_k^*=\widehat W_k^*(\cX_{1:T}):=
  ({\text{TIPUP}}_k (\cX_{1:T}))({\text{TIPUP}}_k (\cX_{1:T})^\top.
\end{align*}

\vspace{0.1in}
\noindent
    {\bf (III and IV) iTOPUP and iTIPUP: \ }
\citet{han2020} proposed an iterative procedure to estimate
$U_k, k=1,...,K$ in \eqref{eq:projection},
based on either TOPUP or TIPUP procedure. Briefly,
at $i$-th iteration, suppose we have obtained
an estimate of the ranks
$\widehat{r}_k^{(i-1)}$ ($k=1,\ldots,K$) and their corresponding
$\widehat U_{k,\widehat r_k^{(i-1)}}^{(i-1)}$ at $(i-1)$-th iteration,
we calculate the
orthogonal projections of $\cX_t, 1\le t\le T$, to obtain
\begin{equation}
  \cZ_{k,t}^{(i)}=\cX_t \times_1 (\widehat U_{1,\widehat r_1^{(i)}}^{(i)})^\top \times_2 \cdots \times_{k-1} (\widehat U_{k-1,\widehat r_{k-1}^{(i)}}^{(i)})^\top \times_{k+1} (\widehat U_{k+1,\widehat r_{k+1}^{(i-1)}}^{(i-1)})^\top \times_{k+2}\cdots\times_K (\widehat U_{K,\widehat r_K^{(i-1)}}^{(i-1)})^\top.
  \label{eq:cZ}
\end{equation}
Note that $\cZ_{k,t}^{(i)}$ uses projection of $\cX_t$ on all modes, except mode-$k$.
The initial ranks
$\widehat{r}_k^{(0)}$ ($k=1,\ldots,K$) and their corresponding
$\widehat U_{k,\widehat r_k^{(0)}}^{(0)}$ can be obtained through the non-iterative
TOPUP and TIPUP procedure. Let ${\cZ}_{k,1:T}^{(i)}=({\cZ}_{k,1}^{(i)},\ldots,{\cZ}_{k,T}^{(i)})$, and
define

\begin{align*}
\widehat W_k^{(i)} & =  \widehat W_k({\cZ}_{k,1:T}^{(i)})
    :={\text{TOPUP}}_k ({\cZ}_{k,1:T}^{(i)}))
    ({\text{TOPUP}}_k ({\cZ}_{k,1:T}^{(i)}))^\top, \\
\widehat W_k^{*(i)} & = \widehat W_k^*({\cZ}_{k,1:T}^{(i)})
      :={\text{TIPUP}}_k ({\cZ}_{k,1:T}^{(i)}))
    ({\text{TIPUP}}_k ({\cZ}_{k,1:T}^{(i)}))^\top. 
\end{align*}

The iterative procedure is motivated by the observation that 
$\cZ_{k,t}^{(j)}$ is a $r_1\ldots r_{k-1}d_k r_{k+1}\ldots r_K$ tensor, much smaller than $\cX_t$, which is
$d_1\ldots d_k$ tensor. Hence $A_k$ can be estimated more accurately if all $A_j(j\neq k)$ are given in advance or can be estimated accurately, since the convergence rate now depends on $(d_k/r_k)\prod_{i=1}^K r_i$ rather than $\prod_{i=1}^K d_i$.

The IC and ER estimators are constructed by replacing $\widehat W$
in (\ref{estimate:IC}) and (\ref{estimate:ER}) with $\widehat W_k$, $\widehat W_k^*$,
$\widehat W_k^{(i)}$ and $\widehat W_k^{*(i)}$. This yields eight different criteria,
summarized in Table~\ref{table:criteria}. Again, we use the same names of the procedures as that in \cite{chen2021factor} and \cite{han2020} to represent 
the various constructions of $\widehat W$.
For the iterative procedures, we start with an initial 
rank estimate $r_k^{(0)}$, $k=1,\ldots,K$ and estimate the ranks through iteration until convergence. See remark below for setting the initial ranks and the stopping criteria. 

\begin{center}
  \begin{table}
    \begin{tabular}{c|c|c}
Estimation method used & IC & ER \\ \hline 
      non-iterative TOPUP &  
      $\widehat r_k ({\rm IC})= \widehat r_k^{(0)} ({\rm IC}) = {\rm IC}(\widehat W_k)$
      & $\widehat r_k ({\rm ER}) = \widehat r_k^{(0)} ({\rm ER}) = {\rm ER}(\widehat W_k)$ \\
non-iterative TIPUP & 
      $\widehat r_k^* ({\rm IC}) = \widehat r_k^{*(0)} ({\rm IC}) = {\rm IC}(\widehat W_k^*)$
      &
      $\widehat r_k^* ({\rm ER}) = \widehat r_k^{*(0)} ({\rm ER}) = {\rm ER}(\widehat W_k^*)$ \\
      $i$-th iteration of iTOPUP &
      $\widehat r_k^{(i)} ({\rm IC}) =  {\rm IC}(\widehat W_k^{(i)})$ &
      $\widehat r_k^{(i)} ({\rm ER}) =  {\rm ER}(\widehat W_k^{(i)})$ \\
      $i$-th iteration of iTIPUP &
      $\widehat r_k^{*(i)} ({\rm IC}) =  {\rm IC}(\widehat W_k^{*(i)})$ &
      $\widehat r_k^{*(i)} ({\rm ER}) =  {\rm ER}(\widehat W_k^{*(i)})$ \\ \hline
    \end{tabular}
    \caption{Estimation criteria}\label{table:criteria}
  \end{table}
\end{center}

\begin{rmk}
  In our theories, we fix $m^*$ in \eqref{estimate:IC} and \eqref{estimate:ER} as
  a finite constant. However, in practice, we may use, for example, $m^*=p/2$. We do not recommend to extend the search up to $p$, as the minimum eigenvalue is likely to be practically 0, especially when $T$ is small and $d_k$ is large.
\end{rmk}

\noindent {\bf The choice of the penalty function $G(\cdot)$ and $H(\cdot)$: \ }
Both criteria essentially try to distinguish the smallest (true)
non-zero eigenvalue from the true zero eigenvalue using noisy estimators
of the eigenvalues. Hence the penalty function is closely related to
the amount of error in the eigenvalue estimation and the strength of the smallest (true) non-zero eigenvalue.
We consider
the following penalty functions $G(\cdot)=g_k(d,T)$:  
\begin{align}
  g_{k,1}(d,T)& =\frac{h_0d^{2-2\nu}}{T} \log\left(\frac{dT}{d+T}\right), \quad  
  g_{k,2}(d,T)  = h_0d^{2-2\nu} \left(\frac{1}{T} +\frac{1}{d} \right) \log\left(\frac{dT}{d+T}\right), \nonumber \\
  g_{k,3}(d,T) & = \frac{h_0d^{2-2\nu}}{T} \log\left(\min\{d,T\}\right),  \quad
g_{k,4}(d,T)  = h_0d^{2-2\nu} \left(\frac{1}{T} +\frac{1}{d} \right) \log\left(\min\{d,T\}\right),
\nonumber \\
g_{k,5}(d,T) & = h_0d^{2-2\nu} \left(\frac{1}{T} +\frac{1}{d} \right) \log\left(\min\{d_k,T\}\right),
\label{penalty:IC}
\end{align}
where $d=\Pi_{k=1}^K d_k$ and $\nu$ is a tuning parameter. Ideally $\nu$ should be chosen to be the strength of the weakest factor (see Assumption \ref{asmp:strength} in Section \ref{section:asmp}),
though in practice we usually do not know its precise value. 
A more thorough discussion on this issue will be given later in Remark~\ref{rmk:penalty}.
Note that only $g_{k,5}$ involves $k$.

For the eigen-ratio criterion, we consider the following penalty function $H(\cdot)=h_{k}(d,T)$:
\begin{align}
  h_{k,1}(d,T)&=c_0h_0, \quad h_{k,2}(d,T)=\frac{h_0d^2}{T^2}, \quad
h_{k,3}(d,T)=\frac{h_0d^2}{T^2d_k^2} \nonumber \\
h_{k,4}(d,T)&=\frac{h_0d^2}{T^2d_k^2}+\frac{h_0d_k^2}{T^2}, \quad
h_{k,5}(d,T)=\frac{h_0d^2}{T^2d_k}+\frac{h_0dd_k}{T^2}. \label{penalty:ER}
\end{align}
where $c_0$ is a small constant, e.g. $c_0=0.1$. Note that
the penalty functions scale with $h_0$, because the strength of divergent eigenvalues
increases with $h_0$. Our theoretical analysis indicates that a better penalty function $H(\cdot)$
should also involve the strengths of the factors, similar to $G(\cdot)$. However, the function
$G(\cdot)$ has a much wider allowable range, and in most of the situations a simple constant function $h_{k,1}$ is sufficient. A more detailed discussion
will be given later in Remark \ref{rmk:penalty}. 


\vspace{0.1in}

\begin{figure}[htb!]
\centering
\subfigure[Eigenvalues of TOPUP realted methods for $k=1$]{
\begin{minipage}{0.28\linewidth}
\includegraphics[scale=0.21,page=1]{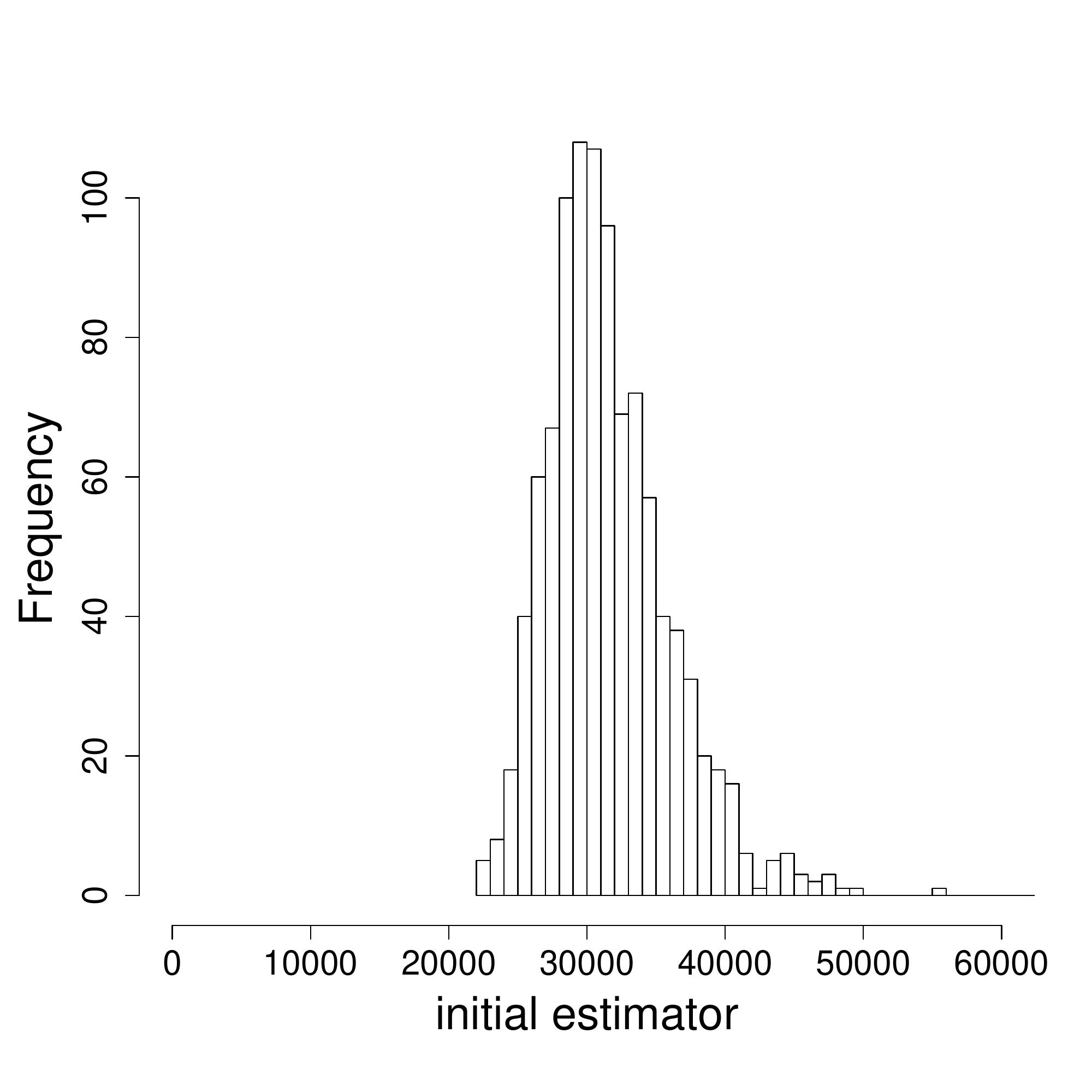}
\end{minipage}
\begin{minipage}{0.28\linewidth}
\includegraphics[scale=0.21,page=2]{eigenplot}
\end{minipage}
\begin{minipage}{0.28\linewidth}
\includegraphics[scale=0.21,page=3]{eigenplot}
\end{minipage}}
\subfigure[Eigenvalues of TIPUP related methods for $k=1$]{
\begin{minipage}{0.28\linewidth}
\includegraphics[scale=0.21,page=4]{eigenplot}
\end{minipage}
\begin{minipage}{0.28\linewidth}
\includegraphics[scale=0.21,page=5]{eigenplot}
\end{minipage}
\begin{minipage}{0.28\linewidth}
\includegraphics[scale=0.21,page=6]{eigenplot}
\end{minipage}}
\caption{Histogram of 6th largest estimated eigenvalue for the row factor ($k=1$) in Model M2 (see Section \ref{section:simulation}) with $T=200$ and $d_1=d_2=40$ for initial estimator, one step estimator and final estimator over 1000 replications. The true value is 0.} 
\label{figure:eigen_shrikage}
\end{figure}

\noindent
{\bf More considerations of the iterative procedure:}
The non-iterative procedures estimate $r_k$ ($k=1,\ldots,K$) individually. The accuracy of $\hat{r}_k$ does not depend on the accuracy of the estimation of the ranks in other directions. On the other hand, the iterative
procedures estimate all the ranks simultaneously, hence the accuracy of estimated rank in one direction depends on that in all other directions.
The iterative algorithm improves the estimation accuracy of the eigenvalues and the principal subspace because the projected tensor $\cZ_{k,t}^{(i)}$ in \eqref{eq:cZ} is of lower dimensional
than $\cX_t$. Figure \ref{figure:eigen_shrikage} numerically shows that the iterative
algorithms improve the accuracy of estimated true zero eigenvalues over the
non-iterative algorithms. In iterative algorithms, one would need to specify all the
ranks $\widehat{r}_k^{(i)}$ ($k=1,\ldots, K$) in each iteration.
Intuitively, an overestimated $\widehat{r}_k^{(i)}>r_k$ would still produce consistent
estimators, since the non-iterative procedure, using $\widehat{r}_k^{(i)}=d_k$ for all other directions,
is consistent. Our theoretical results shown
later confirm that, if $\widehat r_k^{(i-1)}$ used is larger than the true $r_k$, the iterative algorithm warrant the consistency of the IC and ER estimators at $i$-th iteration.
However, an
underestimated $\widehat{r}_k^{(i)}<r_k$ would potentially result in loss of
signal strength hence negatively impacting the estimation in other dimensions.
A precise quantification of the impact requires a more detailed investigation.
But the numerical studies show that for
iteration $i>1$, the performance is relative robust by using the order obtained by the IC or ER criteria. This is partially due to the theoretical justification that, for fixed ranks $r_k$, iterative algorithm only needs one iteration to achieve the ideal convergence rate for the estimation of the eigenvalues (see Theorem \ref{thm:eigen} later). 
In fact, if one has a priori information about a possible maximum (fixed) ranks of the core factor process, one could use such ranks in the iterative algorithms accordingly.

For iTOPUP procedure, we use
$\widehat{r}_k^{(i)}=\widehat r_k^{(i)}({\rm IC})$ or
$\widehat{r}_k^{(i)}=\widehat r_k^{(i)}({\rm ER})$ after the initial iteration $i\geq 1$. However, one needs
to use a more conservative estimator of the rank for the initial step
since $\widehat r_k^{(0)}({\rm IC})$ and $\widehat r_k^{(0)}({\rm ER})$ tend to be inaccurate.
We suggest to use
$\hat{r}_k^{(0)}=\min\{2\widehat r_k^{(0)}({\rm IC}), \widehat r_k^{(0)}({\rm IC})+3\}$ or
$\hat{r}_k^{(0)}=\min\{2\widehat r_k^{(0)}({\rm ER}), \widehat r_k^{(0)}({\rm ER})+3\}$ by default,
unless one has prior knowledge of the number of the factors. iTIPUP procedure is similar.
Although it is safer to use larger initial ranks $r_k^{(0)}$, it is often not necessary to be extremely
conservative, as the initial loss of signal strength of using a rank too small can be corrected later through
iterations. 
      
In the iterative algorithms, iteration is not stopped until the convergence of both the rank estimators and the loading space estimators. Theoretical properties in Section \ref{section:theories} only state consistency results in each iteration step. This stopping rule is mainly suggested by simulation study. In practice, we may stop the algorithm when the estimated number of factors in current iteration is the same as that in previous iteration.

\subsection{Some discussions}

\begin{rmk}
When the means of the factor processes deviate from zeros by a large margin, there is often one or several dominating factors corresponding to these non-zero 
means, as observed
by \cite{brown1989}, while the factors associated with the covariances become weak factors and are more difficult to identify. Assuming that the factor tensor process does not change its dimension after the deterministic means are removed, i.e. the factor tensor process is not a constant in any of its dimensions, then one should always demean the data in practice for the determination of the dimension of the factors, though not necessary for the estimation of the loading spaces, as shown by \cite{chen2021statistical} that aggregating 
the first and second moments of the data may improve the estimation 
accuracy of the factor loading matrices. 


\end{rmk}

\begin{rmk}
When some $r_k=1$, although the factor process has a reduced number of tensor modes, the proposed IC and ER methods should work well in identifying $r_k=1$ cases. If there is no factor structure ($r_1=...=r_K=0$), the proposed IC methods still can select the zero rank. But the ER methods need a slight modification, by constructing a new mock eigenvalue $\hat\lambda_{k,0}$ using the rate of the spiked eigenvalues.
In the current paper, we focus on the case that $r_k$ is fixed and $d_k$ diverges. If $r_k=d_k$ along one (or even all) dimension(s), the theoretical results in Section \ref{section:theories} can be extended. In this case we will need to modify the IC and ER methods using the developed convergence rates of zero eigenvalues. 
\end{rmk}

\begin{rmk}[\bf Improved penalization for IC approaches] \label{rmk:hallin}
The information criterion \eqref{estimate:IC} has the property, exploited by \cite{hallin2007} in the context of dynamic factor models, that a penalty function $G(\cdot)=g_k(d,T)$ leads to a consistent estimate of $r_k$ if and only if $cg_k(d,T)$ does, where $c$ is an arbitrary positive real number. Thus, multiplying the penalty by $c$ has no influence on the asymptotic performance of the identification method. However, for given finite $d$ and $T$, the value of a penalty function $g_k(d,T)$ satisfying \eqref{estimate:IC} can be arbitrarily small or arbitrarily large, and this indeterminacy can affect the actual result quite dramatically. The procedures in \cite{hallin2007} and \cite{alessi2010improved} can also be used in tensor factor model to robustify the IC approach with an empirically optimal choice of $c$. 

Specifically, following \cite{hallin2007} and \cite{alessi2010improved},  we generate a sequence of subsamples of sizes $(d_{1,j},...,d_{K,j}, T_j)$ with $j = 0, . . . , J$ such that $d_{k,0} = 0 < d_{k,1} < d_{k,2} < \cdots < d_{k,J} = d_k$ and $T_0 = 0 < T_1 \le T_2 \le \cdots \le T_J = T$, where $d_k$ is the original data dimension of tensor mode $k$ and $T$ is the original sample size, $1\le k\le K$. For any $j$, we obtain an estimated rank $\widehat r_{k,c,j}$ of $r_k$, which is a non-increasing function of $c$. Assume $r_k > 0$. 
The behavior of $\widehat r_{k,c,j}$, as a function of $j$, is different for different values of $c$. If $c>0$ and small, in practice, as $j$ increases, $\widehat r_{k,c,j}$ would increase to the maximum rank $m^*$ considered in \eqref{estimate:IC}, and we tends to overestimate $r_k$. On the other hand, when $c$ is very large, $\widehat r_{k,c,j}$ tends to zero for any $j$, and $r_k$ is underestimated. Due to the monotonicity of $\widehat r_{k,c,j}$ as a function of $c$, there must exist a range of ``moderate'' values of $c$ such that $\widehat r_{k,c,j}$ is a stable function of the subsample size $(d_{1,j},...,d_{K,j}, T_j)$. The stability can be measured by the empirical variance of $\widehat r_{k,c,j}$ as a function of $j$, 
\begin{align*}
S_{k,c}=\frac1J\sum_{j=1}^J \left(\widehat r_{k,c,j} -\frac1J\sum_{j=1}^J \widehat r_{k,c,j} \right)^2.    
\end{align*}
The optimal $c$ is then chosen to minimizes $S_{k,c}$.
\end{rmk}

\section{Assumptions and Asymptotic Properties} \label{section:theories}

\subsection{Assumptions and notation} \label{section:asmp}

We introduce some notations first. Let $d=\prod_{k=1}^K d_k$ and $d_{-k}=d/d_k$. For a matrix $A = (a_{ij})\in \RR^{m\times n}$, write the SVD as $A=U\Sigma V^\top$, where $\Sigma=\text{diag}(\sigma_1(A), \sigma_2(A), ..., \sigma_{\min\{m,n\}}(A))$, with the singular values $\sigma_1(A)\ge\sigma_2(A)\ge \cdots\ge \sigma_{\min\{m,n\}}(A)\ge 0$ in descending order. The matrix Frobenius norm can be denoted as $\|A\|_{\rm F} = (\sum_{ij} a_{ij}^2)^{1/2}=(\sum_{i=1}^{\min\{m,n\}}\sigma_i^2(A))^{1/2}$. Define the spectral norm
$$ \|A\|_{\rm 2} =  \max_{\|x\|_2=1,\|y\|_2= 1} \|x^\top A y\|_2=\sigma_1(A).$$ The tensor Hilbert Schmidt norm for a tensor $\cA\in\RR^{m_1\times m_2\times \cdots \times m_K}$ is defined as
$$ \|\cA\|_{{\rm HS}}=\sqrt{\sum_{i_1=1}^{m_1}\cdots\sum_{i_K=1}^{m_K}(\cA)_{i_1,...,i_K}^2 }. $$ 
Define the tensor operator norm for an order-4 tensor $\cA\in\RR^{m_1\times m_2\times m_3\times m_4}$,
$$ \| \cA\|_{\rm{op}} =\max\left\{ \sum_{i_1,i_2,i_3,i_4} u_{i_1,i_2} \cdot u_{i_3,i_4}\cdot (\cA)_{i_1,i_2,i_3,i_4}:\|U_1\|_{\rm F}=\|U_2\|_{\rm F}=1 \right\},$$
where $U_1=(u_{i_1,i_2})\in\RR^{m_1\times m_2}$ and $U_2=(u_{i_3,i_4})\in\RR^{m_3\times m_4}$. Define order-4 tensors
\begin{align*}
& \Theta_{k,h} =\sum_{t=h+1}^T \frac{\text{mat}_k(\cM_{t-h})\otimes \text{mat}_k(\cM_{t}) }{T-h} \in\RR^{d_k\times d_{-k}\times d_k \times d_{-k}}, \\
& \Phi_{k,h} =\sum_{t=h+1}^T \frac{\text{mat}_k(\cF_{t-h})\otimes \text{mat}_k(\cF_{t}) }{T-h} \in\RR^{r_k\times r_{-k}\times r_k \times r_{-k}},
\\ 
& \Phi^{\cano}_{k,h} =\sum_{t=h+1}^T \frac{\text{mat}_k(\cM_{t-h}\times_{k=1}^KU_k^\top)
\otimes \text{mat}_k(\cM_{t-h}\times_{k=1}^KU_k^\top)}{T-h}
\in\RR^{r_k\times r_{-k}\times r_k \times r_{-k}},
\end{align*}
where $\otimes$ is the tensor product and $U_k$ is from the SVD form of $A_k=U_k \Lambda_k V_k^\top$.
We view $\Phi^{\cano}_{k,h}$ as the canonical version of the auto-covariance of the factor process. Similarly define
\begin{align*}
& \Theta_{k,h}^* =\sum_{t=h+1}^T \frac{\text{mat}_k(\cM_{t-h}) \text{mat}_k^\top(\cM_{t}) }{T-h} \in\RR^{d_k\times d_k}, \\
& \Phi_{k,h}^* =\sum_{t=h+1}^T \frac{\text{mat}_k(\cF_{t-h}) \text{mat}_k^\top(\cF_{t}) }{T-h} \in\RR^{r_k\times r_k } , \\ 
& \Phi_{k,h}^{*\cano} = U_k^\top\Theta_{k,h}^*U_k
=\sum_{t=h+1}^T \frac{\text{mat}_k(\cM_{t-h}\times_{k=1}^KU_k^\top )
\text{mat}_k^\top(\cM_{t}\times_{k=1}^KU_k^\top ) }{T-h}
\in\RR^{r_k\times r_k}.
\end{align*}
Write $\Phi_{k,1:h_0}=(\Phi_{k,h}, h=1,\ldots,h_0)$ and $\Phi_{k,1:h_0}^*=(\Phi_{k,h}^*, h=1,\ldots,h_0)$.
Denote $\overline\E(\cdot)=\E(\cdot|\{ \cF_1,...,\cF_T\})$. Let $\tau_{k,m}$ be the $m$-th largest singular value of $\overline\E ({\text{TOPUP}}_k (\cX_{1:T}))$, $$\tau_{k,m}=\sigma_{m}(\overline\E ({\text{TOPUP}}_k (\cX_{1:T})))= \sigma_{m}\big({\text{mat}}_1(\Theta_{k,1:h_0})\big).$$ 
Similarly, let 
$$\tau_{k,m}^*=\sigma_{m}(\overline\E ({\text{TIPUP}}_k (\cX_{1:T})))= \sigma_{m}\big({\text{mat}}_1(\Theta_{k,1:h_0}^*)\big).$$ 
For simplicity, we write $U_k=U_{k,r_k}$ and $\widehat U_k=\widehat U_{k,r_k}$.

To facilitate consistency properties of the proposed procedures, we impose the following assumptions.

\begin{assumption}\label{asmp:error}
The error process $\cE_t$ are independent Gaussian tensors, condition on the factor process $\{\cF_t,t\in\mathbb Z\}$. In addition, there exists some constant $\sigma>0$, such that
\begin{equation*}
\overline\E (u^\top \text{vec}(\cE_t))^2\le \sigma^2 \|u\|_2^2, \quad u\in\RR^d.
\end{equation*}
\end{assumption}

\begin{assumption}\label{asmp:mixing}
Assume the factor process $\cF_t$ satisfies the strong $\alpha$-mixing condition such that
\begin{align}\label{cond1}
\alpha(h) \le \exp\left( - c_0 h^{\theta_1} \right)
\end{align}
for some constant $c_0>0$ and $0<\theta_1\le 1$, where
\begin{align*}
\alpha(h) = \sup_t\Big\{\Big|\P(A\cap B) - \P(A)\P(B)\Big|:
A\in \sigma(\cF_s, s\le t), B\in \sigma(\cF_s, s\ge t+h)\Big\}.
\end{align*}
\end{assumption}

\begin{assumption}\label{asmp:factor}
For any $u_k\in\R^{r_k}$ with $\| u_k \|_2=1$ and $1\le k\le K$,
\begin{align}\label{cond2}
\max_t\P\left( \left| \cF_t\times_1 u_1 \times_2 \cdots \times_K u_K \right| \ge x \right) \le c_1 \exp\left\{ -c_2x^{\theta_2} \right\},
\end{align}
where $c_1,c_2$ are some positive constants and $0<\theta_2\le 2$. 
\end{assumption}

\begin{assumption}\label{asmp:strength}
Assume $r_1,...,r_K$ are fixed. There exist some constants $\delta_0,\delta_1$ with $0\le \delta_0\le \delta_1\le 1$, such that $\|A_k\|_2 \asymp d_k^{(1-\delta_0)/2}$ and $\sigma_{r_k}(A_k)\asymp d_k^{(1-\delta_1)/2}$ for all $1\le k\le K$. 
\end{assumption}

\begin{assumption}\label{asmp:rank}
Assume that $h_0$ is fixed, and  \\
(a) (TOPUP related):  $\E[\text{mat}_1(\Phi_{k,1:h_0})]$ is of rank $r_k$ for $1\le k\le K$. \\
(b) (TIPUP related): $\E[\text{mat}_1(\Phi_{k,1:h_0}^{*\cano})]$ is of rank $r_k$ for $1\le k\le K$. 
\end{assumption}

Assumption \ref{asmp:error} is the same assumption used in \citet{chen2021factor} and \citet{han2020}. This assumption corresponds to the white noise assumption of \citet{lam2011,lam2012}. It allows substantial contemporaneous correlation among the entries of $\cE_t$. Note that the normality assumption, which ensures fast convergence rates in our analysis, is imposed for technical convenience. In fact we only need to impose the sub-Gaussian condition. Assumption \ref{asmp:mixing} allows a very general class of time series models, including causal ARMA processes with continuously distributed innovations; see also \cite{tong1990non, bradley2005, tsay2005analysis, fan2008nonlinear, rosenblatt2012markov, tsay2018nonlinear}, among others. The restriction $\theta_1\le 1$ is introduced only for presentation convenience. Assumption \ref{asmp:factor} requires that the tail probability of any orthonormal projection of $\cF_t$ decay exponentially fast. In particular, when $\theta_2=2$, $\cF_t$ is sub-Gaussian. 

Assumption \ref{asmp:strength} is similar to the signal strength condition of \citet{lam2012}, and the pervasive condition on the factor loadings (e.g., \citet{stock2002} and \citet{bai2003}). It plays a key role in identifying the common factors and idiosyncratic noises in \eqref{eq:tensorfactor}. Indices $\delta_0,\delta_1$ are measures of the strength of factors, or the rate of signal strength growth as the dimension $d_k$ grows. When $\delta_0=\delta_1=0$, the factors are called strong factors; otherwise, the factors are called weak factors. In particular, $\delta_0$ represents the strength of the strongest factors and $\delta_1$ the strength of the weakest factors. 

\begin{rmk}[Signal cancellation] \label{rmk:cancellation} 
Assumption \ref{asmp:rank} guarantees that there is no redundant tensor direction in $\cF_t$ when combined with $A_k$'s. It is related to certain signal cancellation phenomenon which is rare for TOPUP procedures but
may occur among TIPUP related procedures.
Consider the case of $k=1$ and $K=2$. We write the factor process in the canonical form as
$\cF_t^{\cano} = U_1^\top \cM_t U_2 = (f^*_{i,j,t})_{d_1\times d_2}$, and
$\phi^{\cano}_{i_1,j_1,i_2,j_2,h} = \sum_{t=h+1}^T f^*_{i_1,j_1,t-h}f^*_{i_2,j_2,t}/(T-h)$ as
the time average cross product between fibers $f^*_{i_1,j_1,1:T}$ and $f^*_{i_2,j_2,1:T}$ of the factor process (in canonical form). Then $\|\Theta_{1,h}\|_{\rm HS}^2 = \sum_{i_1,j_1,i_2,j_2}\big(\phi^{\cano}_{i_1,j_1,i_2,j_2,h}\big)^2$ and 
$\|\Theta_{1,h}^*\|_{\rm F}^2 =\| \Phi_k^{*\cano} \|_{\rm F}^2=\sum_{i_1,i_2}\big(\sum_{j=1}^{r_2} \phi^{\cano}_{i_1,j,i_2,j,h}\big)^2$. 
Note that the summation $\sum_{j=1}^{r_2} \phi^{\cano}_{i_1,j,i_2,j,h}$ is subject to potential cancellation among its terms for $h>0$. 
In the extreme cases, 
$\E[\text{mat}_1(\Phi_{k,1:h_0}^{*\cano})]$ may not have full rank $r_k$ and thus the signal strength $\tau_{k,r_k}^*$ can be much smaller than the order $d^{1-\delta_1}$. In Assumption \ref{asmp:rank}(b), we rule out the possibility of such severe signal cancellation. In practice, \citet{han2020} suggest to examine the patterns of the estimated singular values under different lag $h$ values. 
If there is no severe signal cancellation, we would expect that the pattern of $h_0^{-1/2}\tau_{k,r_k,h_0}$ would be similar to that of $h_0^{-1/2}\tau_{k,r_k,h_0}^*$ under different $h_0$. Here we emphasize that $\tau_{k,r_k,h_0}$ and $\tau_{k,r_k,h_0}^*$ depend on $h_0$, though 
in other places when $h_0$ is fixed we will omit $h_0$ in the notation. Severe signal cancellation would make the patterns different, since $h_0^{-1/2}\tau_{k,r_k,h_0}^*$ suffers 
signal cancellation but $h_0^{-1/2}\tau_{k,r_k,h_0}$ does not.
See the discussion in \citet{han2020}. On the other hand, Assumption~\ref{asmp:rank}(a) is sufficient to guarantee that $\E[\Theta_{k,1:h0}]$ and $A_k$ have the same rank $r_k$. 
\end{rmk}

Instead of Assumptions \ref{asmp:mixing} to 
\ref{asmp:rank}, 
\citet{chen2021factor} and \citet{han2020} imposed conditions on $\|\Theta_{k,0}\|_{\text{op}},\|\Theta_{k,0}^*\|_{\rm 2},\tau_{k,r_k}$ and $\tau_{k,r_k}^*$ in order to allow $r_k$ to increase with $d_k$. The following proposition establishes a connection between these two types of assumptions when the rank $r_k$ is fixed.

\begin{proposition}\label{prop:zt}
Suppose that Assumptions \ref{asmp:error} to
\ref{asmp:rank}
hold. Let $1/\vartheta=1/\theta_1+2/\theta_2$. Then, in an event $\Omega_0$ with probability at least $1-T\exp(-C_1 T^\vartheta)-\exp(-C_2T)$, 
\begin{align*}
\|\Theta_{k,0}\|_{\text{op}} \asymp \|\Theta_{k,0}^*\|_{\rm 2} \asymp d^{1-\delta_0} \quad \text{and }\ \ \tau_{k,r_k} \asymp \tau_{k,r_k}^* \asymp d^{1-\delta_1},    
\end{align*}
where $C_1,C_2>0$.
\end{proposition}

\subsection{Theoretical properties for IC and ER estimators} \label{section:properties}

In this section, we shall present theoretical properties of the IC and ER estimators using non-iterative TOPUP, non-iterative TIPUP, iTOPUP and iTIPUP. We first introduce some quantities related to the estimation errors of the estimated eigenvalues of the four different methods. 
For non-iterative TOPUP, define
\begin{align}
\beta_{k}&=(d^2d_k^{-1}+d^{3/2-3\delta_0/2+\delta_1}  + d_k^{1/2}d^{3/2-2\delta_0+\delta_1} +d_k d^{1-\delta_0})T^{-1},    \label{eq:topup:intial:beta}\\
\gamma_{k}&= d^{3/2-3\delta_0/2}T^{-1/2} +(d^{2-\delta_0} + d^2d_k^{-1} + d^{3/2-3\delta_0/2+\delta_1} + d_k^{1/2}d^{3/2-2\delta_0+\delta_1})T^{-1},  \label{eq:topup:intial:gamma}
\end{align}

For non-iterative TIPUP, define
\begin{align}
\beta_{k}^*&= (d^{1-\delta_0+\delta_1}+ d_k d^{1-2\delta_0+\delta_1})T^{-1},    \label{eq:tipup:intial:beta}\\
\gamma_{k}^*&= (d^{3/2-3\delta_0/2}+ d^{3/2-\delta_0}d_k^{-1/2})T^{-1/2} + (d^{1-\delta_0+\delta_1}+ 
d_k d^{1-2\delta_0+\delta_1}){T}^{-1} ,       \label{eq:tipup:intial:gamma}
\end{align}

Similarly, we use $\tilde\beta_{k}$, $\tilde\gamma_{k}$ and $\tilde\beta_{k}^*$, $\tilde\gamma_{k}^*$ for iTOPUP and iTIPUP, respectively, where
\begin{align}
\tilde\beta_{k}&=(d_kd^{1-\delta_0/2}+d_k d^{1-2\delta_0+\delta_1})T^{-1} ,    \label{eq:topup:iterative:beta}\\
\tilde\gamma_{k}&= d^{3/2-3\delta_0/2}T^{-1/2} + d_k d^{1-2\delta_0+\delta_1}T^{-1}, \label{eq:topup:iterative:gamma} \\     
\tilde\beta_{k}^*&= d_k d^{1-2\delta_0+\delta_1}T^{-1}   ,    \label{eq:tipup:iterative:beta}\\
\tilde\gamma_{k}^*&= d^{3/2-3\delta_0/2}T^{-1/2} + d_k d^{1-2\delta_0+\delta_1}T^{-1} .      \label{eq:tipup:iterative:gamma}
\end{align}
It is clear that $\beta_k$, $\gamma_k$, $\beta_k^*$ and $\gamma_k^*$ dominate
$\tilde\beta_k$, $\tilde\gamma_k$, $\tilde\beta_k^*$ and $\tilde\gamma_k^*$, respectively. Actually, $\beta_k$ and $\gamma_k$ (resp. $\beta_k^*, \gamma_k^*$, or $\tilde\beta_k, \tilde\gamma_k$, or $\tilde\beta_k^*, \tilde\gamma_k^*$) correspond to the $\beta_n$ and $\gamma_n$ sequence in Proposition \ref{prop:generic}.

We impose the following set of conditions to ensure that the estimation error of the divergent eigenvalue is much smaller than the true smallest non-zero eigenvalue, and the estimation error of the zero eigenvalues is relatively small. The $\gamma$'s above are part of the estimation errors of the divergent eigenvalues, and the $\beta$'s are the estimation errors of the (true) zero eigenvalues, for the four different estimation method. Note that $d^{2-2\delta_1}$ is the growth rate of the smallest non-zero eigenvalues corresponding to the weakest factors.
See also Theorem \ref{thm:eigen} below.

\begin{assumption}[Rate condition]\label{asmp:rate}\ \\
\vspace{-1.5em}
\begin{enumerate}
\item[(a)] $\max_{1\le k\le K}\left\{a_k\right\}=o(d^{2-2\delta_1})$ 
\item[(b)] $\max_{1\le k\le K}\left\{b_k\right\}=o(d^{2+2\delta_0-4\delta_1})$,
\end{enumerate}
The sequences $a_k$ and $b_k$ will be one of the $\gamma_k$ and $\beta_k$ sequences defined above, respectively, based on the estimators.
\end{assumption}

We will impose the following sufficient conditions on the penalty function $g_k(\cdot)$ and $h_k(\cdot)$. 
\begin{assumption}[Sufficient condition on the penalty functions]\label{asmp:penalty}\ \\
\vspace{-1.5em}
\begin{enumerate}

\item[(a)] $b_k\prec \min_k\{g_k(d,T)\} \le \max_k\{g_k(d,T)\} \prec d^{2-2\delta_1}$,
\item[(b)] $d^{2\delta_1-2}b_{k}^{2}\prec\!\!\prec
\min_{k}\{h_k(d,T)\})\leq \max_{k}\{h_k(d,T)\}\prec\!\!\prec d^{2+2\delta_0-4\delta_1}$,
\end{enumerate}
where $\varpi_n\prec\varrho_n$ indicates that there is a constant $C$ such that $\varpi_n<C\varrho_n$ uniformly,
and $\varpi_n\prec\!\!\prec\varrho_n$ indicates $\varpi_n=o(\varrho_n)$.
The sequence $b_k$ will be specified for different estimators.
\end{assumption}

\begin{rmk}[Penalty functions]\label{rmk:penalty}
The penalty functions $g_k$ and $h_k$ enter the consistency theorem below through Assumption \ref{asmp:penalty}. They do not have direct impact on the convergence rate of the rank estimators, as long as the condition is satisfied. Indirectly their choice interacts with the required sample size $T$ and dimension $d$. Roughly speaking, Assumption \ref{asmp:penalty}(a) dictates that the penalty $g_k(d,T)$ should be 
less than the smallest diverging
eigenvalue, but large enough to correctly truncate the estimated true zero eigenvalues. 
The $b_k$ sequence in the assumption is taken to be one of the $\beta_k, \beta_k^*,\tilde\beta_k$ and $\tilde\beta_k^*$ defined above, according to the procedure used.
In practice we generally do not know the the factor strengths $\delta_0$ and $\delta_1$, hence may not
always be able to specify a $g_k(d,T)$ that satisfies the condition. However, the range between the 
upper and lower bounds is quite wide in most of the cases, especially with the additional $T^{-1}$ term in the 
lower bound. All of the suggested $g_k(d,T)$ listed in \eqref{penalty:IC} satisfy the condition, if $\nu=\delta_1$.  Our experiments shows that setting $\nu=0$ in \eqref{penalty:IC} is sufficient 
in most of the cases. Only when the true $\delta_1$ is very large (extreme weak factors), the results become
sensitive to the selection of $\nu$. In such cases, a data driven procedure similar to that in \cite{hallin2007}
for vector factor models may be used to estimate $\delta_1$. Its property for tensor factor model may 
need further investigation. One can also study the pattern of the rank estimates under different $\nu$. 

The condition imposed on the penalty function $h_k(d,T)$ in Assumption~\ref{asmp:penalty}(b) is even weaker.
The upper bound goes to infinity 
but the lower bound goes to zero, except when 
both $\delta_0$ and $\delta_1$ are large, and $T$ is of smaller order than $d$. Hence in most of the cases 
a (small) constant function is sufficient. If $r_k$ is the true rank, the function $h_k(d,T)$ is designed to adjust the ratio of eigenvalues $\hat\lambda_{k,j+1}/\hat\lambda_{k,j}$, $j>r_k$ to be bounded below and to be around 1 (as we add $h_k(d,T)$ on both the numerator and denominator) so that they do not accidentally be smaller than  $\hat\lambda_{k,r_k+1}/\hat\lambda_{k,r_k}$ (a number that goes to 0). Hence intuitively we do not expect the impact of $h_k(d,T)$ to be large, which is confirmed by our empirical study.
The suggested functions in \eqref{penalty:ER} all satisfy the Assumption~\ref{asmp:penalty}(b), except the extreme weak factor cases, for which the ER estimators do not perform well under any penalty function. 

Assumption~\ref{asmp:penalty} only provides broard guidance {\it asymptotically}. There is no general unique 
optimal penalty function. 
Note that if $g_k(d,T)$ is an appropriate penalty function, then $cg_k(d,T)$ is appropriate as well asymptotically. 
The same property holds for the approaches of \citet{bai2002,bai2007}, \citet{amengual2007}, \citet{hallin2007} and \citet{li2017}. 
This creates potential problems in practice with given $d$ and $T$. Under certain circumstances, the empirical performance of IC estimators may heavily depend on the threshold function chosen among many alternatives; see the discussion in \citet{hallin2007}. 
\end{rmk}

The following is a set of different sample size conditions for different settings. They ensure sufficiently large sample size $T$ so that the non-iterative (true rank) factor loading space estimator based on TOPUP or TIPUP is consistent (for (a) and (b)), or has a relatively small error (for (c) and (d)).

\begin{assumption}[Condition on the sample size]\label{asmp:sample_size}\ \\
\vspace{-1.5em}
\begin{enumerate}
\item[(a)] $(d^{\delta_1-\delta_0/2} + d^{\delta_1}d_k^{-1/2})T^{-1/2}=o(1),  \ \ 1\le k\le K. $
\item[(b)] $(d_k^{1/2}d^{\delta_1-\delta_0/2-1/2} + d^{\delta_1-1/2})T^{-1/2} =o(1),  \ \ 1\le k\le K.$ 
\item[(c)] $(d^{\delta_1-\delta_0/2} + d^{\delta_1}d_k^{-1/2})T^{-1/2}+ d_k^{1/2}d^{\delta_1-1/2}T^{-1}\le C,  \ \ 1\le k\le K. $
\item[(d)] $(d_k^{1/2}d^{3\delta_1-5\delta_0/2-1/2} + d^{3\delta_1-2\delta_0-1/2})T^{-1/2}\le C,  \ \ 1\le k\le K.$ 
\end{enumerate}
\end{assumption}

\begin{center}
\begin{table}[H]
    \begin{tabular}{|cc||c|c|c|c|c|c|}\hline
\multicolumn{2}{|c||}{}     & \multicolumn{6}{c|}{Assumptions} \\ \hline
  IC Estimator & $\hat{r}_k$   & Model & Rank & Rate 1 & Rate 2 & Penalty $g_k(\cdot)$ & Size    \\ 
   & & (I)-(IV) & (V)  & (VI)(a) & (VI)(b) & (VII)(a) & (VIII) \\ \hline
        TOPUP & $\hat{r}(IC)$ & yes & (a) & $a_k=\gamma_k$ & - & $b_k=\beta_k$ & (a)  \\
        TIPUP & $\hat{r}^*(IC)$ & yes & (b) & $\gamma^*_k$ & - & $\beta_k^*$ & (b)  \\
        iTOPUP & $\hat{r}^{(i)}(IC)$ & yes & (a) & $\tilde\gamma_k$ & - & $\tilde\beta_k$ & (c) \\
        iTIPUP & $\hat{r}^{*(i)}(IC)$ & yes & (b) & $\tilde\gamma_k^*$ & - & $\tilde\beta_k^*$ & (d) \\ \hline \hline
    ER Estimator & & Model & Rank  & Rate 1 & Rate 2 & Penalty $h_k(\cdot)$ & Size   \\
     &  & (I)-(IV) & (V) & (VI)(a) & (VI)(b) & (VII)(b) & (VIII) \\ \hline
        TOPUP & $\hat{r}(ER)$ & yes & (a) & $a_k=\gamma_k$ & $b_k=\beta_k$ & $b_k=\beta_k$ & (a)  \\
        TIPUP & $\hat{r}^{*}(ER)$ &yes & (b) & $\gamma_k^*$ & $\beta_k^*$ & $\beta_k^*$ & (b)  \\
        iTOPUP & $\hat{r}^{(i)}(ER)$ &yes & (a) & $\tilde\gamma_k$ & $\tilde\beta_k$ & $\tilde\beta_k$ & (c)   \\
        iTIPUP & $\hat{r}^{*(i)}(ER)$ &yes & (b) & $\tilde\gamma_k^*$ & $\tilde\beta_k^*$ & $\tilde\beta_k^*$ & (d) \\
        \hline
    \end{tabular}
    \caption{Summary of conditions needed for Theorem~\ref{thm:all}}
    \label{tab:all_conditions}
\end{table}
\end{center}

Theorem \ref{thm:all} presents the asymptotic properties of the
IC and ER estimators in \eqref{estimate:IC} and \eqref{estimate:ER} based on non-iterative TOPUP, non-iterative TIPUP, iTOPUP and iTIPUP.

\begin{theorem}\label{thm:all} Let $1/\vartheta= 1/\theta_1+ 2/\theta_2$ as in Proposition~\ref{prop:zt}. For the various rank determination estimators, if their corresponding conditions listed in Table~\ref{tab:all_conditions} hold, then 
\[
\P(\widehat r_k=r_k, 1\le k\le K)\ge 1 - \sum_{k=1}^K e^{-d_k}-T\exp(-C_1 T^\vartheta)-\exp(-C_2T),
\]
with $C_1, C_2>0$.
\end{theorem}

In addition to the consistency of the rank estimators, we also have the following more detailed properties of the estimated eigenvalues.
Let $\hat\lambda_{k,j}$ be the eigenvalues of $\widehat W_k$ (defined in Section \ref{section:model}) such that $\hat\lambda_{k,1}\ge \hat\lambda_{k,2}\ge...\ge\hat\lambda_{k,d_k}$, $1\le k\le K$. Also let $\lambda_{k,j}$ be the eigenvalues of population version $\E\widehat W_k$ such that $\lambda_{k,1}\ge ... \ge \lambda_{k,r_k}> \lambda_{k,r_k+1}=...=\lambda_{k,d_k}=0$. Similarly, define $\hat\lambda_{k,j}^*$, $\lambda_{k,j}^*$, $\hat\lambda_{k,j}^{(i)}$, $\lambda_{k,j}^{(i)}$, $\hat\lambda_{k,j}^{*(i)}$ and $\lambda_{k,j}^{*(i)}$ as the eigenvalues of $\widehat W_k^*$, $\E\widehat W_k^*$, $\widehat W_k^{(i)}$, $\E\widehat W_k^{(i)}$, $\widehat W_k^{*(i)}$, $\E\widehat W_k^{*(i)}$, $i\ge 1$, respectively.

\begin{theorem}\label{thm:eigen}
Suppose the same conditions (\ref{asmp:error}-\ref{asmp:rank}, \ref{asmp:sample_size}) in Theorem \ref{thm:all} hold.  In an event with probability approaching 1 (as $T\to\infty$ and $d\to\infty$), the following holds. \\
(i). For estimating the true zero eigenvalues, we have, for $j\geq r_k$ and $i\ge 1$, 

\[
\hat\lambda_{k,j}=O(\beta_k), \quad \hat\lambda_{k,j}^*=O(\beta_k^*), \quad \hat\lambda_{k,j}^{(i)}=O(\tilde\beta_k), \mbox{\ and \ } \hat\lambda_{k,j}^{*(i)}=O(\tilde\beta_k^*) 
\]
(ii). For estimating the non-zero eigenvalues, we have, for all $1\le j\le r_k$ and $i\ge 1$,
\begin{align*}
|\hat\lambda_{k,j}-\lambda_{k,j}| &= O( T^{-1/2}d^{2+\delta_1-5\delta_0/2} + T^{-1/2}d_k^{-1/2}d^{2+\delta_1-2\delta_0}+\gamma_k)    ,\\
|\hat\lambda_{k,j}^*-\lambda_{k,j}^*| &= O( T^{-1/2}d_k^{1/2} d^{3/2+\delta_1-5\delta_0/2} + T^{-1/2}d^{3/2+\delta_1-2\delta_0}+\gamma_k^*)    ,\\
|\hat\lambda_{k,j}^{(i)}-\lambda_{k,j}^{(i)}| &= O( T^{-1/2}d_k^{1/2}d ^{3/2+\delta_1-5\delta_0/2} +\tilde\gamma_k)    ,\\
|\hat\lambda_{k,j}^{*(i)}-\lambda_{k,j}^{*(i)}| &= O( T^{-1/2}d_k^{1/2} d^{3/2+\delta_1-5\delta_0/2} +\tilde\gamma_k^*).
\end{align*}
\end{theorem}

\begin{rmk}[Strong factor cases] To illustrate the theorem, we consider the 
strong factor case $\delta_0=\delta_1=0$. Here the quantities \eqref{eq:topup:intial:beta}-\eqref{eq:tipup:iterative:gamma} can be simplified to
\begin{center}
\begin{tabular}{ll}
$\beta_k \asymp T^{-1}d_k^{-1}d^2+T^{-1}d_k^{1/2}d^{3/2}$,  & $\gamma_k\asymp T^{-1}d^2+T^{-1/2}d^{3/2}$, \\
 $\beta_k^* \asymp \tilde\beta_k \asymp \tilde\beta_k^* \asymp T^{-1}d_kd$, &  
$\gamma_k^* \asymp \tilde\gamma_k \asymp \tilde\gamma_k^* \asymp T^{-1}d_kd+T^{-1/2}d^{3/2}. $
\end{tabular}
\end{center}
The sample size conditions in Assumption \ref{asmp:sample_size} all reduces to $T\to\infty$. 
And the penalty function conditions (Assumption \ref{asmp:penalty}) is equivalent to (a) $b_k\prec g_k(d,T)\prec d^2$, (b) $d^{-2}b_k^2\prec\!\!\prec h_k(d,T)
\prec\!\!\prec d^2$, $1\le k\le K$. 
Thus, we shall expect similar performance for non-iterative TIPUP, iTOPUP and iTIPUP, but the non-iterative TOPUP may be worse.
\end{rmk}

\begin{rmk}[The vector factor models] \label{rmk:vector}
Theorem \ref{thm:all} and \ref{thm:eigen} hold for vector factor models by setting $K=1$ and $d=d_1$. In such a case, TOPUP is the same as TIPUP. More specifically, assuming all factors have the same strength ($\delta_0=\delta_1$), a common assumption used
in the literature, Theorem \ref{thm:eigen} reduces to $|\hat\lambda_{k,j}-\lambda_{k,j}|=O_{\P}(T^{-1/2}d^{2-3\delta_0/2})$ for $1\le j\le r_1$ and $\hat\lambda_{k,j}=O_{\P}(T^{-1}d^{2-\delta_0})$ for $j> r_1$ for the vector factor model case. This is the same as the convergence rate of the estimated eigenvalues derived in \cite{lam2012}, though our improved technical proof removed the restrictive conditions that $T=O(d)$ and all the non-zero eigenvalues are distinct. 
In addition, Theorem~\ref{thm:all} provides the rate of convergence of the rank estimators. 
\end{rmk}

\begin{rmk}
Note that our model setting is different from that used in \cite{bai2002} and \cite{hallin2007} where covariance matrix or spectral density matrix are used, instead of the auto-co-moment we use here. It is possible to extend our approach to identify the number of factors in these models, by setting $h_0=0$ in the construction of $\widehat{W}$ and using an extension of Proposition \ref{prop:generic} discussed in Remark~\ref{spike}. The main difference is that, in our model and with auto-co-moments, we are trying to separate non-zero and zero eigenvalues in 
the underlying $W$, while in approximate factor model and $h_0=0$, one would be trying to separate spiked and non-spiked eigenvalues. Hence a detailed analysis 
of the corresponding $\gamma_n$ and $\beta_n$ in Proposition \ref{prop:generic} will be needed.
\end{rmk}

\begin{rmk}[Sample size requirement comparison] \label{rmk:sample_size}  
The sample size required for the non-iterative estimators as shown in Assumptions~\ref{asmp:sample_size}(a,b)
is of higher order than that for the iterative estimators as in Assumptions~\ref{asmp:sample_size}(c,d). This is because, when the true ranks are used,
TOPUP and TIPUP require a larger sample size to consistently estimate the true loading spaces $A_k$
than the iTOPUP and iTIPUP procedures which require only a sufficiently ``good'' 
initial estimator of the loading space, but not necessarily a consistent one \citep{han2020}.
Similarly, the required sample size condition for TIPUP in Assumption \ref{asmp:sample_size}(b) is much weaker than that for TOPUP in Assumption \ref{asmp:sample_size}(a). In this regard, iterative procedures are better than
the non-iterative ones, and TIPUP based procedures are better than TOPUP based ones. 

\end{rmk}

\begin{rmk}[Convergence rate comparison] \label{rmk:iterative} 
The convergence rates of the estimated eigenvalues in the iterative methods 
are faster than that in the non-iterative methods, 
especially when there are weak factors in the model. Moreover, the rate of TIPUP related procedures is also faster than that of TOPUP related procedures. This can be seen by comparing $\beta_k,\gamma_k$, $\beta_k^*$, $\gamma_k^*$ with $\tilde\beta_k,\tilde\gamma_k$, $\tilde\beta_k^*$, $\tilde\gamma_k^*$.

For example, consider the case that all $d_k$ are of the same order and $K>1$. The following table shows the comparison of the convergence rate ($\beta$'s) of the estimated true zero eigenvalues, where $\prec$ and $\prec\!\!\prec$ are defined in Assumption \ref{asmp:penalty}.

\begin{table}[tb]
\begin{center}
\renewcommand\arraystretch{1.3}
\begin{tabular}{|l|l||l|}\hline
$\delta_0$ condition & $\delta_1$ condition & comparison \\ \hline\hline
\multirow{4}{*}{$\delta_0>1/K$}& $\delta_1\ge 3\delta_0/2$ & $\tilde\beta_k^*\asymp \tilde\beta_k \prec\!\!\prec \beta_k^* \prec\!\!\prec \beta_k$ \\ \cline{2-3}
& $\delta_0/2+1/K < \delta_1 < 3\delta_0/2$ & $\tilde\beta_k^* \prec\!\!\prec \tilde\beta_k \prec\!\!\prec \beta_k^* \prec\!\!\prec \beta_k$ \\ \cline{2-3}
& $\delta_1=\delta_0/2+1/K \ge \delta_0$ & $\tilde\beta_k^* \prec\!\!\prec \tilde\beta_k \asymp \beta_k^* \prec\!\!\prec \beta_k$ \\ \cline{2-3} 
& $\delta_0 \le \delta_1 < \delta_0/2+1/K$ & $\tilde\beta_k^* \prec\!\!\prec \beta_k^* \prec\!\!\prec \tilde\beta_k \prec\!\!\prec \beta_k$ \\ \hline
\multirow{2}{*}{$\delta_0\le 1/K$}& $\delta_1\ge 3\delta_0/2$ & $\tilde\beta_k^*\asymp \tilde\beta_k \asymp \beta_k^* \prec\!\!\prec \beta_k$ \\ \cline{2-3}
& $\delta_1 < 3\delta_0/2$ & $\tilde\beta_k^* \asymp \beta_k^* \prec\!\!\prec \tilde\beta_k  \prec\!\!\prec \beta_k$ \\ \hline
\end{tabular}
\end{center}
    \caption{Comparison of convergence rate for estimating the true zero eigenvalues}
    \label{tab:comparison_beta}
\end{table}

\begin{table}[tb]
\begin{center}
\renewcommand\arraystretch{1.3}
\begin{tabular}{|l|l||l|}\hline
$\delta_0$ condition & $\delta_1$ and $T$ condition & comparison \\ \hline\hline
\multirow{2}{*}{$\delta_0>1/K$}   &    $d^{1+2\delta_0-1/K} + d^{2\delta_1-\delta_0+1/K} \prec T$       &     $\tilde\gamma_k^*\asymp \tilde\gamma_k \asymp \gamma_k \prec\!\!\prec \gamma_k^*$ \\ \cline{2-3}
& $T \prec\!\!\prec d^{1+2\delta_0-1/K}$ or $T \prec\!\!\prec d^{2\delta_1-\delta_0+1/K} $       &      $\tilde\gamma_k^*\asymp \tilde\gamma_k \prec\!\!\prec \gamma_k^* \prec\!\!\prec \gamma_k$ \\ \hline
\multirow{2}{*}{$\delta_0\le 1/K$}  &   $d^{1+2\delta_0}+d^{2\delta_1-\delta_0+1/K} \prec T$    &      $\tilde\gamma_k^*\asymp \tilde\gamma_k \asymp \gamma_k^* \asymp \gamma_k$ \\ \cline{2-3}
& $T \prec\!\!\prec d^{1+2\delta_0}$ or $T \prec\!\!\prec d^{2\delta_1-\delta_0+1/K} $    &    $\tilde\gamma_k^*\asymp \tilde\gamma_k \asymp \gamma_k^* \prec\!\!\prec \gamma_k$ \\  \hline
\end{tabular}
\end{center}
    \caption{Comparison of the $\gamma$'s in the convergence rate for estimating the true non-zero eigenvalues,
   under Assumption \ref{asmp:sample_size}(a)-(d). }
    \label{tab:comparison_gamma}
\end{table}
From Tables~\ref{tab:comparison_beta},  it is clear that $\tilde\beta_k^*$ is always the smallest, and $\beta_k$ is the largest. Similarly, Table~\ref{tab:comparison_gamma} shows that $\tilde\gamma_k^*$ and $\tilde\gamma_k$ are always the smallest among these four $\gamma$'s.

For the IC estimators with a fixed penalty functions $g_k(d,T)$, a faster convergence rate of the 
eigenvalue estimators make the sufficient condition Assumption \ref{asmp:penalty}(a) easier to satisfy with 
smaller sample size $T$ and/or dimension $d_k$, $1\le k\le K$. Similarly, for the ER estimators, faster rates for estimating
the true zero eigenvalue increase the gap between the estimated divergent eigenvalues and the estimated true zero eigenvalue, leading to better performance of the ER estimators. 
\end{rmk}

Combining the discussion in Remarks~\ref{rmk:sample_size} and \ref{rmk:iterative}, we can conclude 
that in general the iterative procedues are better than the non-iterative ones and the TIPUP based 
procedures are better than the TOPUP ones, assuming no signal cancellation when TIPUP is used (see Remark~\ref{rmk:cancellation}).


\section{Simulation Study}\label{section:simulation}

In this section, we compare the empirical performance of the proposed methods and their variants under various simulation setups. We consider the identification of the number of factors based on 
the non-iterative TIPUP and TOPUP methods (denoted as initial estimators), the one step iterative methods (denoted as one-step estimators) and the iterative procedures after convergence (denoted as final estimator). 
In the iterative algorithm, at $i$-th iteration ($i>1$), we use the order obtained by the IC or ER criteria. Iteration is stopped when both the rank estimators and the loading space estimators converge. We also check the performance of different choices of penalty function $g_k(d,T)$ and $h_k(d,T)$. Specifically, we consider penalty functions \eqref{penalty:IC} and \eqref{penalty:ER}, and denote them as IC1-IC5, ER1-ER5, respectively. 
The empirical performance of IC1-IC5 (resp. ER1-ER5) are very similar, thus we only present IC2 and ER1 in this section. The detailed comparison are shown in Appendix B.

The simulation study consists of three parts. The first part is designed to
investigate the overall performance of our methods and their comparisons under models with different factor strength. As the strength of the weakest factors is unknown, we by default set $\nu=0$ for all the penalty function $g_k(d,T)$ in \eqref{penalty:IC}. In the second part, we investigate the case in which some factors have a dominantly strong explanatory power. The third part is the case in which we use $\nu=\delta_1$ in \eqref{penalty:IC} for all IC estimators, when some factors are weak.
For each case, we compute the proportion of correct identification of the rank of the factor processes 
or the root mean squared errors (RMSEs) of the rank estimates 
from 1000 simulated data sets. In Section \ref{section:simulation_tuning}, we study the selection of the optimal constant $c$ in IC criteria, using the method proposed in Remark \ref{rmk:hallin}.

The simulation uses the following matrix factor model:
\[
X_t=A_1 F_t A_2^\top+ E_t.
\]
Here, $E_t$ is white 
and is generated according to $E_t=\Psi_1^{1/2} Z_t\Psi_2^{1/2}$,
where $\Psi_1, ~\Psi_2$ are the column and row covariance matrices with
the diagonal elements being $1$ and all off diagonal elements being $0.2$. All of the elements in the $d_1\times d_2$ matrix $Z_t$ are i.i.d $N(0,1)$. This type of model of $E_t$ has been proposed and studied in the literature, see, for example \cite{Hoff2011,Hafner&2020,linton2020estimation}. The entries $f_{ijt}$ in the factor matrix $F_t$ were drawn from independent univariate AR(1) model $f_{ijt}=\phi_{ij}f_{ij(t-1)}+\epsilon_{ijt}$ with standard $N(0,1)$ innovation. 




\subsection{Part I: Determining strong and weak factors, using $\nu=0$ in $g_k(\cdot)$} 

In the first part, the following three models are studied:
\begin{enumerate}
\item[(M1).] Set $r_1=r_2=5$. The univariate $f_{ijt}$ follows AR(1) with AR coefficient $\phi_{ij}$, where
\begin{equation}\label{eq:phi.matrix}
(\phi_{ij})=\left(\begin{matrix}
0.8 & 0.5 & 0.5 & 0.3 & 0.3 \\
0.5 & 0.8 & 0.5 & 0.3 & 0.3 \\
0.3 & 0.5 & 0.8 & 0.5 & 0.3 \\
0.3 & 0.3 & 0.5 & 0.8 & 0.5 \\
0.3 & 0.3 & 0.5 & 0.5 & 0.8
\end{matrix}\right);
\end{equation}
All elements of $A_1$ and $A_2$ are i.i.d N(0,1). 
\item[(M2).] Set $r_1=r_2=5$. The univariate $f_{ijt}$ follows AR(1) with AR coefficient $\phi_{ij}$, where $\phi_{ij}$ is defined in \eqref{eq:phi.matrix}.
The elements of the first two columns of $A_1$ and $A_2$ are i.i.d N(0,1) 
and the elements of the last three columns of $A_1$ and $A_2$ are i.i.d $N(0.1) /d_1^{0.2}$ and $N(0.1) /d_2^{0.2}$, respectively. 
\item[(M3).] Same setting as in Model M2, except the elements of $A_1$ and $A_2$ are i.i.d $N(0,1) /d_1^{0.3}$ and $N(0.1) /d_2^{0.3}$. 
\end{enumerate}
All the factors in Models M1 are strong factors. Model M2 is the case in which four ($2\times 2$) factors are strong ($\delta_0=0$), twelve factors are weak factor with strength $0.2$ 
and the rest nine factors are weak factor with strength $\delta_1=0.4$. Model M3 is the case in which all the factors are very weak factors with strength $\delta_0=\delta_1=0.6$. Models M2 and M3 are designed to examine the effects of weak factors on the estimators. We choose a set of data dimensions to be $(d_1,d_2)=(20,20),(40,40),(80,80)$ and 
the sample size to be 
$T=100,300,500,1000$. Again, in this first part of simulation, we fix $h_0=1$ and set $\nu=0$ in the penalty function in \eqref{penalty:IC}, under the assumption that all factors are strong, even though 
some of factors simulated are weak (e.g., true $\delta_1=0.4$ in Model M2).

\begin{table}[htbp]
\centering
\caption{Proportion of correct identification of rank $r$ using IC2 and ER1 estimators based on both TOPUP and TIPUP procedures for Model M1, over 1000 replications}
\label{tab:r5.strong}
\scriptsize
{\begin{tabular}{ccccc|cccc|cccccc}\hline \hline
 &\multicolumn{4}{c}{initial estimator} & \multicolumn{4}{c}{one step estimator} & \multicolumn{4}{c}{final estimator}\\ \hline
 &\multicolumn{2}{c|}{IC} & \multicolumn{2}{c|}{ER} 
 &\multicolumn{2}{c|}{IC} & \multicolumn{2}{c|}{ER} 
 &\multicolumn{2}{c|}{IC} & \multicolumn{2}{c}{ER} \\ \hline
$T$ & TOP & TIP & TOP & TIP & TOP & TIP & TOP & TIP & TOP & TIP & TOP & TIP\\ \hline
&\multicolumn{5}{l}{ $d_1=d_2=20$}\\ \hline
100 & 0.10 & 1 & 1 & 1 & 0.75 & 1 & 1 & 1 & 0.85 & 1 & 1 & 1\\
300 & 0.99 & 1 & 1 & 1 & 1 & 1 & 1 & 1 & 1 & 1 & 1 & 1 \\
500 & 1 & 1 & 1 & 1 & 1 & 1 & 1 & 1 & 1 & 1 & 1 & 1 \\
1000 & 1 & 1 & 1 & 1 & 1 & 1 & 1 & 1 & 1 & 1 & 1 & 1 \\
\hline
&\multicolumn{5}{l}{ $d_1=d_2=40$}\\ \hline
100 & 0 & 1 & 1 & 1 & 0.47 & 1 & 1 & 1 & 0.48 & 1 & 1 & 1\\
300 & 0.03 & 1 & 1 & 1 & 0.98 & 1 & 1 & 1 & 1 & 1 & 1 & 1 \\
500 & 0.46 & 1 & 1 & 1 & 1 & 1 & 1 & 1 & 1 & 1 & 1 & 1 \\
1000 & 1 & 1 & 1 & 1 & 1 & 1 & 1 & 1 & 1 & 1 &  1 & 1 \\
\hline
&\multicolumn{5}{l}{ $d_1=d_2=80$}\\ \hline
100 & 0 & 1 & 1 & 1 & 0.66 & 1 & 1 & 1 & 0.98 & 1 & 1 & 1 \\
300 & 0 & 1 & 1 & 1 & 0.98 & 1 & 1 & 1 & 1 & 1 & 1 & 1 \\
500 & 0.03 & 1 & 1 & 1 & 1 & 1 & 1 & 1 & 1 & 1 & 1 & 1 \\
1000 & 0.70 & 1 & 1 & 1 & 1 & 1 & 1 & 1 & 1 & 1 & 1 & 1 \\
\hline
\end{tabular}}
\end{table}


For Model M1, the results in Table \ref{tab:r5.strong} show clearly that,
using TOPUP and IC, the initial estimator behaves very poorly even for large sample sizes. On the other hand, the one step estimator uniformly and significantly outperforms the non-iterative initial estimator.
In addition, the final estimator performs the best over all choices of $d_1$, $d_2$ and $T$. 
We also observe that the performance improves as the dimension increases, except that $d_1=40$ is not as good as $d_1=20$ when $T=100$. This improvement is due to the fact that when with strong factors ($\delta_0=\delta_1=0$), larger dimension $(d_1,d_2)$ provides more data points and information on the rank $r_k$.
With the same settings, IC criterion based on TIPUP determines the ranks perfectly, indicating that 
it is uniformly better than IC criterion based on TOPUP. 
This is partially due to the fact that non-iterative and iterative TIPUP procedures estimate the loading matrices and the eigenvalues more accurately than the corresponding TOPUP procedures. More interestingly, with the same setting, 
the ER estimators based on both TOPUP and TIPUP procedures perform perfectly. 

\begin{table}[htbp]
\centering
\caption{Proportion of correct identification of rank $r$ using IC2 and ER1 estimators based on both TOPUP and TIPUP procedures for Model M2, over 1000 replications}
\label{tab:r5.strongweak}
\scriptsize
{\begin{tabular}{ccccc|cccc|cccccc}\hline \hline
 &\multicolumn{4}{c}{initial estimator} & \multicolumn{4}{c}{one step estimator} & \multicolumn{4}{c}{final estimator}\\ \hline
 &\multicolumn{2}{c|}{IC} & \multicolumn{2}{c|}{ER} 
 &\multicolumn{2}{c|}{IC} & \multicolumn{2}{c|}{ER} 
 &\multicolumn{2}{c|}{IC} & \multicolumn{2}{c}{ER} \\ \hline
$T$ & TOP & TIP & TOP & TIP & TOP & TIP & TOP & TIP & TOP & TIP & TOP & TIP\\ \hline
&\multicolumn{5}{l}{ $d_1=d_2=20$}\\ \hline
100 & 0.80 & 0.01 & 0.29 & 1 & 0.73 & 0.01 & 0.68 & 1 & 0.60 & 0 & 0.79 & 1\\
300 & 0.73 & 0.35 & 0.01 & 1 & 0.66 & 0.34 & 0.74 & 1 & 0.66 & 0.20 & 0.78 & 1 \\
500 & 0.58 & 0.79 & 0 & 1 & 0.51 & 0.79 & 0.80 & 1 & 0.38 & 0.71 & 0.88 & 1 \\
1000 & 0.46 & 1 & 0 & 1 & 0.41 & 1 & 0.87 & 1 & 0.29 & 1 & 0.99 & 1 \\
\hline
&\multicolumn{5}{l}{ $d_1=d_2=40$}\\ \hline
100 & 0.95 & 0.01 & 0 & 0.99 & 0.90 & 0.01 & 0.74 & 1 & 0.87 & 0.01 & 0.81 & 1\\
300 & 0.96 & 0.64 & 0 & 1 & 0.93 & 0.64 & 0.42 & 1 & 0.90 & 0.64 & 0.48 & 1 \\
500 & 0.98 & 1 & 0 & 1 & 0.96 & 1 & 0.34 & 1 & 0.95 & 1 & 0.40 & 1 \\
1000 & 0.99 & 1 & 0 & 1 & 0.99 & 1 & 0.37 & 1 & 0.99 & 1 & 0.43 & 1 \\
\hline
&\multicolumn{5}{l}{ $d_1=d_2=80$}\\ \hline
100 & 0.57 & 0 & 0 & 0.97 & 0.45 & 0 & 0.77 & 1 & 0.35 & 0 & 0.84 & 1 \\
300 & 0.75 & 0.71 & 0 & 1 & 0.61 & 0.71 & 0.51 & 1 & 0.56 & 0.71 & 0.61 & 1 \\
500 & 0.90 & 1 & 0 & 1 & 0.84 & 1 & 0.50 & 1 & 0.81 & 1 & 0.59 & 1 \\
1000 & 1 & 1 & 0 & 1 & 1 & 1 & 0.61 & 1 & 1 & 1 & 0.71 & 1 \\
\hline
\end{tabular}}
\end{table}

For Model M2, Tables \ref{tab:r5.strongweak} reports 
the proportion of correct rank identification using IC and ER estimators based on TOPUP and TIPUP procedures. It is seen that, for small sample sizes ($T=100$ and $300$), the performance of IC estimators deteriorate when we use iterative procedures, which may indicate that the sample size $T$ and dimension $d$ do not meet the required Assumption \ref{asmp:penalty}(a) in Theorem \ref{thm:all}.
In addition, for $T=100$, the IC estimators using TIPUP procedures do not work at all, though they are better when $T=300$. This is due to the existence of weak factors and the fact that we use the default $\nu=0$ in \eqref{penalty:IC}. When the sample size is small, the estimators tend to identify the strong factors while miss the weak factors as their corresponding eigenvalues are relatively small and comparable to the penalty function $g_k(\cdot)$.
The IC estimators using TOPUP procedures performed much better for small sample sizes. The performance also becomes worse as $d_1$ and $d_2$ increases, also due to the existence of weak factors. The ER estimators based on TIPUP procedures show almost perfect accuracy, even there are weak factors in the model. We note that, even with weak factors, the performance remains almost the same with larger $(d_1,d_2)$. Again, accuracy improves by using the iterative procedure. The ER estimator based on TIPUP procedures are much better than all the other estimators for Model M2. 

\begin{table}[htbp]
\centering
\caption{Proportion of estimated rank pair ($\hat r_1$, $\hat r_2$) for the IC2 and ER1 estimators for Model M2 over 1000 replications, based on TOPUP and TIPUP procedures. $T=300$ and $(d_1,d_2)=(80,80)$. The true rank pair is (5,5).}
\label{tab:strongweak:pair}
\small
{\begin{tabular}{ccc|cc|cc}\hline \hline
 &\multicolumn{2}{c}{initial estimator} & \multicolumn{2}{c}{one step estimator} & \multicolumn{2}{c}{final estimator}\\ \hline
($\hat r_1$, $\hat r_2$) & IC2-TOP & IC2-TIP & IC2-TOP & IC2-TIP & IC2-TOP & IC2-TIP \\ \hline
(4,4) & 0 & 0 & 0.02 & 0 & 0.15 & 0\\
(4,5) & 0.01 & 0 & 0.02 & 0 & 0.02 & 0\\
(5,4) & 0.24 & 0.29 & 0.35 & 0.29 & 0.27 & 0.29\\
(5,5) & 0.75 & 0.71 & 0.61 & 0.71 & 0.56 & 0.71 \\
\hline
($\hat r_1$, $\hat r_2$) & ER1-TOP & ER1-TIP & ER1-TOP & ER1-TIP & ER1-TOP & ER1-TIP \\ \hline
(2,2) & 1 & 0 & 0.15 & 0& 0.19 & 0\\
(2,5) & 0 & 0 & 0.20 & 0& 0.13 & 0\\
(5,2) & 0 & 0 & 0.14 & 0 & 0.07 & 0\\
(5,5) & 0 & 1 & 0.51 & 1 & 0.61 & 1 \\
\hline
\end{tabular}}
\end{table}

Table \ref{tab:strongweak:pair} shows the more detailed identification results using 
IC2 and ER1 estimators for Model M2 over 1000 replications, based on TOPUP and TIPUP procedures. The sample size is
$T=300$ and the data dimension is $(d_1,d_2)=(80,80)$. The true rank pair is (5,5) for Model M3. From the table,
it is seen that the IC procedures tend to under-estimate the number of factors, with the correspoding iterative procedures perform
the worst. The ER estimators using the TOPUP procedures are likely to pick up only the strong factors, as the gap between strong factors ($\delta_1=0$) and weak factors ($\delta_1=0.4$) may be larger than that between weak factors and true zero eigenvalue estimations. We 
note that the outstanding performance of the ER estimators using the TIPUP procedures is quite different from 
the performance of a similar ER estimator in vector factor models under similar mixed strong and week factor 
cases (see e.g. \citet{lam2012}). The main reason is that the other tensor modes provide additional information and in certain sense serve as additional samples.
Then, for each $k\le K$, the signals of all divergent eigenvalues depend on $d$ instead of $d_k$, leading to larger gap between weak factors and true zero eigenvalue estimations.

For Model M3 with all very weak factors ($\delta_0=\delta_1=0.6$), Table \ref{tab:r5.weak} reports RMSEs of the ER1 estimators. The results of using IC estimators (not shown here) are significantly worse than that of the ER estimators due to the difficulty of IC estimators in dealing with weak factors when $\nu=0$ is used. It is seen from the table that ER estimators based on TIPUP procedure outperform that based on TOPUP procedure. We also see that the iterative algorithms improves the performance very significantly under this very weak factor case. 
The performance varies with the change of $(d_1,d_2)$ in a non-standard way, as the performance with $(40,40)$ seems to be better than that with $(20,20)$ and $(80,80)$. We note that in weak factor cases, a large $d_k$ will potentially reduce the accuracy of the estimator of $r_k$, since the signal level on the $k$-th dimension becomes weaker. On the other hand, a larger $d_i$ ($i\neq k$) in the other dimension potentially improves the estimation of $r_k$, since we have more ``repeated'' observations to be used for estimating $r_k$.


Table \ref{tab:weak:pair} shows the relative frequency of different estimated ranks of the ER1 estimator based on both TIPUP and TOPUP procedures, for the case of $T=300$ and $(d_1,d_2)=(80,80)$ under Model M3 with the true rank $(5,5)$. It is seen that the ER estimators tend to overestimate the number of factors, when all the factors are weak. All ER1-TOPUP estimators essentially identify $(6,6)$ as the rank. The non-iterative ER1 estimator based on TIPUP procedure can overestimate the ranks by a large margin.
However, the iterations can gradually correct the over-estimation.

\begin{table}[htbp]
\centering
\caption{Root mean squared errors (RMSEs) of the ER1 estimators based on TOPUP and TIPUP procedures for Model M3, averaging over 1000 replications}
\label{tab:r5.weak}
\small
{\begin{tabular}{ccc|cc|cccccc}\hline \hline
 &\multicolumn{2}{c}{initial estimator} & \multicolumn{2}{c}{one step estimator} & \multicolumn{2}{c}{final estimator}\\ \hline
$T$ & ER1-TOP & ER1-TIP & ER1-TOP & ER1-TIP & ER1-TOP & ER1-TIP\\ \hline
&\multicolumn{5}{l}{ $d_1=d_2=20$}\\ \hline
100 & 2.15 & 1.50 & 1.84 & 1.14 & 1.66 & 0.45 \\
300 & 1.17 & 0.60 & 1.07 & 0.43 & 0.86 & 0.12 \\
500 & 1.02 & 0.30 & 1.04 & 0.17 & 0.60 & 0.04 \\
1000 & 1.00 & 0.07 & 0.95 & 0.04 & 0.19 & 0.03 \\
\hline
&\multicolumn{5}{l}{ $d_1=d_2=40$}\\ \hline
100 & 1.01 & 1.19 & 1.03 & 0.86 & 1.02 & 0.35 \\
300 & 1.00 & 0.27 & 1.00 & 0.23 & 0.76 & 0.13 \\
500 & 1.00 & 0.17 & 1.00 & 0.14 & 0.49 & 0.04 \\
1000 & 1.00 & 0.04 & 0.99 & 0 & 0.11 & 0 \\
\hline
&\multicolumn{5}{l}{ $d_1=d_2=80$}\\ \hline
100 & 2.71 & 2.78 & 2.06 & 2.45 & 1.90 & 1.37 \\
300 & 1.00 & 0.99 & 1.01 & 0.65 & 0.95 & 0.40 \\
500 & 1.00 & 0.32 & 1.00 & 0.27 & 0.90 & 0.23 \\
1000 & 1.00 & 0.15 & 1.00 & 0.13 & 0.61 & 0.11 \\
\hline
\end{tabular}}
\end{table}

\begin{table}[htbp]
\centering
\caption{Proportion of the estimated rank pair ($\hat r_1$, $\hat r_2$) of the ER1 estimators for Model M3 over 1000 replications, when $T=300$ and $(d_1,d_2)=(80,80)$. The true rank pair is (5,5). }
\label{tab:weak:pair}
\small
{\begin{tabular}{ccc|cc|cc}\hline \hline
 &\multicolumn{2}{c}{initial estimator} & \multicolumn{2}{c}{one step estimator} & \multicolumn{2}{c}{final estimator}\\ \hline
($\hat r_1$, $\hat r_2$) & ER1-TOP & ER1-TIP & ER1-TOP & ER1-TIP & ER1-TOP & ER1-TIP \\ \hline
(5,5) & 0 & 0.51 & 0 & 0.72 & 0.07 & 0.83 \\
(6,5) & 0 & 0.07 & 0 & 0.05 & 0 & 0.05 \\
(6,6) & 1 & 0.09 & 0.98 & 0.11 & 0.84 & 0.11 \\
(7,5) & 0 & 0.15 & 0 & 0.06 & 0 & 0 \\
Others & 0 & 0.18 & 0.02 & 0.06 & 0.09 & 0.01 \\
\hline
\end{tabular}}
\end{table}

In summary, the first part of simulation shows that the ER estimators and IC estimator based on TIPUP procedures perform very well when all the factors are strong. The ER estimators significantly outperform the IC estimators when some or all factors are weak. Different from the results shown in \citet{lam2012} for vector factor models with both strong and weak factors, in tensor factor models, the ER estimators (based on TIPUP procedure) are able to determine the correct number of factors in many cases. The results also show that the iterative procedure significantly improves the performance except the IC estimators based on TOPUP in Model M2, which may due to that the sample size $T$ and dimension $d$ do not meet the required Assumption \ref{asmp:penalty}(a) in Theorem \ref{thm:all}. It also shows that the estimators based on TIPUP perform better than that based on TOPUP in general. Hence, when some factors are weak, the 
iterative ER estimators based on TIPUP are the choice. 

\subsection{Part II: The case of dominating strong factors}
The second part of our simulation examines the effects of dominate strong factors on the IC and ER estimators. The data are generated from the following model,
\begin{enumerate}
\item[(M4).] Set $r_1=r_2=2$. The univariate $f_{ijt}$ follows AR(1) with AR coefficient $\phi_{11}=0.98$ and $\phi_{12}=\phi_{21}=\phi_{22}=0.15$; The elements of the loading matrices $A_1$ and $A_2$ are i.i.d $N(0,1)$. 
\end{enumerate}
We fix $d_1=d_2=40$ and $T = 200$. Again, we use IC2 and ER1 for demonstration, and assume $\nu=0$ in
the penalty function in \eqref{penalty:IC}. Although all of the four factors are strong factors, the strongly
imbalanced signal strength in $\cF_t$ makes one of factors dominating the others in explanatory power. Table \ref{tab:dom.strong} reports the relative frequencies of estimated rank pairs over 1000 replications. It is seem that the ER estimators are very likely to pick up only the dominate factor, although iterations significantly improves the accuracy. The IC estimators performs much better in this case. And
over all, estimators based on TIPUP perform better than the corresponding estimators using TOPUP. 
Overall, IC-TIPUP performs the best in this case. 

\begin{table}[htbp]
\centering
\caption{Proportion of estimated rank pair ($\hat r_1$, $\hat r_2$) for the IC2 and ER1 estimators for Model M4 over 1000 replications, based on TOPUP and TIPUP procedures, $T=200$ and $(d_1,d_2)=(40,40)$. The true rank pair is (2,2).}
\label{tab:dom.strong}
\small
{\begin{tabular}{ccc|cc|cccc}\hline \hline
 &\multicolumn{2}{c}{initial estimator} & \multicolumn{2}{c}{one step estimator} & \multicolumn{2}{c}{final estimator}\\ \hline
($\hat r_1$, $\hat r_2$) & IC2-TOP & IC2-TIP & IC2-TOP & IC2-TIP & IC2-TOP & IC2-TIP \\ \hline
(1,1) & 0 & 0.02 & 0 & 0.02 & 0 & 0.04\\
(1,2) & 0.05 & 0.08 & 0.02 & 0.08 & 0 & 0.06\\
(2,1) & 0.01 & 0.05 & 0.01 & 0.05 & 0 & 0.03 \\
(2,2) & 0.22 & 0.85 & 0.31 & 0.85 & 0.46 & 0.87 \\
(3,3) & 0.54 & 0 & 0.54 & 0 & 0.47 & 0 \\
Others & 0.18 & 0 & 0.12 & 0 & 0.07 & 0 \\
\hline
($\hat r_1$, $\hat r_2$) & ER1-TOP & ER1-TIP & ER1-TOP & ER1-TIP & ER1-TOP & ER1-TIP \\ \hline
(1,1) & 1 & 0.878 & 0.88 & 0.26 & 0.53 & 0.39\\
(1,2) & 0 & 0.065 & 0 & 0.17 & 0.33 & 0.09\\
(2,1) & 0 & 0.017 & 0.04 & 0.14 & 0.06 & 0.06\\
(2,2) & 0 & 0.040 & 0.08 & 0.43 & 0.08 & 0.46\\
\hline
\end{tabular}}
\end{table}

\subsection{Part III: Using the correct penalty function in the IC estimators}

The third part of the simulation considers the impact of penalty function selection. We consider Model M2 and M3 again, with $d_1=d_2=40$, but use the true weakest factor strength $\delta_1$ as $\nu$ in the penalty function $g_k(T,d)$ in \eqref{penalty:IC} instead assuming strong factor and use $\nu=0$ as in Section 5.1. To compare with Tables \ref{tab:r5.strongweak} for M2 and \ref{tab:r5.weak} for M3, we report the proportion of correct rank identification using IC2 estimators in Table \ref{tab:r5.strongweak:delta1} for M2 and RMSEs of the IC2 estimators in Table \ref{tab:r5.weak.delta1} for M3.

Comparing Tables \ref{tab:r5.strongweak} and \ref{tab:r5.strongweak:delta1}, it is seen that the performance of 
IC2 estimators using TIPUP improve by using the correct $\nu$ in the penalty function. 
We notice that 
the initial and one step IC estimators using TOPUP with the correct $\nu$ in the penalty term 
actually under-perform the ones with the incorrect $\nu$. The reason for this unusual behavior of TOPUP is 
unclear.
It might be that the magnitude of the estimation of true zero eigenvalue is much larger in this mixed 
weak and strong factor case, as shown in Figure \ref{figure:eigen_shrikage}. Hence, reducing the penalty 
term by using $\nu=\delta_1=0.4$ resulting in severe over-estimation. However, the TIPUP estimators estimate
the true zero eigenvalues more accurately, hence are able to take advantage of the more accurate
penalty term.

For Model M3 with all weak factors ($\delta_0=\delta_1=0.6$), all IC estimators using the 
wrong penalty function $g_k(T,d)$ with $\nu=0$ in \eqref{penalty:IC} identified $(1,1)$ rank pair
in all 1000 simulations for all sample sizes. Comparing it with the result shown in Table~\ref{tab:r5.weak.delta1}, the importance of using the right $\nu$ in the penalty function 
is obvious in this all-weak factor case. Table~\ref{tab:r5.weak.delta1} also shows that 
IC estimator using TIPUP outperforms that using TOPUP, when the right penalty function is used. Comparing Tables \ref{tab:r5.weak} and \ref{tab:r5.weak.delta1}, the iterative IC estimators out-perform the ER estimators when $T=300,500,1000$. 
It shows the great potential of IC estimators using a proper $\nu$ in the penalty function
under the weak factor cases. As mentioned earlier, more investigation is needed to determine the proper value for $\nu$.

\begin{table}[htbp]
\centering
\caption{Proportion of correct identification of rank $r$ 
using IC2 estimators based on TOPUP and TIPUP procedures for Model M2, over 1000 replications, when $\delta_1$ is given and $(d_1,d_2)=(40,40)$}
\label{tab:r5.strongweak:delta1}
\small
{\begin{tabular}{ccc|cc|cccc}\hline \hline
 &\multicolumn{2}{c}{initial estimator} & \multicolumn{2}{c}{one step estimator} & \multicolumn{2}{c}{final estimator}\\ \hline
$T$ & IC2-TOP & IC2-TIP & IC2-TOP & IC2-TIP & IC2-TOP & IC2-TIP \\ \hline
100 & 0 & 1 & 0.01 & 1 & 0.65 & 1 \\
300 & 0 & 1 & 0.29 & 1 & 0.90 & 1 \\
500 & 0 & 1 & 0.62 & 1 & 0.96 & 1 \\
1000 & 0.43 & 1 & 0.95 & 1 & 1 & 1 \\
\hline
\end{tabular}}
\end{table}

\begin{table}[htbp]
\centering
\caption{Root mean squared errors (RMSEs) of the IC estimators based on TOPUP and TIPUP procedures for Model M3, averaging over 1000 replications, when $\delta_1$ is given and $(d_1,d_2)=(40,40)$}
\label{tab:r5.weak.delta1}
\small
{\begin{tabular}{ccc|cc|cccc}\hline \hline
& &\multicolumn{2}{c}{initial estimator} & \multicolumn{2}{c}{one step estimator} & \multicolumn{2}{c}{final estimator}\\ \hline
$T$& & IC2-TOP & IC2-TIP & IC2-TOP & IC2-TIP & IC2-TOP & IC2-TIP \\ \hline
100 && 1.01 & 0.91 & 0.91 & 0.92 & 0.98 & 0.92   \\
300 && 1.00 & 0.03 & 0.87 & 0.03 & 0.23 & 0.03  \\
500 && 1.00 & 0 & 0.43 & 0 & 0 & 0   \\
1000 && 1.00 & 0 & 0.05 & 0 & 0 & 0  \\ \hline
\end{tabular}}
\end{table}

\subsection{Selection of the optimal $c$ in IC criteria}\label{section:simulation_tuning}

\begin{figure}[htb!]
\centering
\subfigure[Tensor mode $k=1$]{
\includegraphics[scale=0.4]{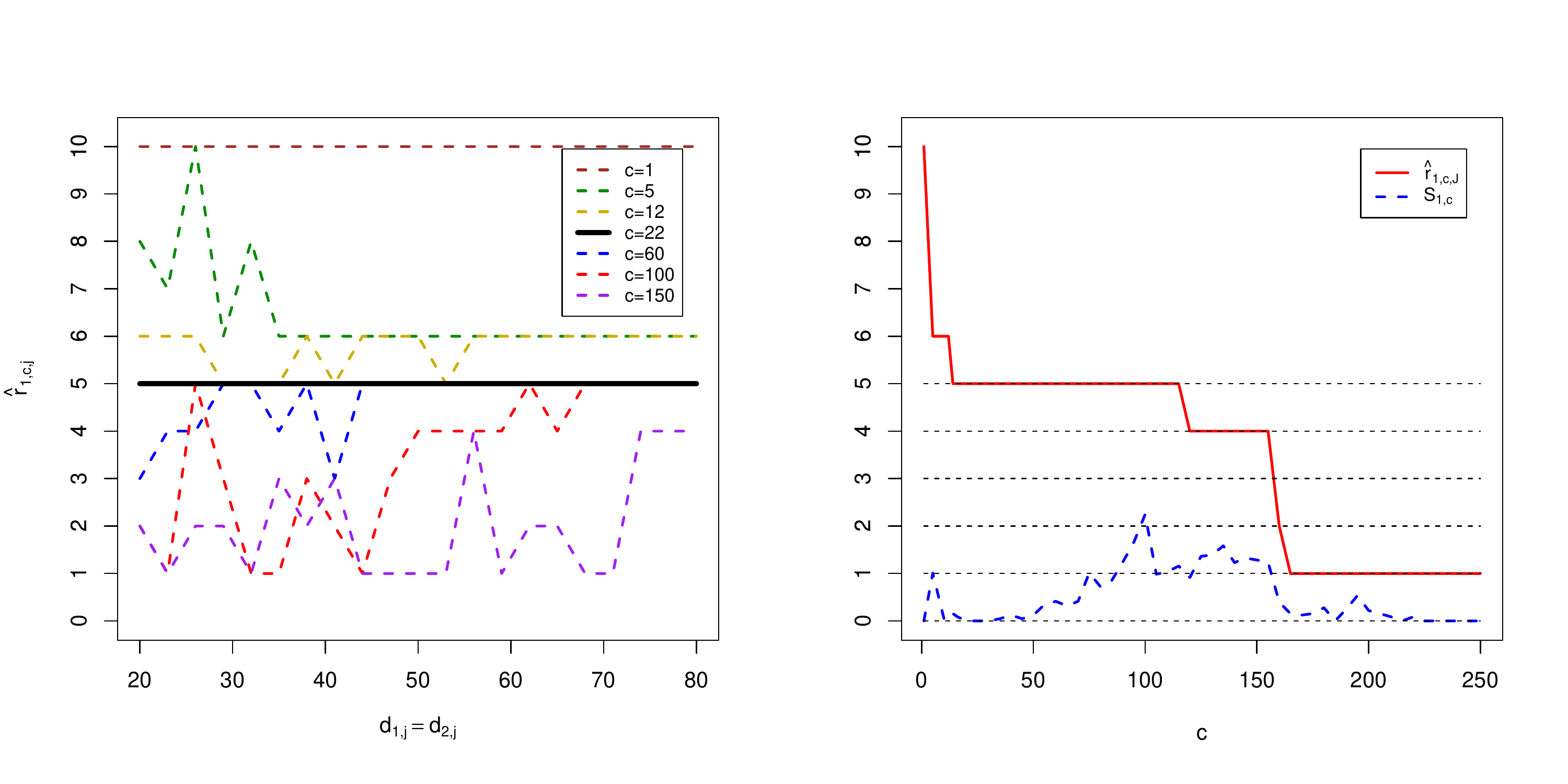}}
\subfigure[Tensor mode $k=2$]{
\includegraphics[scale=0.4]{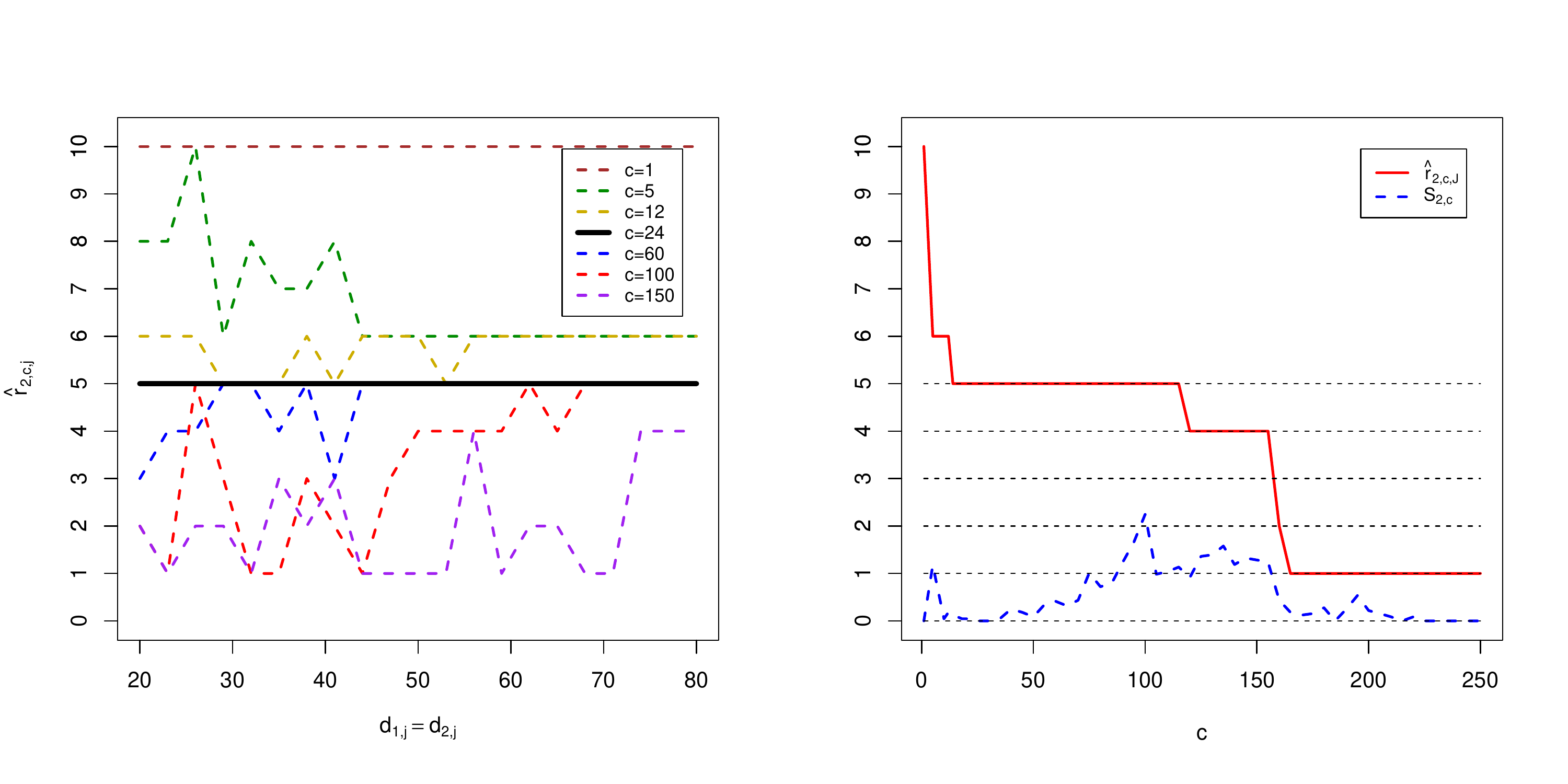}}
\caption{Robustified IC criterion using iTOPUP related methods} 
\label{figure:hallin_method}
\end{figure}

To study the empirical property of the optimal constant
$c$ discussed in Remark \ref{rmk:hallin},
we simulate data from Model M1
with $r_1=r_2=5$, and set $d_1=d_2=80$, $T=300$ and set the upper bound of the rank as $m^*=10$.
In Figure \ref{figure:hallin_method}, we show respectively the behavior of $\widehat r_{k,c,j}$ as a function of $(d_{1,j},d_{2,j})$, and of $\widehat r_{k,c,J}$ and of $S_{k,c}$ as functions of $c$, when setting $T_1=T_2=\cdots=T_J=T$ and $d_{1,j}=d_{2,j}$. 
The rank $r_k$ of the factor process can be determined by considering the mapping $c \to S_{k,c}$ and choosing $\widehat r_{k,c,J}=\widehat r_{k,\widehat c,J}$, where $\widehat c$ belongs to an interval of $c$ implying $S_{k,c} \approx 0$ and therefore the value of $\widehat r_{k,c,J}$ is a constant function of $c$. Similar to \cite{hallin2007} and \cite{alessi2010improved}, we see that the second stability interval always delivers an estimated number $\widehat r_{k,c,J}$ which is closer to the true $r_k = 5$ than the number suggested by the other intervals. That is, the smallest values of $c$ for which $\widehat r_{k,c,j}$ is also close to a constant function of $j$, $j\le J$. Note that the first stability interval always corresponds to the predefined upper bound $m^*$ and it is thus a non-admissible solution.

\section{Real Data Analysis} \label{section:data}
In this section, we illustrate the proposed procedures using the Fama–French 10 by 10 monthly return series
as an example. According to ten levels of market capital (size) and ten levels of book to equity ratio (BE), stocks are grouped into 100 portfolios. The sampling period used in this excises is from January 1964 to December 2015 for a total of 624 months. 
There are overall 62,400 observed monthly returns used in this analysis. The data is from \\
\centerline{http://mba.tuck.dartmouth.edu/pages/faculty/ken.french/data\_library.html.}  \\
Similar to \citet{wang2019}, we subtract from 
each of the series their corresponding monthly excess market return. 

The data was used by \citet{wang2019} to demonstrate the estimation of a matrix factor model using a non-iterative TOPUP procedure. They used an estimator similar to our non-iterative TOPUP based ER estimator without a penalty term to estimate the number of factors and found the estimator suggested $(1,1)$ as the rank -- a single factor in the model. In the end 
they demonstrate the model using $(2,2)$ as the ranks for demonstration of the estimation procedures.

\begin{figure}[htb!]
\centering
\subfigure[TIPUP]{
\includegraphics[scale=0.54,page=1]{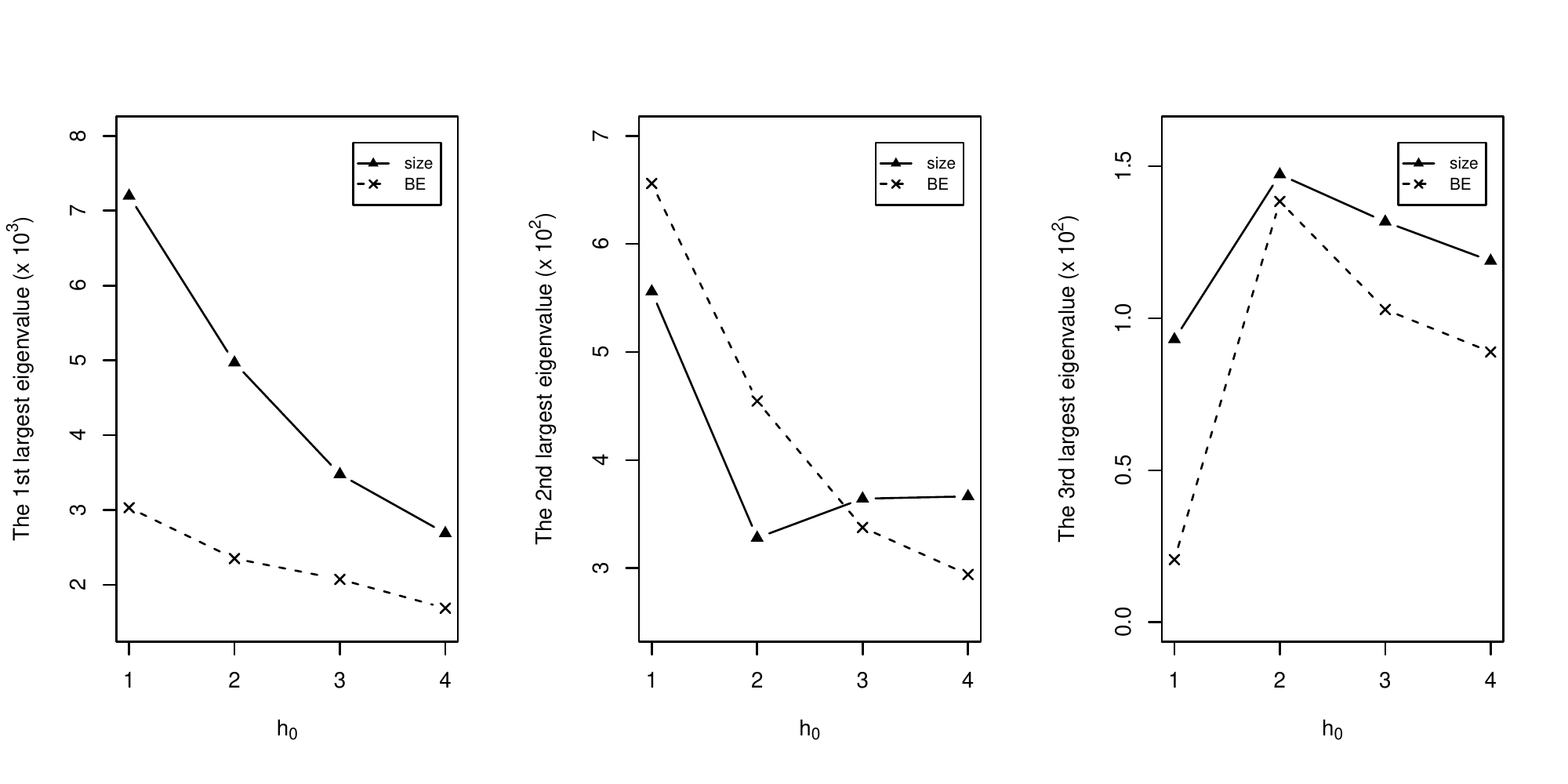}}
\subfigure[TOPUP]{
\includegraphics[scale=0.54,page=2]{eigen_finance}}
\caption{The 1st to 3rd largest estimated eigenvalues using the non-iterative TIPUP and TOPUP procedures, $h_0^{-1}\hat\tau_{k,m_k}^{*2}$ and $h_0^{-1}\hat\tau_{k,m_k}^{2}$ ($m_k\le 3,k\le 2$), under different maximum 
lag values $h_0$. The line marked as size is for the row factors, corresponding to the levels of market capital (size), and the line marked as BE is for the column factors, corresponding to the levels of book to equity ratio (BE)}
\label{figure:eigen_fama_french}
\end{figure}

Figure \ref{figure:eigen_fama_french} shows the 1st to 3rd largest eigenvalues of using the non-iterative TIPUP and TOPUP procedures, $h_0^{-1}\hat\tau_{k,m_k}^{*2}$ and $h_0^{-1}\hat\tau_{k,m_k}^{2}$ ($m_k\le 3$), under different lag values $h_0$. The pattern of eigenvalues using the iterative TIPUP and TOPUP are similar, thus is omitted. It can be seen from panel (a) of Figure \ref{figure:eigen_fama_french} that, using TIPUP procedure, the 1st and 2nd largest eigenvalues reach their maximum value at $h_0=1$, and tends to decrease as $h_0$ increases. However, the 3nd largest eigenvalue reaches its maximum at $h_0=2$. In contrast, from panel (b) of Figure \ref{figure:eigen_fama_french}, using TOPUP
procedure, the 1st to 3nd largest eigenvalues reach the maximum at $h_0=1$. The difference of the patterns of estimated singular values indicates possible severe signal cancellation when using $h_0=1$, according to the suggestions in \cite{han2020}. Hence, 
we choose $h_0=2$, and consider IC and ER estimators based on the iterative TIPUP procedure. 

\begin{figure}[htbp!]
\centering
\includegraphics[scale=0.33,page=1]{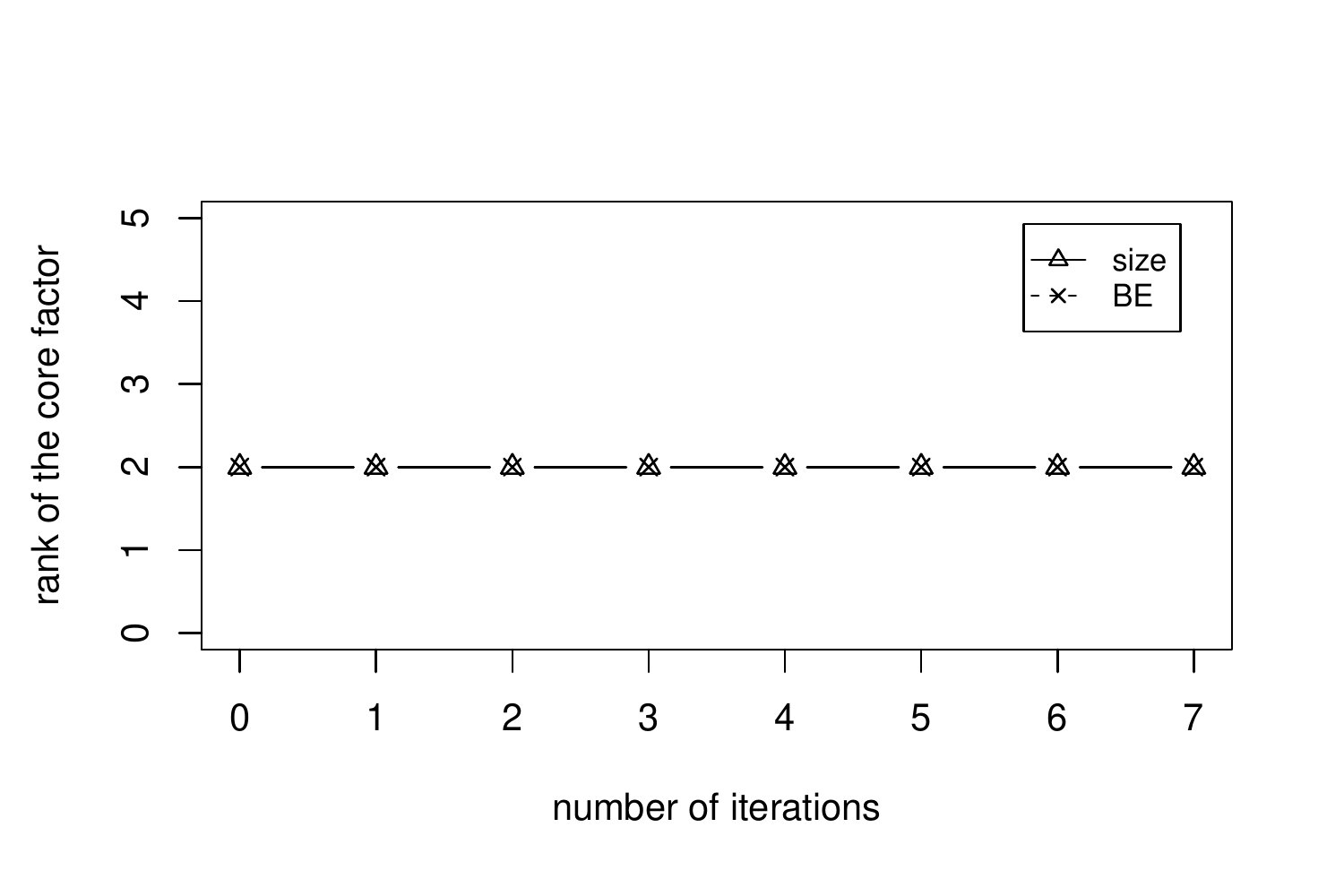} \ \ 
\includegraphics[scale=0.33,page=2]{factor_num_finance}
\caption{Estimated rank $\hat r_k$ of the core factor process against the number of iterations for the Fama-French 10 by 10 series, using IC2 (left) and ER1 (right) estimator based on TIPUP procedure.}
\label{figure:factor_num_finance}
\end{figure}

Figure \ref{figure:factor_num_finance} shows the estimated rank $\hat r_k$ of the core factor process 
with different number of iterations, using IC2(TIPUP) with $\nu=0$ (left figure) and ER1(TIPUP) (right figure).
It is seen that the iterative algorithms converge very quickly.

\begin{table}[th]
    \centering
    \begin{tabular}{c|cccccc}
    $\nu$                & 0 & 0.05 & 0.1 & 0.15 & 0.2 & 0.25 \\ \hline
IC1, IC3         &   (3,3) & (4,3) & (4,3) & (4,4) & (4,4) & (4,4) \\
IC2, IC4, IC5    &  (2,2) & (2,2) & (2,2) & (2,2) & (3,3) & (3,3) 
    \end{tabular}
    \caption{Estimated ranks using different IC criteria and $\nu$ for the Fama-French return series}
    \label{tab:IC_out}
\end{table}

Table~\ref{tab:IC_out} shows the estimated rank pairs using different IC(TIPUP) estimators and different $\nu$ 
parameter for the penalty function. It is seem that IC1 and IC3 tend to select larger models. 
These rank estimates do not change when we use $h_0=3$ and $4$. 

\begin{figure}[htbp]
\centering
\includegraphics[scale=0.6,page=1]{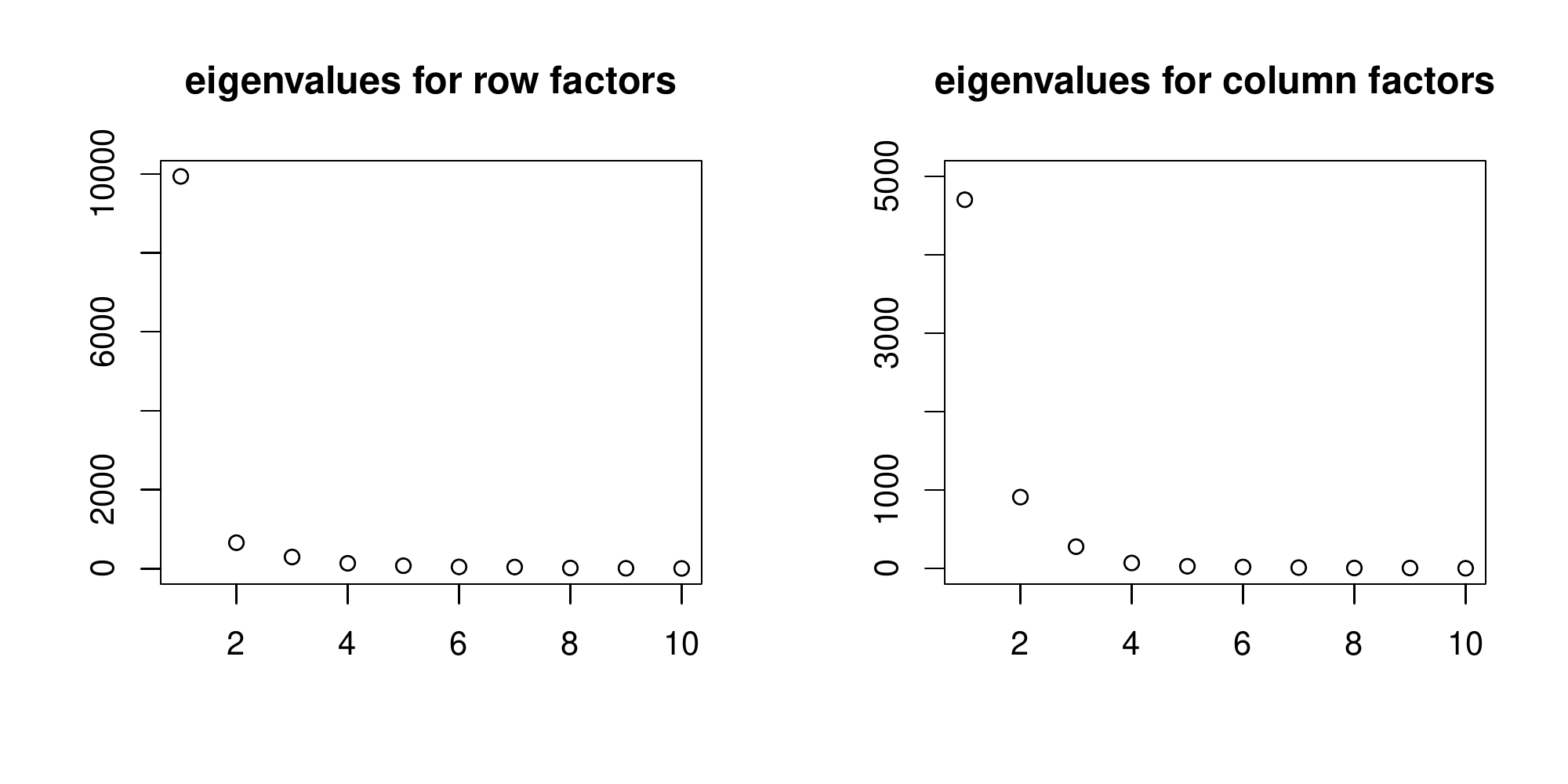}
\caption{The eigenvalues $\tau_{k,m_k}^{*2}$, $1\le m_k\le d_k$, for each $1\le k\le 2$, using initial TIPUP procedure}
\label{figure:eigengap_finance}
\end{figure}

On the other hand, ER1-ER5 in \eqref{penalty:ER} produce exactly the same rank estimate $(1,3)$ using $h_0=2$. But these rank estimates change to $(1,1)$ when we use $h_0=3$ and $4$. Figure \ref{figure:eigengap_finance} shows the estimated eigenvalues $\tau_{k,m_k}^{*2}$, $1\le m_k\le d_k$, using the non-iterative initial TIPUP procedure, for $k=1$ (size factor) and $k=2$ (BE factor). It is seen that for the size factor, the 
largest eigenvalue is more than 20 folds larger than the second largest eigenvalue. 
As simulation results in Section 5.2 show, in such an unbalanced case, the ER 
estimator may find it difficult to find the gap between the true non-zero eigenvalues and the 
true zero eigenvalues, based
on the ratio of the eigenvalues. On
the other hand, the IC estimator may fare better in such cases since it is based on the level of estimation error of the true zero eigenvalue. 

Overall, it seems that $(2,2)$ or $(3,3)$ are possibly good choices. More detailed analysis, include 
goodness-of-fit measures, prediction performance and result interpretation, is needed.


%

\section{Discussions} \label{section:conclusion}

In this paper, we develop two rank identification estimators, in an attempt to fill a gap on modelling tensor factor model in the literature. Non-iterative and iterative IC and ER estimators, based on similar ideas of the TOPUP and TIPUP  procedures of \citet{chen2021factor} and \citet{han2020} are considered. Theoretical analysis shows that the iterative estimators are much better than the non-iterative estimators. We show that in general the estimators based on TIPUP procedures 
are better than that based on the TOPUP procedure, due to its fast convergence rate of the estimated eigenvalues
under proper conditions. 
However, in situations when TIPUP procedures also lead to significant signal cancellation, extra care needs to be 
taken, including increasing the maximum lag $h_0$ in the procedure. 

Simulation studies are conducted to compare the finite sample performance of the estimators using the 
non-iterative and iterative estimation procedures. The results show that the ER estimators based on both
TIPUP and TOPUP procedures, and the IC estimators based on TIPUP procedures generally perform very well when all the factors are strong. 
The ER estimators are better than the IC estimators when some factors are weak, unless one chooses the precise
tuning parameter $\nu$ in the IC penalty function, which is a difficult task. 
When some dominant factors have unrealistically high explanatory power, the ER estimators may not perform well. But the IC estimators still work very well, since the factors are strong. 
In summary, IC estimator based on iTIPUP shall be used to estimate the number of strong factors, while ER estimators based on iTIPUP are likely to capture weak factors.

\bibliographystyle{imsart-nameyear}
\bibliography{tensorfactor}


\newpage

\appendix



%




\section{Proofs}

It suffices to consider $K=2$ as the TOPUP and TIPUP begin with mode-$k$ matrix unfolding. We observe a matrix time series with $X_t=A_1 F_t A_2^\top + E_t =G_t+E_t\in \mathbb R^{d_1\times d_2}$. Let $U_1$, $U_2$ be the left $r_1$, $r_2$ singular vectors of $A_1$ and $A_2$, respectively. Recall $\odot$ is kronecker product and $\otimes$ is tensor product. Without loss of generality, we only consider the case $k=1$. Denote $\overline\E(\cdot)=\E(\cdot|\{F_1,...,F_T\})$. For simplicity, write
\begin{align}
\cM(m,\widehat U_{1,m}) &= \frac{1}{d^2}\tr\left\{ \left(I-\widehat U_{1,m} \widehat U_{1,m}^\top \right)\widehat W_k \right\}, \\
\cM^*(m,\hat U_{1,m}) &= \frac{1}{d^2}\tr\left\{ \left(I-\widehat U_{1,m} \widehat U_{1,m}^\top \right)\widehat W_k^* \right\}, \\
\cM^{(i)}(m,\widehat U_{1,m}) &= \frac{1}{d^2}\tr\left\{ \left(I-\widehat U_{1,m} \widehat U_{1,m}^\top \right)\widehat W_k^{(i)} \right\}, \\
\cM^{*(i)}(m,\hat U_{1,m}) &= \frac{1}{d^2}\tr\left\{ \left(I-\widehat U_{1,m} \widehat U_{1,m}^\top \right)\widehat W_k^{*(i)} \right\}.
\end{align}

Although we present Theorem \ref{thm:all} for the IC and ER estimators together, the proofs of the IC and ER estimators using the same estimation procedure are very similar. Thus, we first prove Theorem \ref{thm:all} based on TOPUP, and then move to iTOPUP, TIPUP and iTIPUP in sequence.

\subsection{Proof of Theorem \ref{thm:all} for non-iterative TOPUP}

\begin{lemma}\label{lemma:topup:norm}
\begin{align} \label{eq:topup:norm}
&\left\| \frac{1}{T} \sum_{t=h+1}^T \mat1\left(A_1F_{t-h}A_2^\top \otimes A_1 F_t A_2^\top \right) \right\|_2^2  \le \|\Theta_{1,0}^*\|_2 \cdot\|\Theta_{1,0}\|_{op}
\end{align}

\end{lemma}

\begin{proof}
Note that $U_1 U_1^\top A_1=A_1$, $U_2 U_2^\top A_2=A_2$ and $\| U_1 \|_2=\| U_2 \|_2 =1$. It follows that
\begin{align*}
&\left\| \frac{1}{T}\sum_{t=h+1}^T \mat1\left(A_1F_{t-h}A_2^\top \otimes A_1 F_t A_2^\top \right) \right\|_2  \\
&=\left\| \frac{1}{T}\sum_{t=h+1}^T U_1 U_1^\top\mat1 \left(A_1F_{t-h}A_2^\top U_2 \otimes U_1^\top A_1 F_t A_2^\top U_2 \right) \cdot \left( U_2^\top\odot U_1^\top\odot U_2^\top  \right) \right\|_2 \\
&=\left\| \frac{1}{T}\sum_{t=h+1}^T U_1^\top\mat1 \left(A_1F_{t-h}A_2^\top U_2 \otimes U_1^\top A_1 F_t A_2^\top U_2 \right)  \right\|_2
\end{align*}
There exist $u\in\RR^{r_1}$ and $\cW\in\RR^{r_2\times r_1\times r_2}$ with $\|u\|_2^2=\|\vec1(\cW)\|_2^2=1$, such that
\begin{align*}
&\left\| \frac{1}{T}\sum_{t=h+1}^T U_1^\top\mat1 \left(A_1F_{t-h}A_2^\top U_2 \otimes U_1^\top A_1 F_t A_2^\top U_2 \right)  \right\|_2^2 \\
&=\frac{1}{T^2}\left(\sum_{t=h+1}^T \sum_{i_1,j_1,i_2,j_2} u_{i_1} (U_1^\top A_1 F_{t-h} A_2^\top U_2)_{i_1,j_1} (U_1^\top A_1 F_{t} A_2^\top U_2)_{i_2,j_2} w_{j_1,i_2,j_2} \right)^2 \\
&\le \frac{1}{T^2}\sum_{t=h+1}^T \sum_{j_1}\left(\sum_{i_1} u_{i_1} (U_1^\top A_1 F_{t-h} A_2^\top U_2)_{i_1,j_1} \right)^2 \sum_{s=h+1}^T \sum_{j_1} \left(\sum_{i_2,j_2}(U_1^\top A_1 F_{s} A_2^\top U_2)_{i_2,j_2} w_{j_1,i_2,j_2} \right)^2 \\
&= \frac{1}{T^2} \sum_{t=h+1}^T \left\|u^\top(U_1^\top A_1 F_{t-h} A_2^\top U_2) \right\|^2 \sum_{s=h+1}^T \sum_{j_1} \left(\tr(U_1^\top A_1 F_{s} A_2^\top U_2 \cW_{j_1}^{(2,3)} \right)^2 \\
&\le \frac{1}{T} \sum_{t=h+1}^T \left\|u^\top(U_1^\top A_1 F_{t-h} A_2^\top U_2) \right\|^2 \sum_{s=h+1}^T  \|\Theta_{1,0} \|_{op} \sum_{j_1} \|\cW_{j_1}^{(2,3)} \|_F^2 \\
&\le \|\Theta_{1,0} \|_2 \cdot \|\Theta_{1,0} \|_{op} \cdot\|u\|_2^2 \cdot \|\cW\|_F^2 .
\end{align*}
Then \eqref{eq:topup:norm} follows.
\end{proof}

\begin{lemma}\label{lemma:topup:rate}
Let $U_{1\perp}\in \RR^{d_1\times (d_1-r_1)}$ be the orthonormal complement of $U_1$. Denote $U_{1\perp}=(U_{1j})_{r_1+1\le j\le d_1}$ and $U_1=(U_{1j})_{1\le j\le r_1}$, where $U_{1j}\in \RR^{d_1}$. Denote $\eta_d=d^{\delta_1-\delta_0/2} + d^{\delta_1}d_1^{-1/2}$. Define $\hat U_1$ be the estimated top $r_1$ left singular space of $\text{TOPUP}_1(X_{1:T})$. Suppose Assumptions \ref{asmp:error}, \ref{asmp:mixing}, \ref{asmp:factor}, \ref{asmp:strength} and \ref{asmp:rank}(a) hold. Then, in an event $\Omega_{11}\cap \Omega_0$ with $\P(\Omega_{11})\ge 1-e^{-d_2}/2$ and $\P(\Omega_0)\ge 1-T\exp(-C_1 T^\vartheta)-\exp(-C_2T)$ with $C_1,C_2>0, 1/\vartheta=1/\theta_1+2/\theta_2$, there exist a matrix $\tilde U_1\in\RR^{d_1\times r_1}$ with $\tilde U_1 \tilde U_1^\top =\hat U_1 \hat U_1^\top$, such that
\begin{align}
\| \hat U_1 -U_1\tilde U_1^\top \hat U_1 \|_2 &\le C\left(\frac{\eta_d}{\sqrt{T}}\right), \label{eq:topup:rate:a} \\
\|U_1 U_1^\top \hat U_{1j}\|_2 & \le C \left(\frac{\eta_d}{\sqrt{T}}\right), \label{eq:topup:rate:b}
\end{align}
for all $r_1+1\le j\le d_1$ and some $C>0$.
\end{lemma}

\begin{proof}
Under Assumptions \ref{asmp:error}, \ref{asmp:mixing}, \ref{asmp:factor}, \ref{asmp:strength}, \ref{asmp:rank}(a) and Proposition \ref{prop:zt}, as the derivation of Theorem 1 in \citet{han2020}, in an event $\Omega_{11}\cap \Omega_0$ with $\P(\Omega_{11})\ge 1-e^{-d_2}/2$ and $\P(\Omega_0)\ge 1-T\exp(-C_1 T^\vartheta)-\exp(-C_2T)$, 
\begin{equation}\label{eq:topup:rate1}
\| \hat U_1 \hat U_1^\top -U_1  U_1^\top \|_2 \le C \left(\frac{\eta_d}{\sqrt{T}}\right).
\end{equation}
Applying Lemma \ref{lemma:perturbation} and Theorem 1 in \citet{chen2021factor}, for sufficient large $T$, there exists a matrix $Q\in\RR^{(d_1-r_1)\times r_1}$ such that
\begin{equation*}
\|Q\|_2  \le \| \hat U_1 \hat U_1^\top -U_1  U_1^\top \|_2
\end{equation*}
and $\tilde U_1=(U_1+U_{1\perp} Q)(I+Q^\top Q)^{-1/2}\in R^{d_1\times r_1}$ is an estimator for $U_1$ with $\tilde U_1 \tilde U_1^\top =\hat U_1 \hat U_1^\top$. Elementary calculation shows that
\begin{align*}
\|\tilde U_1-U_1\|_2 &=\|(U_1(I-(I+Q^\top Q)^{1/2})+U_{1\perp} Q)(I+Q^\top Q)^{-1/2}\|_2 \\
&\le  \|I-(I+Q^\top Q)^{1/2} \|_2 +\| Q\|_2 \le 2\|Q\|_2.
\end{align*}
It follows that
\begin{align*}
\| \hat U_1 -U_1\tilde U_1^\top \hat U_1 \|_2 =\| \tilde U_1 \tilde U_1^\top \hat U_1 -U_1\tilde U_1^\top \hat U_1 \|_2  \le  \|\tilde U_1-U_1\|_2 \|  \tilde U_1^\top \hat U_1 \|_2 \le 2\| \hat U_1 \hat U_1^\top -U_1  U_1^\top \|_2.
\end{align*}
In view of \eqref{eq:topup:rate1}, we have \eqref{eq:topup:rate:a}.

For \eqref{eq:topup:rate:b}, notice that in the event $\Omega_{11}\cap\Omega_0$, for all $r_1+1\le j\le d_1$, 
\begin{align*}
\|U_1 U_1^\top \hat U_{1j}\|_2 = \| \hat U_{1\perp} \hat U_{1\perp}^\top \hat U_{1j} - U_{1\perp} U_{1\perp}^\top \hat U_{1j}\|_2 \le  \| \hat U_{1\perp} \hat U_{1\perp}^\top  - U_{1\perp} U_{1\perp}^\top \|_2 \le C\left(\frac{\eta_d}{\sqrt{T}}\right).
\end{align*}
\end{proof}

\begin{lemma}\label{lemma:topup:spectral}
Suppose Assumptions \ref{asmp:error}, \ref{asmp:mixing}, \ref{asmp:factor}, \ref{asmp:strength} and \ref{asmp:rank}(a) hold. Then, in an event $\Omega_{12}\cap \Omega_0$ with $\P(\Omega_{12})\ge 1-e^{-d_2}/2$ and $\P(\Omega_0)\ge 1-T\exp(-C_1 T^\vartheta)-\exp(-C_2T)$ with $C_1,C_2>0, 1/\vartheta=1/\theta_1+2/\theta_2$, we have
\begin{align}
&\left\| \frac{1}{T-h} \sum_{t=h+1}^T \mat1(A_1 F_{t-h} A_2^\top\otimes E_t )  \right\|_2\le C \left( \frac{d^{1-\delta_0/2} }{\sqrt{T}} \right), \label{eq:topup:spectral:1}\\
&\left\| \frac{1}{T-h} \sum_{t=h+1}^T \mat1(A_1 F_{t-h} A_2^\top\otimes U_1^\top E_t U_2 )  \right\|_2 \le C \left( \frac{d^{1/2-\delta_0/2} }{\sqrt{T}} \right), \label{eq:topup:spectral:2} \\
&\left\| \frac{1}{T-h} \sum_{t=h+1}^T \mat1(E_{t-h}\otimes A_1 F_{t} A_2^\top )  \right\|_2 \le C \left(\frac{(\sqrt{d_1}+\sqrt{d/d_1}) d^{1/2-\delta_0/2} }{\sqrt{T}} \right), \label{eq:topup:spectral:3} \\
&\left\| \frac{1}{T-h} \sum_{t=h+1}^T \mat1(E_{t-h}U_2\otimes A_1 F_{t} A_2^\top )  \right\|_2 \le C \left(\frac{(d_1^{1/2} d^{1/2-\delta_0/2} }{\sqrt{T}} \right), \label{eq:topup:spectral:4} \\
&\left\| \frac{1}{T-h} \sum_{t=h+1}^T U_1^\top\mat1(E_{t-h}U_2\otimes A_1 F_{t} A_2^\top )  \right\|_2 \le C \left(\frac{d^{1/2-\delta_0/2} }{\sqrt{T}} \right), \label{eq:topup:spectral:5} \\ 
&\left\| \frac{1}{T-h} \sum_{t=h+1}^T \mat1(E_{t-h}\otimes E_{t} )  \right\|_2 \le C \left(\frac{d }{\sqrt{d_1T}} \right), \label{eq:topup:spectral:6} \\
&\left\| \frac{1}{T-h} \sum_{t=h+1}^T \mat1(E_{t-h} U_2\otimes E_{t} )  \right\|_2 \le C \left(\frac{d^{1/2} }{\sqrt{T}} \right), \label{eq:topup:spectral:7} \\
&\left\| \frac{1}{T-h} \sum_{t=h+1}^T \mat1(E_{t-h}\otimes U_1^\top E_{t} U_2 )  \right\|_2 \le C \left(\frac{d^{1/2} }{\sqrt{T}} \right), \label{eq:topup:spectral:8} \\
&\left\| \frac{1}{T-h} \sum_{t=h+1}^T \mat1(E_{t-h}U_2\otimes U_1^\top E_{t} U_2 )  \right\|_2 \le C \left(\frac{d_1^{1/2} }{\sqrt{T}} \right), \label{eq:topup:spectral:9} \\
&\left\| \frac{1}{T-h} \sum_{t=h+1}^T U_1^\top\mat1(E_{t-h}\otimes U_1^\top E_{t}U_2 )  \right\|_2 \le C \left(\frac{d^{1/2} }{\sqrt{d_1T}} \right), \label{eq:topup:spectral:10} \\
&\left\| \frac{1}{T-h} \sum_{t=h+1}^T U_1^\top\mat1(E_{t-h} U_2\otimes U_1^\top E_{t}U_2 )  \right\|_2 \le C \left(\frac{1}{\sqrt{T}} \right), \label{eq:topup:spectral:11}
\end{align}
for some constant positive $C$ depending on $K$ only.
\end{lemma}

\begin{proof}
Note that $U_1 U_1^\top A_1=A_1$ and $U_2 U_2^\top A_2=A_2$. Let
$$\|\Delta_1\|_2:=\left\| \frac{1}{T-h} \sum_{t=h+1}^T \mat1(A_1 F_{t-h} A_2^\top\otimes E_t )  \right\|_2=\left\| \frac{1}{T-h} \sum_{t=h+1}^T U_1^\top \mat1(A_1 F_{t-h} A_2^\top U_2\otimes E_t )  \right\|_2.$$
By Theorem 1 in \citet{chen2021factor},
\begin{align*}
\overline\E\|\Delta_1\|_2 \le \frac{\sigma \sqrt{2T}(\sqrt{r_1}+\sqrt{r_2d_1d_2})}{T-h} \|\Theta_{1,0}^* \|_2^{1/2}.
\end{align*}
Elementary calculation shows that
\begin{eqnarray*}
&&\left| \left\| \sum_{t=h+1}^T \mat1(A_1 F_{t-h} A_2^\top \otimes E_t  ) \right\|_{2} - \left\| \sum_{t=h+1}^T \mat1(A_1 F_{t-h} A_2^\top  \otimes E_t^*  )  \right\|_{2} \right| \\
&\le& \left\| \sum_{t=h+1}^T \mat1(A_1 F_{t-h}A_2^\top  \otimes (E_t-E_t^*)  )  \right\|_{2} \\
&\le& \left\| (\mat1(A_1 F_1 A_2^\top\otimes I_{d_1}),...,\mat1(A_1 F_{T-h} A_2^\top \otimes I_{d_1})) \begin{pmatrix}
I_{d_2}\odot I_{d_1}\odot (E_{h+1}-E_{h+1}^{*}) \\
\vdots\\
I_{d_2}\odot I_{d_1}\odot(E_{T}-E_{T}^{*})
\end{pmatrix} \right\|_{2} \\
&\le& \sqrt{T} \|\Theta_{1,0}^*\|_{2}^{1/2}  \left\|\begin{pmatrix}
E_{h+1}-E_{h+1}^{*} \\
\vdots\\
E_{T}-E_{T}^{*}
\end{pmatrix} \right\|_{\rm F}.
\end{eqnarray*}
That is, $\left\| \sum_{t=h+1}^T \mat1(A_1 F_{t-h}A_2^\top  \otimes E_t  ) \right\|_{2}$ is a $\sigma\sqrt{T}\|\Theta_{1,0}^*\|_{2}^{1/2}$ Lipschitz function in $(E_1,...,E_T)$. 
Then, by Gaussian concentration inequalities for Lipschitz functions,
\begin{align*}
&\P\left( \left\| \sum_{t=h+1}^T \frac{\mat1(A_1 F_{t-h}A_2^\top  \otimes E_t  )}{T-h} \right\|_{2} - \frac{\sigma(2T)^{1/2}(\sqrt{r_1}+\sqrt{d_1d_2r_2})}{T-h} \|\Theta_{1,0}^*\|_{2}^{1/2} \ge \frac{\sigma\sqrt{T} }{T-h} \|\Theta_{1,0}^*\|_{2}^{1/2} x\right) \\
&\le 2e^{-\frac{x^2}{2 } }.
\end{align*}
This implies that with $x\asymp \sqrt{d}$, in an event $\Omega_{a}$ with at least probability $1-e^{-d_2}/6$,
\begin{align}
\|\Delta_1\|_2 \le  \frac{C_0\sqrt{Td}}{T-h} \|\Theta_{1,0}^*\|_{2}^{1/2},
\end{align}
where $C_0>0$ depends on $K$ only. Then, using Proposition \ref{prop:zt}, in the event $\Omega_a\cap\Omega_0$, \eqref{eq:topup:spectral:1} follows. Similar arguments yield \eqref{eq:topup:spectral:2} in the event $\Omega_a\cap\Omega_0$.

Let
$$\|\Delta_2\|_2:=\left\| \frac{1}{T-h} \sum_{t=h+1}^T \mat1(E_{t-h}\otimes A_1 F_{t} A_2^\top )  \right\|_2=\left\| \frac{1}{T-h} \sum_{t=h+1}^T \mat1(E_{t-h}\otimes U_1^\top A_1 F_{t} A_2^\top U_2)  \right\|_2.$$
By Theorem 1 in \citet{chen2021factor},
\begin{align*}
\overline\E\|\Delta_2\|_2 \le \frac{\sigma \sqrt{2T}(\sqrt{d_1}+\sqrt{d_2r_1r_2})}{T-h} \|\Theta_{1,0}^* \|_{op}^{1/2}.
\end{align*}
Then, using Proposition \ref{prop:zt}, in an event $\Omega_b\cap\Omega_0$ with $\P(\Omega_b)\ge1-e^{-d_2}/6$, \eqref{eq:topup:spectral:3} follows from the same argument as the above step. Similarly, in the event $\Omega_b\cap\Omega_0$, we can obtain \eqref{eq:topup:spectral:4} and \eqref{eq:topup:spectral:5}.

We split the sum into two terms over the index sets, $S_1=\{(h,2h]\cup(3h,4h]\cup\cdots\} \cap(h,T]$ and its complement $S_2$ in $(h,T]$, so that $\{E_{t-h},t\in S_a\}$ is independent of $\{E_t, t\in S_a\}$ for each $a=1,2$. Let $n_a=|S_a|$. By Lemma \ref{lm-GH}(ii), for any $x>0$,
\begin{align*}
\P\left(\left\| \sum_{t\in S_a} \mat1(E_{t-h} \otimes E_t  ) \right\|_{\rm S}  \ge d_1\sqrt{d_2}+2d_2 \sqrt{d_1 n_a}+x^2+\sqrt{n_a}x+3\sqrt{d_1d_2}x          \right) \le   2e ^{-x^2/2}.
\end{align*}
With $x\asymp \sqrt{d_2}$ and some constant $C_{1}$ depending on $K$ only, we have
\begin{align}
\P\left( \|\Delta_3\|_2 \ge \frac{C_1 d_1\sqrt{d_2}+C_1 d_2\sqrt{d_1 T}}{T} \right)\le e^{-d_2}/6.
\end{align}
Then, as in the derivation of $\|\Delta_3\|_2$ in the proof of Theorem 1 in \citet{chen2021factor}, in an event $\Omega_c$ with $\P(\Omega_c)\ge 1-e^{-d_2}/6$, \eqref{eq:topup:spectral:6} follows. Similar arguments yield \eqref{eq:topup:spectral:7}, \eqref{eq:topup:spectral:8}, \eqref{eq:topup:spectral:9}, \eqref{eq:topup:spectral:10} and \eqref{eq:topup:spectral:11} in the event $\Omega_c$. Set $\Omega_{12}=\Omega_a\cap\Omega_b\cap\Omega_c$, then $\P(\Omega_{12})\ge 1 -e^{-d_2}/2$.
\end{proof}

\begin{lemma}\label{lemma:topup:upf}
Suppose Assumptions \ref{asmp:error}, \ref{asmp:mixing}, \ref{asmp:factor}, \ref{asmp:strength} and \ref{asmp:rank}(a) hold. In an event $\Omega_1$ with $\P(\Omega_1)\ge 1-e^{-d_2}-T\exp(-C_1 T^\vartheta)-\exp(-C_2T)$ with $C_1,C_2>0, 1/\vartheta=1/\theta_1+2/\theta_2$, for any fixed $m$ with $m> r_1$, $\cM(r_1,\hat U_1)-\cM(m,\hat U_{1,m})\le C\beta_{d,T}$, where $C>0$,
\begin{equation*}
\beta_{d,T}=\frac{1}{Td_1 } +\frac{d_1 }{Td^{1+\delta_0}} + \frac{d_1^{1/2}\eta_d}{Td^{1/2+3\delta_0/2}} +\frac{ \eta_d}{T^{3/2}d^{1/2+\delta_0/2}},
\end{equation*}
and $\eta_d=d^{\delta_1-\delta_0/2} + d^{\delta_1}d_1^{-1/2}$.
\end{lemma}

\begin{proof}
Let $\Omega_1=\Omega_{11}\cap\Omega_{12}\cap\Omega_0$, where $\Omega_{11}, \Omega_{12}$ are defined in Lemma \ref{lemma:topup:rate} and \ref{lemma:topup:spectral}, respectively, and $\Omega_0$ is the event in Proposition \ref{prop:zt}. Let $U_{1,r_1}=U_{1}$ and $U_{1,m}=(U_{11},U_{12},...,U_{1m})$.
\begin{align*}
\left| \cM(m,\hat U_{1,m})-\cM(r_1,\hat U_1) \right| &\le \left| \cM(m,\hat U_{1,m})-\cM(r_1, U_1) \right| + \left| \cM(r_1,\hat U_1)-\cM(r_1, U_1) \right| \\
&\le 2 \max_{r_1< m\le m_1} \left| \cM(m,\hat U_{1,m})-\cM(r_1, U_1) \right|.
\end{align*}
As $m_1$ is fixed, it is sufficient to prove for each $m$ with $m> r_1$, in the event $\Omega_1$,
\begin{equation}
\left|\cM(m,\hat U_{1,m})-\cM(r_1, U_1)\right| \le C\beta_{d,T}.
\end{equation}
In the following, we shall only work on $\Omega_1$.

Elementary calculation shows that
\begin{align*}
&d^2\cdot\cM(m,\hat U_{1,m})-d^2\cdot\cM(r_1, U_1) \\
=& \sum_{h=1}^{h_0}\frac{1}{(T-h)^2}\sum_{s,t=h+1}^T \tr\left\{ \left(U_1 U_1^\top - \hat U_{1,m} \hat U_{1,m}^\top \right) \mat1(X_{t-h}\otimes X_t) {\mat1}^\top (X_{s-h}\otimes X_s)  \right\} \\
=& \sum_{h=1}^{h_0}\frac{1}{(T-h)^2}\sum_{s,t=h+1}^T \tr\left\{  {\mat1}^\top (A_1 F_{s-h} A_2^\top\otimes X_s) \left(U_1 U_1^\top - \hat U_{1,m} \hat U_{1,m}^\top \right) \mat1(A_1 F_{t-h} A_2^\top\otimes X_t)  \right\} \\
&+ \sum_{h=1}^{h_0}\frac{1}{(T-h)^2}\sum_{s,t=h+1}^T \tr\left\{  {\mat1}^\top ( E_{s-h} \otimes X_s) \left(U_1 U_1^\top - \hat U_{1,m} \hat U_{1,m}^\top \right) \mat1(E_{t-h} \otimes X_t)  \right\} \\
&+ \sum_{h=1}^{h_0}\frac{1}{(T-h)^2}\sum_{s,t=h+1}^T \tr\left\{  {\mat1}^\top (A_1 F_{s-h} A_2^\top\otimes X_s) \left(U_1 U_1^\top - \hat U_{1,m} \hat U_{1,m}^\top \right) \mat1(E_{t-h} \otimes X_t)  \right\} \\
&+ \sum_{h=1}^{h_0}\frac{1}{(T-h)^2}\sum_{s,t=h+1}^T \tr\left\{  {\mat1}^\top (E_{s-h} \otimes X_s) \left(U_1 U_1^\top - \hat U_{1,m} \hat U_{1,m}^\top \right) \mat1(A_1 F_{t-h} A_2^\top\otimes X_t)  \right\} \\
:=&\I+\II+\III+\IV.
\end{align*}
Note that $\III=\IV$.

We first consider $\II$.
\begin{align*}
\II=&\sum_{h=1}^{h_0}\frac{1}{(T-h)^2}\sum_{s,t=h+1}^T \tr\left\{  {\mat1}^\top ( E_{s-h} \otimes X_s) \left(U_1 U_1^\top - \hat U_{1,m} \hat U_{1,m}^\top \right) \mat1(E_{t-h} \otimes X_t)  \right\}    \\
=& \sum_{h=1}^{h_0}\frac{1}{(T-h)^2}\sum_{s,t=h+1}^T \tr\left\{  {\mat1}^\top ( E_{s-h} \otimes E_s) \left(U_1 U_1^\top - \hat U_{1,m} \hat U_{1,m}^\top \right) \mat1(E_{t-h} \otimes E_t)  \right\}\\
&+\sum_{h=1}^{h_0}\frac{1}{(T-h)^2}\sum_{s,t=h+1}^T \tr\left\{  {\mat1}^\top ( E_{s-h} \otimes A_1 F_s A_2^\top) \left(U_1 U_1^\top - \hat U_{1,m} \hat U_{1,m}^\top \right) \mat1(E_{t-h} \otimes A_1 F_t A_2^\top)  \right\} \\
&+2\sum_{h=1}^{h_0}\frac{1}{(T-h)^2}\sum_{s,t=h+1}^T \tr\left\{  {\mat1}^\top ( E_{s-h} \otimes A_1 F_s A_2^\top) \left(U_1 U_1^\top - \hat U_{1,m} \hat U_{1,m}^\top \right) \mat1(E_{t-h} \otimes E_t)  \right\} \\
:=&\II_1+\II_2+2\II_3.
\end{align*}
We consider each term in turn. Note that
\begin{align}\label{eq:split:u1}
\hat U_{1} \hat U_{1}^\top - U_1 U_1^\top=& \left( \hat U_1 -U_1\tilde U_1^\top \hat U_1\right) \left( \hat U_1 -U_1\tilde U_1^\top \hat U_1\right)^\top + U_1\tilde U_1^\top \hat U_1 \left( \hat U_1 -U_1\tilde U_1^\top \hat U_1\right)^\top \notag\\
&+\left( \hat U_1 -U_1\tilde U_1^\top \hat U_1\right) \left( U_1\tilde U_1^\top \hat U_1 \right)^\top,
\end{align} 
where $\tilde U_1$ is defined in Lemma \ref{lemma:topup:rate}. Then, by Lemma \ref{lemma:topup:rate} and \ref{lemma:topup:spectral},
\begin{align*}
\II_1=&\sum_{h=1}^{h_0}\frac{1}{(T-h)^2}\sum_{s,t=h+1}^T \tr\left\{  {\mat1}^\top ( E_{s-h} \otimes E_s) \left(U_1 U_1^\top - \hat U_{1} \hat U_{1}^\top \right) \mat1(E_{t-h} \otimes E_t)  \right\} \\
&+\sum_{h=1}^{h_0}\sum_{j=r_1+1}^{m}\frac{1}{(T-h)^2} \tr\left\{  {\mat1}^\top ( E_{s-h} \otimes E_s) \hat U_{1j} \hat U_{1j}^\top  \mat1(E_{t-h} \otimes E_t)  \right\} \\
\le& r_1\sum_{h=1}^{h_0}\left\| \frac{1}{(T-h)}\sum_{t=h+1}^T \mat1(E_{t-h}\otimes E_t)      \right\|_2^2     \left( \left\| \hat U_1 -U_1\tilde U_1^\top \hat U_1 \right\|_2^2 + \left\| \hat U_1 -U_1\tilde U_1^\top \hat U_1 \right\|_2 \cdot \left\| U_1\tilde U_1^\top \hat U_1 \right\|_2   \right) \\
&+(m-r_1) \sum_{h=1}^{h_0}\left\| \frac{1}{(T-h)}\sum_{t=h+1}^T \mat1(E_{t-h}\otimes E_t)      \right\|_2^2 \\
=& O \left( \frac{d^2}{d_1T} \right),
\end{align*}
by the fact $\eta_{d}=o(\sqrt{T})$.
\begin{align*}
\II_2=& \sum_{h=1}^{h_0}\frac{1}{(T-h)^2}\sum_{s,t=h+1}^T \tr\left\{  {\mat1}^\top ( E_{s-h} \otimes A_1 F_s A_2^\top) \left(U_1 U_1^\top - \hat U_{1} \hat U_{1}^\top \right) \mat1(E_{t-h} \otimes A_1 F_t A_2^\top)  \right\} \\
&+\sum_{h=1}^{h_0}\sum_{j=r_1+1}^m\frac{1}{(T-h)^2}\sum_{s,t=h+1}^T \tr\left\{  {\mat1}^\top ( E_{s-h} \otimes A_1 F_s A_2^\top) \hat U_{1j} \hat U_{1j}^\top  \mat1(E_{t-h} \otimes A_1 F_t A_2^\top)  \right\} \\
=& O\left( \frac{(d_1+d/d_1)d^{1-\delta_0}}{T} \right).
\end{align*}
\begin{align*}
\II_3=& \sum_{h=1}^{h_0}\frac{1}{(T-h)^2}\sum_{s,t=h+1}^T \tr\left\{  {\mat1}^\top ( E_{s-h} \otimes U_1^\top A_1 F_s A_2^\top U_2) \left(U_1 U_1^\top - \hat U_{1} \hat U_{1}^\top \right) \mat1(E_{t-h} \otimes U_1^\top E_t U_2 )  \right\} \\
&+ \sum_{h=1}^{h_0}\sum_{j=r_1+1}^m\frac{1}{(T-h)^2}\sum_{s,t=h+1}^T \tr\left\{  {\mat1}^\top ( E_{s-h} \otimes U_1^\top A_1 F_s A_2^\top U_2) \hat U_{1j} \hat U_{1j}^\top \mat1(E_{t-h} \otimes U_1^\top E_t U_2 )  \right\} \\
\le& (m+O_{\P}(1))\sum_{h=1}^{h_0} \frac{1}{(T-h)^2} \left\|\sum_{s=h+1}^T   {\mat1} ( E_{s-h} \otimes U_1^\top A_1 F_s A_2^\top U_2) \right\|_2 \left\|\sum_{s,t=h+1}^T  \mat1(E_{t-h} \otimes U_1^\top E_t U_2) \right\|_2 \\
=& O\left(\frac{(\sqrt{d_1}+\sqrt{d/d_1})d^{1/2-\delta_0/2}}{\sqrt{T}} \cdot \frac{\sqrt{d}}{\sqrt{T}} \right) = O\left(\frac{(\sqrt{d_1}+\sqrt{d/d_1})d^{1-\delta_0/2}}{ T } \right) 
\end{align*}
by Lemma \ref{lemma:topup:rate} and \ref{lemma:topup:spectral}.
Combing the bounds of $\II_1,\II_2$ and $\II_3$, we have
\begin{equation} \label{eq:lemma:upf:2}
\II= O \left(\frac{d^2}{d_1 T} +\frac{d_1 d^{1-\delta_0}}{T}\right).
\end{equation}

Next, we consider $\III$.
\begin{align*}
\III=& \sum_{h=1}^{h_0}\frac{1}{(T-h)^2}\sum_{s,t=h+1}^T \tr\left\{  {\mat1}^\top (A_1 F_{s-h} A_2^\top\otimes X_s) \left(U_1 U_1^\top - \hat U_{1,m} \hat U_{1,m}^\top \right) \mat1(E_{t-h} \otimes X_t)  \right\} \\
=& \sum_{h=1}^{h_0}\frac{1}{(T-h)^2}\sum_{s,t=h+1}^T \tr\left\{  {\mat1}^\top (A_1 F_{s-h} A_2^\top\otimes A_1 F_s A_2^\top) \left(U_1 U_1^\top - \hat U_{1,m} \hat U_{1,m}^\top \right) \mat1(E_{t-h} \otimes A_1 F_t A_2^\top)  \right\}  \\
&+ \sum_{h=1}^{h_0}\frac{1}{(T-h)^2}\sum_{s,t=h+1}^T \tr\left\{  {\mat1}^\top (A_1 F_{s-h} A_2^\top\otimes E_s) \left(U_1 U_1^\top - \hat U_{1,m} \hat U_{1,m}^\top \right) \mat1(E_{t-h} \otimes E_t)  \right\}  \\
&+  \sum_{h=1}^{h_0}\frac{1}{(T-h)^2}\sum_{s,t=h+1}^T \tr\left\{  {\mat1}^\top (A_1 F_{s-h} A_2^\top\otimes E_s) \left(U_1 U_1^\top - \hat U_{1,m} \hat U_{1,m}^\top \right) \mat1(E_{t-h} \otimes A_1 F_t A_2^\top)  \right\} \\
&+ \sum_{h=1}^{h_0}\frac{1}{(T-h)^2}\sum_{s,t=h+1}^T \tr\left\{  {\mat1}^\top (A_1 F_{s-h} A_2^\top\otimes A_1 F_s A_2^\top) \left(U_1 U_1^\top - \hat U_{1,m} \hat U_{1,m}^\top \right) \mat1(E_{t-h} \otimes E_t)  \right\} \\
:=& \III_1+\III_2+ \III_3 + \III_4.
\end{align*}
Again, we bound each term in turn. By Lemma \ref{lemma:topup:norm}, \ref{lemma:topup:rate}, \ref{lemma:topup:spectral},
\begin{align*}
&\III_1\\
=& \sum_{h=1}^{h_0}\frac{1}{(T-h)^2}\sum_{s,t=h+1}^T \tr\left\{  {\mat1}^\top (A_1 F_{s-h} A_2^\top U_2\otimes A_1 F_s A_2^\top) \left(U_1 U_1^\top - \hat U_{1} \hat U_{1}^\top \right) \mat1(E_{t-h} U_2\otimes A_1 F_t A_2^\top)  \right\} \\
&+ \sum_{h=1}^{h_0}\sum_{j=r_1+1}^m \frac{1}{(T-h)^2}\sum_{s,t=h+1}^T \tr\left\{  {\mat1}^\top (A_1 F_{s-h} A_2^\top U_2 \otimes A_1 F_s A_2^\top)  \hat U_{1j} \hat U_{1j}^\top  \mat1(E_{t-h} U_2 \otimes A_1 F_t A_2^\top)  \right\} \\
\le& \sum_{h=1}^{h_0}\frac{r_1}{(T-h)^2}\sum_{s,t=h+1}^T \left\| {\mat1} (A_1 F_{s-h} A_2^\top U_2 \otimes A_1 F_s A_2^\top) \right\|_2 \left\| \mat1(E_{t-h} U_2 \otimes A_1 F_t A_2^\top)  \right\|_2 \left\| U_1 U_1^\top - \hat U_{1} \hat U_{1}^\top  \right\|_2 \\
&+ \sum_{h=1}^{h_0} \frac{m-r_1}{(T-h)^2}\sum_{s,t=h+1}^T \left\| {\mat1} (A_1 F_{s-h} A_2^\top U_2 \otimes A_1 F_s A_2^\top) \right\|_2 \left\| \mat1(E_{t-h} U_2 \otimes A_1 F_t A_2^\top)  \right\|_2 \left\| U_1 U_1^\top \hat U_{1j} \right\|_2 \\
=& O \left( \frac{d_1^{1/2} d^{3/2-3\delta_0/2}\eta_d}{T} \right),
\end{align*}
where $\eta_d =d^{\delta_1-\delta_0/2}+d^{\delta_1}d_1^{-1/2}$.
\begin{align*}
&\III_2= \sum_{h=1}^{h_0}\frac{1}{(T-h)^2}\sum_{s,t=h+1}^T \tr\left\{  {\mat1}^\top (A_1 F_{s-h} A_2^\top U_2 \otimes E_s) \left(U_1 U_1^\top - \hat U_{1,m} \hat U_{1,m}^\top \right) \mat1(E_{t-h} U_2 \otimes E_t)  \right\}  \\
&\le  \sum_{h=1}^{h_0}\frac{r_1}{(T-h)^2}\sum_{s,t=h+1}^T \left\|  {\mat1} (A_1 F_{s-h} A_2^\top U_2 \otimes E_s) \right\|_2  \left\| \mat1(E_{t-h} U_2 \otimes E_t)  \right\|_2 \left\|U_1^\top (U_1 U_1^\top - \hat U_{1,m} \hat U_{1,m}^\top) \right\|_2 \\
&= O \left( \frac{d^{1-\delta_0/2}}{\sqrt{T}}\cdot \frac{\sqrt{d}}{\sqrt{T}} \cdot \frac{\eta_d}{\sqrt{T}} \right) = O \left( \frac{d^{3/2-\delta_0/2}\eta_d}{T^{3/2}} \right).
\end{align*}
Similarly,
\begin{align*}
\III_3=& \sum_{h=1}^{h_0}\frac{1}{(T-h)^2}\sum_{s,t=h+1}^T \tr\left\{  {\mat1}^\top (A_1 F_{s-h} A_2^\top U_2 \otimes U_1^\top E_s U_2) \left(U_1 U_1^\top - \hat U_{1,m} \hat U_{1,m}^\top \right) \right. \\
&\qquad\qquad\qquad \left. \cdot\mat1(E_{t-h} U_2 \otimes U_1^\top A_1 F_t A_2^\top U_2)  \right\} \\
=& O \left( \frac{d_1^{1/2} d^{1-\delta_0} \eta_d}{T^{3/2}}\right),
\end{align*}
and
\begin{align*}
\III_4=& \sum_{h=1}^{h_0}\frac{1}{(T-h)^2}\sum_{s,t=h+1}^T \tr\left\{  {\mat1}^\top (A_1 F_{s-h} A_2^\top U_2\otimes U_1^\top A_1 F_s A_2^\top U_2 ) \left(U_1 U_1^\top - \hat U_{1,m} \hat U_{1,m}^\top \right) \right. \\
&\qquad\qquad\qquad \left. \cdot \mat1(E_{t-h} U_2 \otimes U_1^\top E_t U_2)  \right\}  \\
=& O \left( \frac{d_1^{1/2} d^{1-\delta_0} \eta_d}{T} \right).
\end{align*}
Combing the bounds of $\III_1,\III_2,\III_3$ and $\III_4$, we have
\begin{equation} \label{eq:lemma:upf:3}
\III = O \left(\frac{d_1^{1/2}d^{3/2-3\delta_0/2}\eta_d}{T} +\frac{ d^{3/2-\delta_0/2}\eta_d}{T^{3/2}}\right).
\end{equation}
In view of \eqref{eq:lemma:upf:2} and \eqref{eq:lemma:upf:3},
\begin{equation} \label{eq:lemma:upf:23}
|\II+\III+\IV| \le C_1 \left(\frac{d^2}{d_1 T} +\frac{d_1 d^{1-\delta_0}}{T} + \frac{d_1^{1/2}d^{3/2-3\delta_0/2}\eta_d}{T} +\frac{ d^{3/2-\delta_0/2}\eta_d}{T^{3/2}} \right) \le C_2 \left( d^2\beta_{d,T}\right).
\end{equation}

Next, we consider $\I$.
\begin{align*}
\I=&\sum_{h=1}^{h_0}\frac{1}{(T-h)^2}\sum_{s,t=h+1}^T \tr\left\{  {\mat1}^\top (A_1 F_{s-h} A_2^\top\otimes X_s) \left(U_1 U_1^\top - \hat U_{1,m} \hat U_{1,m}^\top \right) \mat1(A_1 F_{t-h} A_2^\top\otimes X_t)  \right\} \\
=& \sum_{h=1}^{h_0}\frac{1}{(T-h)^2}\sum_{s,t=h+1}^T \tr\left\{  {\mat1}^\top (A_1 F_{s-h} A_2^\top\otimes X_s) \left(I - \hat U_{1,m} \hat U_{1,m}^\top \right) \mat1(A_1 F_{t-h} A_2^\top\otimes X_t)  \right\} \\
=& \sum_{h=1}^{h_0}\frac{1}{(T-h)^2}\tr\left\{ \sum_{s=h+1}^T {\mat1}^\top (A_1 F_{s-h} A_2^\top\otimes X_s) \left(I - \hat U_{1,m} \hat U_{1,m}^\top \right) \sum_{t=h+1}^T\mat1(A_1 F_{t-h} A_2^\top\otimes X_t)  \right\} \\
\ge&0,
\end{align*}
using the fact that $I-  \hat U_{1,m} \hat U_{1,m}^\top$ is positive semi-definite, $x^\top \hat U_{1,m} \hat U_{1,m}^\top x \le x^\top x$.

Using the definition of our optimization target, we have $\cM(m,\hat U_{1,m}) < \cM(r_1,\hat U_{1})\le \cM(r_1, U_{1})$. It follows that $\I+\II+\III+\IV\le 0$. As $\I\ge0$, by \eqref{eq:lemma:upf:23}, $\I=O(d^2\beta_{d,T})$. In summary,
\begin{equation*}
\left| \cM(m,\hat U_{1,m})- \cM(r_1, U_1) \right| \le C \left( \beta_{d,T}\right).
\end{equation*}

\end{proof}

\begin{lemma}\label{lemma:topup:lessf}
Suppose Assumptions \ref{asmp:error}, \ref{asmp:mixing}, \ref{asmp:factor}, \ref{asmp:strength} and \ref{asmp:rank}(a) hold. In an event $\Omega_1$ with $\P(\Omega_1)\ge 1- e^{-d_2}-T\exp(-C_1 T^\vartheta)-\exp(-C_2T)$ with $C_1,C_2>0, 1/\vartheta=1/\theta_1+2/\theta_2$, for any $m$ with $m< r_1$, there exist constant $c_m>0$ and $C>0$, such that 
$\cM(m,\hat U_{1,m})-\cM(r_1,\hat U_1)\ge (c_m+o(1)) d^{-2\delta_1}+C\gamma_{d,T}$, where 
\begin{equation*}
\gamma_{d,T}=\frac{1}{Td^{\delta_0}} + \frac{1}{Td_1} + \frac{1}{T^{1/2}d^{1/2+3\delta_0/2}} + \frac{d_1^{1/2}\eta_d}{Td^{1/2+3\delta_0/2}},
\end{equation*}
and $\eta_d=d^{\delta_1-\delta_0/2} + d^{\delta_1}d_1^{-1/2}$.
\end{lemma}

\begin{proof}
Let $\Omega_1=\Omega_{11}\cap\Omega_{12}\cap\Omega_0$, where $\Omega_{11}, \Omega_{12}$ are defined in Lemma \ref{lemma:topup:rate} and \ref{lemma:topup:spectral}, respectively, and $\Omega_0$ is the event in Proposition \ref{prop:zt}. In the following, we shall only work on $\Omega_1$.

Let $U_{1,r_1}=U_{1}$ and $U_{1,m}=(U_{11},U_{12},...,U_{1m})$.
\begin{align*}
&d^2\cdot\cM(m,\hat U_{1,m})-d^2\cdot\cM(r_1, \hat U_1) \\
=& \sum_{h=1}^{h_0}\frac{1}{(T-h)^2}\sum_{s,t=h+1}^T \tr\left\{ \left(\hat U_1 \hat U_1^\top - \hat U_{1,m} \hat U_{1,m}^\top \right) \mat1(X_{t-h}\otimes X_t) {\mat1}^\top (X_{s-h}\otimes X_s)  \right\} \\
=& \sum_{h=1}^{h_0}\sum_{j=m+1}^{r_1}\frac{1}{(T-h)^2}\sum_{s,t=h+1}^T \tr\left\{  {\mat1}^\top (A_1 F_{s-h} A_2^\top\otimes X_s) \hat U_{1j} \hat U_{1j}^\top  \mat1(A_1 F_{t-h} A_2^\top\otimes X_t)  \right\} \\
&+ \sum_{h=1}^{h_0}\sum_{j=m+1}^{r_1}\frac{1}{(T-h)^2}\sum_{s,t=h+1}^T \tr\left\{  {\mat1}^\top ( E_{s-h} \otimes X_s) \hat U_{1j} \hat U_{1j}^\top \mat1(E_{t-h} \otimes X_t)  \right\} \\
&+ \sum_{h=1}^{h_0}\sum_{j=m+1}^{r_1}\frac{1}{(T-h)^2}\sum_{s,t=h+1}^T \tr\left\{  {\mat1}^\top (A_1 F_{s-h} A_2^\top\otimes X_s) \hat U_{1j} \hat U_{1j}^\top \mat1(E_{t-h} \otimes X_t)  \right\} \\
&+ \sum_{h=1}^{h_0}\sum_{j=m+1}^{r_1}\frac{1}{(T-h)^2}\sum_{s,t=h+1}^T \tr\left\{  {\mat1}^\top (E_{s-h} \otimes X_s) \hat U_{1j} \hat U_{1j}^\top \mat1(A_1 F_{t-h} A_2^\top\otimes X_t)  \right\} \\
:=&\I+\II+\III+\IV.
\end{align*}
Note that $\III=\IV$.

We first consider $\I$.
\begin{align*}
\I=& \sum_{h=1}^{h_0}\sum_{j=m+1}^{r_1}\frac{1}{(T-h)^2}\sum_{s,t=h+1}^T \tr\left\{  {\mat1}^\top (A_1 F_{s-h} A_2^\top\otimes A_1 F_s A_2^\top) \hat U_{1j} \hat U_{1j}^\top  \mat1(A_1 F_{t-h} A_2^\top\otimes A_1 F_t A_2^\top)  \right\} \\
&+ \sum_{h=1}^{h_0}\sum_{j=m+1}^{r_1}\frac{1}{(T-h)^2}\sum_{s,t=h+1}^T \tr\left\{  {\mat1}^\top (A_1 F_{s-h} A_2^\top\otimes E_s) \hat U_{1j} \hat U_{1j}^\top  \mat1(A_1 F_{t-h} A_2^\top\otimes E_t)  \right\} \\
&+ 2\sum_{h=1}^{h_0}\sum_{j=m+1}^{r_1}\frac{1}{(T-h)^2}\sum_{s,t=h+1}^T \tr\left\{  {\mat1}^\top (A_1 F_{s-h} A_2^\top\otimes E_s) \hat U_{1j} \hat U_{1j}^\top  \mat1(A_1 F_{t-h} A_2^\top\otimes A_1 F_t A_2^\top)  \right\} \\
:=& \I_1+\I_2+2\I_3.
\end{align*}
For $\I_1$,
\begin{align*}
\I_1&= \sum_{h=1}^{h_0}\sum_{j=m+1}^{r_1}\frac{1}{(T-h)^2}\sum_{s,t=h+1}^T \tr\left\{  \hat U_{1j}^\top  \mat1(A_1 F_{t-h} A_2^\top\otimes A_1 F_t A_2^\top)  {\mat1}^\top (A_1 F_{s-h} A_2^\top\otimes A_1 F_s A_2^\top) \hat U_{1j} \right\} \\
&= \sum_{h=1}^{h_0}\sum_{j=m+1}^{r_1} \tr\left\{  \hat U_{1j}^\top U_1 U_1^\top \mat1 \left(\Theta_{1,h} \right) {\mat1}^\top \left(\Theta_{1,h} \right) U_1  U_1^\top \hat U_{1j} \right\} \\
&\ge \sum_{j=m+1}^{r_1} \lambda_{\min} \left(\sum_{h=1}^{h_0} U_1^\top \mat1 \left(\Theta_{1,h} \right) {\mat1}^\top \left(\Theta_{1,h} \right) U_1 \right) \tr( \hat U_{1j}^\top U_1 U_1^\top \hat U_{1j} ) \\
&=\sigma_{r_1}^2 \left( U_1^\top \overline\E \mat1(\text{TOPUP}_1)\right) \sum_{j=m+1}^{r_1} \hat U_{1j}^\top U_1 U_1^\top \hat U_{1j}.
\end{align*}
Let $W_1\Lambda_0 W_2^\top $ be singular value decomposition of $U_1^\top \hat U_1$, where
$$\Lambda_0=\left(\begin{matrix} \phi_1 & &\\ & \ddots & \\ & & \phi_{r_1}\end{matrix} \right),$$
with $\phi_1\ge \phi_2 \ge ... \ge \phi_{r_1}$. Note that $W_1 W_1^\top = W_1^\top W_1=W_2 W_2^\top =W_2^\top W_2=I_{r_1}$. By Lemma \ref{lemma:sintheta},
\begin{equation*}
\phi_{r_1}^2=\sigma_{\min}(U_1^\top \hat U_1)=1- \|\sin\varTheta(\hat U_1,U_1)\|_2^2\ge 1- \|\hat U_1 \hat U_1^\top -U_1 U_1^\top \|_2^2 .
\end{equation*}
It follows that
\begin{align}\label{eq:topup:lessf:I1}
\hat U_{1j}^\top U_1 U_1^\top \hat U_{1j}&= e_j^\top\hat U_{1}^\top U_1 U_1^\top \hat U_{1} e_j = e_j^\top W_2 \Lambda_0 W_1^\top W_1 \Lambda_0 W_2^\top e_j = e_j^\top W_2 \Lambda_0^2 W_2^\top e_j \notag\\
&\ge \lambda_{\min}(\Lambda_0^2)  e_j^\top W_2 W_2^\top e_j =\phi_{r_1}^2 \notag\\
&\ge 1- \|\hat U_1 \hat U_1^\top -U_1 U_1^\top \|_2^2,
\end{align}
where $e_j$ is a $r_1\times 1$ vector with 1 at $j$-th element and 0 otherwise. Thus, we can obtain
\begin{align}
\I_1&\ge \sigma_{r_1}^2 \left( U_1^\top \overline\E \mat1(\text{TOPUP}_1)\right) (r_1-m) ( 1- \|\hat U_1 \hat U_1^\top -U_1 U_1^\top \|_2^2) .
\end{align}
By Proposition \ref{prop:zt}, in the event $\Omega_0$, $\sigma_{r_1} \left( U_1^\top \overline\E \mat1(\text{TOPUP}_1)\right)=\tau_{1,r_1}\asymp d^{1-\delta_1}$, with $0\le \delta_0\le \delta_1<1$. Hence, there exists a constant $c_m>0$ such that in the event $\Omega_1$,
\begin{align}\label{eq:topup:plim}
\I_1 \ge (c_m+o(1))d^{2-2\delta_1}.
\end{align}

By Lemma \ref{lemma:topup:spectral},
\begin{align*}
\I_2=&\sum_{h=1}^{h_0}\sum_{j=m+1}^{r_1}\frac{1}{(T-h)^2}\sum_{s,t=h+1}^T \tr\left\{  {\mat1}^\top (A_1 F_{s-h} A_2^\top\otimes E_s) \hat U_{1j} \hat U_{1j}^\top  \mat1(A_1 F_{t-h} A_2^\top\otimes E_t)  \right\} \\
=& O\left(\frac{d^{2-\delta_0}}{T} \right),
\end{align*}
and
\begin{align*}
\I_3=&\sum_{h=1}^{h_0}\sum_{j=m+1}^{r_1}\frac{1}{(T-h)^2}\sum_{s,t=h+1}^T \tr\left\{  {\mat1}^\top (A_1 F_{s-h} A_2^\top\otimes U_1^\top E_s U_2) \hat U_{1j} \hat U_{1j}^\top  \mat1(A_1 F_{t-h} A_2^\top\otimes U_1^\top A_1 F_t A_2^\top U_2)  \right\} \\
=& O\left(\frac{d^{3/2-3\delta_0/2}}{T^{1/2}} \right).
\end{align*}
This implies that
\begin{align}\label{eq:topup:lessf:i23}
\I_2+\I_3=& O\left(\frac{d^{2-\delta_0}}{T} + \frac{d^{3/2-3\delta_0/2}}{T^{1/2}} \right).
\end{align}

Next, we consider $\II$. 
\begin{align*}
\II=&\sum_{h=1}^{h_0}\sum_{j=m+1}^{r_1}\frac{1}{(T-h)^2}\sum_{s,t=h+1}^T \tr\left\{  {\mat1}^\top ( E_{s-h} \otimes E_s) \hat U_{1j} \hat U_{1j}^\top \mat1(E_{t-h} \otimes E_t)  \right\} \\
&+ \sum_{h=1}^{h_0}\sum_{j=m+1}^{r_1}\frac{1}{(T-h)^2}\sum_{s,t=h+1}^T \tr\left\{  {\mat1}^\top ( E_{s-h} \otimes A_1 F_s A_2^\top) \hat U_{1j} \hat U_{1j}^\top \mat1(E_{t-h} \otimes A_1 F_t A_2^\top)  \right\} \\
&+ 2\sum_{h=1}^{h_0}\sum_{j=m+1}^{r_1}\frac{1}{(T-h)^2}\sum_{s,t=h+1}^T \tr\left\{  {\mat1}^\top ( E_{s-h} \otimes A_1 F_s A_2^\top) \hat U_{1j} \hat U_{1j}^\top \mat1(E_{t-h} \otimes E_t)  \right\} \\
:=&\II_1+\II_2+2\II_3.
\end{align*}
By Lemma \ref{lemma:topup:spectral},
\begin{align*}
\II_1=O \left( \frac{d^2}{Td_1} \right)
\end{align*}
Using the same decomposition \eqref{eq:split:u1},
\begin{align*}
\II_2=& \sum_{h=1}^{h_0}\sum_{j=m+1}^{r_1}\frac{1}{(T-h)^2}\sum_{s,t=h+1}^T \tr\left\{  {\mat1}^\top ( E_{s-h} \otimes A_1 F_s A_2^\top) \left(\hat U_{1j}-U_1 \tilde U_1^\top \hat U_{1j}\right) \left(\hat U_{1j}-U_1 \tilde U_1^\top \hat U_{1j}\right)^\top \right.\\
&\qquad\qquad\qquad\qquad\qquad\qquad\quad \left.\cdot \mat1(E_{t-h} \otimes A_1 F_t A_2^\top)  \right\} \\
&+ 2 \sum_{h=1}^{h_0}\sum_{j=m+1}^{r_1}\frac{1}{(T-h)^2}\sum_{s,t=h+1}^T \tr\left\{  {\mat1}^\top ( E_{s-h} \otimes A_1 F_s A_2^\top) \left(\hat U_{1j}-U_1 \tilde U_1^\top \hat U_{1j}\right) \left(U_1 \tilde U_1^\top \hat U_{1j}\right)^\top \right. \\
&\qquad\qquad\qquad\qquad\qquad\qquad\quad \left.\cdot \mat1(E_{t-h} \otimes A_1 F_t A_2^\top)  \right\} \\
&+ \sum_{h=1}^{h_0}\sum_{j=m+1}^{r_1}\frac{1}{(T-h)^2}\sum_{s,t=h+1}^T \tr\left\{  {\mat1}^\top ( E_{s-h} \otimes A_1 F_s A_2^\top) \left(U_1 \tilde U_1^\top \hat U_{1j}\right) \left(U_1 \tilde U_1^\top \hat U_{1j}\right)^\top \right. \\
&\qquad\qquad\qquad\qquad\qquad\qquad\quad \left.\cdot \mat1(E_{t-h} \otimes A_1 F_t A_2^\top)  \right\} \\
=& O \left( \frac{d_1d^{1-\delta_0}\eta_d^2}{T^2}+\frac{d^{2-\delta_0}}{Td_1} \right)
\end{align*}
Similarly, 
\begin{align*}
\II_3=&\sum_{h=1}^{h_0}\sum_{j=m+1}^{r_1}\frac{1}{(T-h)^2}\sum_{s,t=h+1}^T \tr\left\{  {\mat1}^\top ( E_{s-h} \otimes U_1^\top  A_1 F_s A_2^\top U_2) \left(\hat U_{1j}-U_1 \tilde U_1^\top \hat U_{1j}\right) \right. \\
&\qquad\qquad\qquad\qquad\qquad\quad \left.\cdot \left(\hat U_{1j}-U_1 \tilde U_1^\top \hat U_{1j}\right)^\top \mat1(E_{t-h} \otimes U_1^\top E_t U_2)  \right\} \\
&+\sum_{h=1}^{h_0}\sum_{j=m+1}^{r_1}\frac{1}{(T-h)^2}\sum_{s,t=h+1}^T \tr\left\{  {\mat1}^\top ( E_{s-h} \otimes U_1^\top  A_1 F_s A_2^\top U_2) \left(\hat U_{1j}-U_1 \tilde U_1^\top \hat U_{1j}\right) \right. \\
&\qquad\qquad\qquad\qquad\qquad\quad \left.\cdot \left(U_1 \tilde U_1^\top \hat U_{1j}\right)^\top \mat1(E_{t-h} \otimes U_1^\top E_t U_2)  \right\} \\
&+\sum_{h=1}^{h_0}\sum_{j=m+1}^{r_1}\frac{1}{(T-h)^2}\sum_{s,t=h+1}^T \tr\left\{  {\mat1}^\top ( E_{s-h} \otimes U_1^\top  A_1 F_s A_2^\top U_2) \left(U_1 \tilde U_1^\top \hat U_{1j}\right) \right. \\
&\qquad\qquad\qquad\qquad\qquad\quad \left.\cdot \left(\hat U_{1j}-U_1 \tilde U_1^\top \hat U_{1j}\right)^\top \mat1(E_{t-h} \otimes U_1^\top E_t U_2)  \right\} \\
&+ \sum_{h=1}^{h_0}\sum_{j=m+1}^{r_1}\frac{1}{(T-h)^2}\sum_{s,t=h+1}^T \tr\left\{  {\mat1}^\top ( E_{s-h} \otimes U_1^\top  A_1 F_s A_2^\top U_2) \left(U_1 \tilde U_1^\top \hat U_{1j}\right) \right. \\
&\qquad\qquad\qquad\qquad\qquad\quad \left.\cdot \left(U_1 \tilde U_1^\top \hat U_{1j}\right)^\top \mat1(E_{t-h} \otimes U_1^\top E_t U_2)  \right\} \\
=& O \left( \frac{d^{3/2-\delta_0/2}}{Td_1^{1/2}}+\frac{d_1^{1/2}d^{1-\delta_0/2}\eta_d^2}{T^2} \right)
\end{align*}
Combing $\II_1,\II_2$ and $\II_3$, we have 
\begin{align}\label{eq:topup:lessf:ii}
\II=& O \left(\frac{d^{2}}{Td_1} + \frac{d_1d^{1-\delta_0}\eta_d^2}{T^{2}} \right).
\end{align}

Next, we consider $\III$.
\begin{align*}
\III=&\sum_{h=1}^{h_0}\sum_{j=m+1}^{r_1}\frac{1}{(T-h)^2}\sum_{s,t=h+1}^T \tr\left\{  {\mat1}^\top (A_1 F_{s-h} A_2^\top\otimes E_s) \hat U_{1j} \hat U_{1j}^\top \mat1(E_{t-h} \otimes E_t)  \right\} \\
&+ \sum_{h=1}^{h_0}\sum_{j=m+1}^{r_1}\frac{1}{(T-h)^2}\sum_{s,t=h+1}^T \tr\left\{  {\mat1}^\top (A_1 F_{s-h} A_2^\top\otimes A_1 F_s A_2^\top) \hat U_{1j} \hat U_{1j}^\top \mat1(E_{t-h} \otimes A_1 F_t A_2^\top)  \right\} \\
&+ \sum_{h=1}^{h_0}\sum_{j=m+1}^{r_1}\frac{1}{(T-h)^2}\sum_{s,t=h+1}^T \tr\left\{  {\mat1}^\top (A_1 F_{s-h} A_2^\top\otimes E_s) \hat U_{1j} \hat U_{1j}^\top \mat1(E_{t-h} \otimes A_1 F_t A_2^\top)  \right\} \\
&+ \sum_{h=1}^{h_0}\sum_{j=m+1}^{r_1}\frac{1}{(T-h)^2}\sum_{s,t=h+1}^T \tr\left\{  {\mat1}^\top (A_1 F_{s-h} A_2^\top\otimes A_1 F_s A_2^\top) \hat U_{1j} \hat U_{1j}^\top \mat1(E_{t-h} \otimes E_t)  \right\} \\
:=&\III_1+\III_2+\III_3+\III_4.
\end{align*}
Again, we bound each term in turn. Using the decomposition \eqref{eq:split:u1}, adopting same arguments in the proof of $\II$, we have,
\begin{align*}
\III_1&= O \left(\frac{d^{3/2-\delta_0/2}}{T}  \right), \\
\III_2&= O \left(\frac{d^{3/2-3\delta_0/2}}{T^{1/2}} + \frac{d_1^{1/2}d^{3/2-3\delta_0/2}\eta_d}{T} \right),\\
\III_3&= O \left(\frac{d^{1-\delta_0}}{T } + \frac{d_1^{1/2}d^{1-\delta_0}\eta_d}{T^{3/2}} \right),\\
\III_4&= O \left(\frac{d^{1-\delta_0}}{T^{1/2}} + \frac{d_1^{1/2}d^{1-\delta_0}\eta_d}{T} \right) .
\end{align*}
This implies that
\begin{align}\label{eq:topup:lessf:iii}
\III=& O \left(\frac{d^{3/2-\delta_0/2}}{T^{1/2}} + \frac{d_1^{1/2}d^{3/2-\delta_0/2}\eta_d}{T} \right).
\end{align}
Employing \eqref{eq:topup:plim}, \eqref{eq:topup:lessf:i23}, \eqref{eq:topup:lessf:ii} and \eqref{eq:topup:lessf:iii}, in the event $\Omega_1$, there exists a constant $c_m>0$,
\begin{align}
\cM(m,\hat U_{1,m})-\cM(r_1, \hat U_1) \ge (c_m+ o(1)) d^{-2\delta_1} + C\left( \frac{1}{Td^{\delta_0}} + \frac{1}{Td_1} + \frac{1}{T^{1/2}d^{1/2+3\delta_0/2}} + \frac{d_1^{1/2}\eta_d}{Td^{1/2+3\delta_0/2}} \right).
\end{align}

\end{proof}

\begin{proof}[Proof of Theorem \ref{thm:all} for non-iterative TOPUP]
Let $\Omega_1=\Omega_{11}\cap\Omega_{12}\cap\Omega_0$, where $\Omega_{11}, \Omega_{12}$ are defined in Lemma \ref{lemma:topup:rate} and \ref{lemma:topup:spectral}, respectively, and $\Omega_0$ is the event in Proposition \ref{prop:zt}. Then $\Omega_1$ is the same event in Lemma \ref{lemma:topup:upf} and \ref{lemma:topup:lessf}, with $\P(\Omega_1)\ge 1 -e^{-d_2}-T\exp(-C_1 T^\vartheta)-\exp(-C_2T)$ and $C_1,C_2>0, 1/\vartheta=1/\theta_1+2/\theta_2$ . In the following, we shall only work on $\Omega_1$.

(i) (IC estimator using non-iterative TOPUP) We shall prove in the event $\Omega_1$, $\cM(m,\hat U_{1,m})+m\cdot d^{-2}g_1(d,T)\ge \cM(r_1,\hat U_{1})+r_1\cdot d^{-2}g_1(d,T) $ for all $m\neq r_1$ and $m\le m_1$. For $m<r_1$,
It is sufficient to show $\cM(m,\hat U_{1,m})-\cM(r_1,\hat U_{1})\ge (r_1-m)d^{-2}g_1(d,T)>0 $. This is followed by Lemma \ref{lemma:topup:lessf} and the second part of Assumption \ref{asmp:penalty}(a). For $m> r_1$, as $\cM(m,\hat U_{1,m})\le \cM(r_1,\hat U_{1})$, it is sufficient to prove $\cM(r_1,\hat U_{1})-\cM(m,\hat U_{1,m})\le (m-r_1)d^{-2}g_1(d,T)$. By Lemma \ref{lemma:topup:upf}, $\cM(r_1,\hat U_{1})-\cM(m,\hat U_{1,m})\le C\left(\beta_{d,T} \right)$. Then the first case is followed by the second part of Assumption \ref{asmp:penalty}(a). 

(ii) (ER estimator using non-iterative TOPUP) By Lemma \ref{lemma:topup:upf}, for all $j>r_1$, $\hat\lambda_j\le C_{11}(d^2\beta_{d,T})$. Lemma \ref{lemma:topup:lessf} implies that $\hat\lambda_j\ge (c_m+o(1)) d^{2-2\delta_1}+C_{12}(d^2 \gamma_{d,T})$ for all $1\le j \le r_1$. Employing similar arguments in the proof of Lemma \ref{lemma:topup:lessf}, we can obtain $\hat\lambda_1\le (c_m^*+o(1)) d^{2-2\delta_0}+ C_{13} (d^2 \gamma_{d,T})$, for some constant $c_m^*>0$. Provided that $h_1(d,T)=o(d^{2+2\delta_0-4\delta_1})=o(d^{2-2\delta_1})$ and $\gamma_{d,T}=o(d^{-2\delta_1})$,
\begin{align*}
& \frac{\hat\lambda_{j+1}+h_1(d,T)}{\hat\lambda_j+h_1(d,T)} \ge \frac{h_1(d,T)}{h_1(d,T)+ C_{14}(d^2\beta_{d,T}) }, \quad\text{for } j>r_1,\\
& \frac{\hat\lambda_{j+1}+h_1(d,T)}{\hat\lambda_j+h_1(d,T)}\ge (c_1+o(1)) d^{2\delta_0-2\delta_1}, \quad\text{for } 1\le j<r_1,   \\
& \frac{\hat\lambda_{j+1}+h_1(d,T)}{\hat\lambda_j+h_1(d,T)} \le c_2 d^{2\delta_1}(d^{-2} h_1(d,T)+ C_{15} (\beta_{d,T})), \quad\text{for } j=r_1,
\end{align*}
where $c_1$ and $c_2$ are some positive constants. Under the condition that $\beta_{d,T}+d^{-2}h_1(d,T)=o(d^{2\delta_0-4\delta_1})$, we have $d^{2\delta_1}(d^{-2} h_1(d,T)+\beta_{d,T})=o(d^{2\delta_0-2\delta_1})$. Furthermore, under the condition that $h_1(d,T)\gg d^{2+2\delta_1}\beta_{d,T}^2$, we can show that
\begin{align*}
d^{2\delta_1}(d^{-2} h_1(d,T)+\beta_{d,T}) \ll  \frac{h_1(d,T)}{h_1(d,T)+d^2\beta_{d,T} }.   
\end{align*}
Thus, (ii) is followed by Assumption \ref{asmp:penalty}(b). 
\end{proof}

\subsection{Proof of Theorem \ref{thm:all} for iTOPUP}

\begin{lemma}\label{lemma:topup:spectral:iterative}
Suppose Assumptions \ref{asmp:error}, \ref{asmp:mixing}, \ref{asmp:factor}, \ref{asmp:strength} and \ref{asmp:rank}(a) hold. Assume that $\|\hat U_{2} \hat U_2^\top - U_2 U_2^\top \|_2\le c_0\sqrt{d_1/d_2}$ for some positive constant $c_0$. Then, in an event $\Omega_{22}\cap\Omega_0$ with $\P(\Omega_{22})\ge 1-e^{-d_2}/2$ and $\P(\Omega_0)\ge 1- T\exp(-C_1 T^\vartheta)-\exp(-C_2T)$ with $C_1,C_2>0, 1/\vartheta=1/\theta_1+2/\theta_2$, for all fixed $m$ such that $m\ge r_2$, we have
\begin{align}
&\left\| \frac{1}{T-h} \sum_{t=h+1}^T \mat1(A_1 F_{t-h} A_2^\top \hat U_{2,m}\otimes E_t \hat U_{2,m} )  \right\|_2 \le C \left( \frac{d_1^{1/2}d^{1/2-\delta_0/2} }{\sqrt{T}} \right), \label{eq:topup:spectral:iter1}\\
&\left\| \frac{1}{T-h} \sum_{t=h+1}^T \mat1(A_1 F_{t-h} A_2^\top \hat U_{2,m}\otimes U_1^\top E_t \hat U_{2,m} )  \right\|_2 \le C \left( \frac{d^{1/2-\delta_0/2} }{\sqrt{T}} \right), \label{eq:topup:spectral:iter2} \\
&\left\| \frac{1}{T-h} \sum_{t=h+1}^T \mat1(E_{t-h} \hat U_{2,m}\otimes A_1 F_{t} A_2^\top \hat U_{2,m} )  \right\|_2 \le C \left(\frac{d_1^{1/2} d^{1/2-\delta_0/2} }{\sqrt{T}} \right), \label{eq:topup:spectral:iter3} \\
&\left\| \frac{1}{T-h} \sum_{t=h+1}^T U_1^\top\mat1(E_{t-h}\hat U_{2,m}\otimes A_1 F_{t} A_2^\top \hat U_{2,m} )  \right\|_2 \le C \left(\frac{d^{1/2-\delta_0/2} }{\sqrt{T}} \right), \label{eq:topup:spectral:iter5} \\ 
&\left\| \frac{1}{T-h} \sum_{t=h+1}^T \mat1(E_{t-h} \hat U_{2,m}\otimes E_{t} \hat U_{2,m} )  \right\|_2 \le C \left(\frac{d_1^{1/2} }{\sqrt{T}} \right), \label{eq:topup:spectral:iter6} \\
&\left\| \frac{1}{T-h} \sum_{t=h+1}^T \mat1(E_{t-h} \hat U_{2,m}\otimes U_1^\top E_{t} \hat U_{2,m} )  \right\|_2 \le C \left(\frac{d_1^{1/2} }{\sqrt{T}} \right), \label{eq:topup:spectral:iter8} \\
&\left\| \frac{1}{T-h} \sum_{t=h+1}^T U_1^\top\mat1(E_{t-h} \hat U_{2,m}\otimes U_1^\top E_{t} \hat U_{2,m} )  \right\|_2 \le C \left(\frac{1}{\sqrt{T}} \right), \label{eq:topup:spectral:iter11}
\end{align}
for some constant positive $C$ depending on $K$ only.
\end{lemma}
\begin{proof}
Recall $\hat U_2=\hat U_{2,r_2}=[\hat U_{21},...,\hat U_{2r_2}] $. Write $U_{2,m}=[U_2, U_{2}^c]$ for $m>r_2$. Let $U_{2\perp}$ be complement part of $U_2$, namely $[U_2,U_{2\perp}]\in\bO_{d_2}$. If $m\le r_2$, then $\|\hat U_{2,m}^\top U_{2\perp} \|_2\le \| \hat U_2^\top U_{2\perp} \|_2=\| \hat U_2 \hat U_2^\top - U_2 U_2^\top \|_2$. If $m>r_2$, then $\hat U_{2,m} \hat U_{2,m}^\top = \hat U_{2} \hat U_{2}^\top +\sum_{j=r_2+1}^m \hat U_{2j} \hat U_{2j}^\top=  \hat U_{2} \hat U_{2}^\top + \hat U_{2}^c \hat U_{2}^{c\top}$. Furthermore, for any matrix $B\in\R^{a\times d_2}$, 
$$\|B\hat U_{2,m}\|_2 = \| [B\hat U_{2}, B\hat U_{2}^c] \|_2 \le \sqrt{2}\left( \| B\hat U_{2}\|_2 +\| B\hat U_{2}^c \|_2 \right) .$$
Note that $$\|U_{2} U_{2}^\top \hat U_{2}^c \|_2=\|U_{2}^\top \hat U_{2}^c \|_2 \le \|U_{2}^\top \hat U_{2\perp} \|_2 = \| \hat U_2 \hat U_2^\top - U_2 U_2^\top \|_2.$$ 
Adopting similar procedures in the proof of Theorem 1 in \citet{han2020}, we can show Lemma \ref{lemma:topup:spectral:iterative}.
\end{proof}

\begin{lemma}\label{lemma:topup:upf:iterative}
Suppose Assumptions \ref{asmp:error}, \ref{asmp:mixing}, \ref{asmp:factor}, \ref{asmp:strength} and \ref{asmp:rank}(a) hold. Let $r_1\le r_1^{(j)}\le m_1<d_1$ for all $1\le j\le i-1$, and $m_1=O(r_1)$. There exists an event $\Omega_2$ such that $\P(\Omega_2)\ge 1 -e^{-d_2}-T\exp(-C_1 T^\vartheta)-\exp(-C_2T)$, $C_1,C_2>0, 1/\vartheta=1/\theta_1+2/\theta_2$ and $\Omega_2$ is independent of iteration number $i$. Then, at $i$-th iteration, in the event $\Omega_2$, for any fixed $m$ with $m> r_1$, $\cM^{(i)}(r_1,\hat U_1) - \cM^{(i)}(m,\hat U_{1,m}) \le C\beta_{d,T}$, where $C>0$,
\begin{equation*}
\beta_{d,T}=
\frac{d_1 }{Td^{1+\delta_0/2}}+ \frac{d_1^{1/2}\eta_1^{(i)}}{Td^{1/2+3\delta_0/2}},
\end{equation*}
and $\eta_k^{(i)}=d_k^{1/2}d^{\delta_1-\delta_0/2-1/2} + d_k^{1/2}d^{\delta_1-1}$.
\end{lemma}
\begin{proof}
Note that by \cite{han2020}, in the event $\Omega_2$, at $i$-th iteration ($i\ge1$), for the iTOPUP estimator, assume that $\| \hat U_{k,r_k}^{(i)} \hat U_{k,r_k}^{(i)\top}- U_k U_k^\top)\|_2 \le C_0(\eta_k^{(i)}T^{-1/2})$, $1\le k\le K$, for some $C_0>0$. The proof is similar to Lemma \ref{lemma:topup:upf}. Thus, it is omitted.
\end{proof}

\begin{lemma}\label{lemma:topup:lessf:iterative}
Suppose Assumptions \ref{asmp:error}, \ref{asmp:mixing}, \ref{asmp:factor}, \ref{asmp:strength} and \ref{asmp:rank}(a) hold. Let $r_1\le r_1^{(j)}\le m_1<d_1$ for all $1\le j\le i-1$, and $m_1=O(r_1)$. There exists an event $\Omega_2$ such that $\P(\Omega_2)\ge 1 -e^{-d_2}-T\exp(-C_1 T^\vartheta)-\exp(-C_2T)$, $C_1,C_2>0, 1/\vartheta=1/\theta_1+2/\theta_2$ and $\Omega_2$ is independent of iteration number $i$. Then, at $i$-th iteration, in the event $\Omega_2$, for any $m$ with $m< r_1$, there exist constants $c_m>0$ and $C>0$, such that
$\cM^{(i)}(m,\hat U_{1,m})-\cM^{(i)}(r_1,\hat U_1)\ge (c_m+o(1)) d^{-2\delta_1}+ C \gamma_{d,T}$, where 
\begin{equation*}
\gamma_{d,T}=\frac{d_1}{Td^{1+\delta_0}} + \frac{1}{T^{1/2}d^{1/2+3\delta_0/2}} + \frac{d_1^{1/2}\eta_1^{(i)}}{Td^{1/2+3\delta_0/2}},
\end{equation*}
and $\eta_k^{(i)}=d_k^{1/2}d^{\delta_1-\delta_0/2-1/2} + d_k^{1/2}d^{\delta_1-1}$.
\end{lemma}
\begin{proof}
Note that by \cite{han2020}, in the event $\Omega_2$, at $i$-th iteration ($i\ge1$), for the iTOPUP estimator, assume that $\| \hat U_{k,r_k}^{(i)} \hat U_{k,r_k}^{(i)\top}- U_k U_k^\top)\|_2 \le C_0(\eta_k^{(i)}T^{-1/2})$, $1\le k\le K$, for some $C_0>0$. The proof is omitted, as it is similar to Lemma \ref{lemma:topup:lessf}. 
\end{proof}

\begin{proof}[Proof of Theorem \ref{thm:all} for iTOPUP]
Employing similar arguments in the proofs of Theorem \ref{thm:all} for non-iterative TOPUP we can show Theorem \ref{thm:all} for iTOPUP.
\end{proof}

\subsection{Proof of Theorem \ref{thm:all} for non-iterative TIPUP }

\begin{lemma}\label{lemma:tipup:rate}
Let $U_{1\perp}\in \RR^{d_1\times (d_1-r_1)}$ be the orthonormal complement of $U_1$. Denote $U_{1\perp}=(U_{1j})_{r_1+1\le j\le d_1}$ and $U_1=(U_{1j})_{1\le j\le r_1}$, where $U_{1j}\in \RR^{d_1}$. Denote $\eta_d^*=d_k^{1/2}d^{\delta_1-\delta_0/2-1/2} + d^{\delta_1-1/2}$. Define $\hat U_1$ be the estimated top $r_1$ left singular space of $\text{TIPUP}_1(X_{1:T})$. Suppose Assumptions \ref{asmp:error}, \ref{asmp:mixing}, \ref{asmp:factor}, \ref{asmp:strength} and \ref{asmp:rank}(b) hold. Then, in an event $\Omega_{31}\cap \Omega_0$ with $\P(\Omega_{31})\ge 1-e^{-d_2}/2$ and $\P(\Omega_0)\ge 1-T\exp(-C_1 T^\vartheta)-\exp(-C_2T)$ with $C_1,C_2>0, 1/\vartheta=1/\theta_1+2/\theta_2$, there exist a matrix $\tilde U_1\in\RR^{d_1\times r_1}$ with $\tilde U_1 \tilde U_1^\top =\hat U_1 \hat U_1^\top$, such that
\begin{align}
\| \hat U_1 -U_1\tilde U_1^\top \hat U_1 \|_2 & \le C \left(\frac{\eta_d^*}{\sqrt{T}}\right), \label{eq:tipup:rate:a} \\
\|U_1 U_1^\top \hat U_{1j}\|_2 & \le C \left(\frac{\eta_d^*}{\sqrt{T}}\right), \label{eq:tipup:rate:b}
\end{align}
for all $r_1+1\le j\le d_1$ and $C>0$.
\end{lemma}
The proof is similar to Lemma \ref{lemma:topup:rate}, thus is omitted.

\begin{lemma}\label{lemma:tipup:spectral}
Suppose Assumptions \ref{asmp:error}, \ref{asmp:mixing}, \ref{asmp:factor}, \ref{asmp:strength} and \ref{asmp:rank}(b) hold. Then, in an event $\Omega_{32}\cap \Omega_0$ with $\P(\Omega_{32})\ge 1-e^{-d_1}/2$ and $\P(\Omega_0)\ge 1-T\exp(-C_1 T^\vartheta)-\exp(-C_2T)$ with $C_1,C_2>0, 1/\vartheta=1/\theta_1+2/\theta_2$, we have
\begin{align}
&\left\| \frac{1}{T-h} \sum_{t=h+1}^T A_1 F_{t-h} A_2^\top E_t^\top   \right\|_2 \le C \left( \frac{d_1^{1/2}d^{1/2-\delta_0/2} }{\sqrt{T}} \right), \label{eq:tipup:spectral:1} \\
&\left\| \frac{1}{T-h} \sum_{t=h+1}^T E_{t-h} A_2 F_t^\top A_1^\top   \right\|_2 \le C \left( \frac{d_1^{1/2}d^{1/2-\delta_0/2} }{\sqrt{T}} \right), \label{eq:tipup:spectral:2} \\
&\left\| \frac{1}{T-h} \sum_{t=h+1}^T A_1 F_{t-h} A_2^\top E_t^\top U_1  \right\|_2 \le C \left( \frac{d^{1/2-\delta_0/2} }{\sqrt{T}} \right), \label{eq:tipup:spectral:3} \\
&\left\| \frac{1}{T-h} \sum_{t=h+1}^T U_1^\top E_{t-h} A_2^\top F_t^\top A_1^\top  \right\|_2 \le C \left( \frac{d^{1/2-\delta_0/2} }{\sqrt{T}} \right), \label{eq:tipup:spectral:4} \\
&\left\| \frac{1}{T-h} \sum_{t=h+1}^T E_{t-h} E_t^\top  \right\|_2 \le C \left( \frac{d^{1/2} }{\sqrt{T}} \right), \label{eq:tipup:spectral:5} \\
&\left\| \frac{1}{T-h} \sum_{t=h+1}^T U_1^\top E_{t-h} E_t^\top U_1  \right\|_2 \le C \left( \frac{d^{1/2} }{\sqrt{d_1T}} \right), \label{eq:tipup:spectral:6} 
\end{align}
for some constant positive $C$ depending on $K$ only.
\end{lemma}

\begin{proof}
By Theorem 2 in \citet{chen2021factor}, under Assumption \ref{asmp:error}, 
\begin{align*}
\overline\E\left\| \frac{1}{T-h} \sum_{t=h+1}^T A_1 F_{t-h} A_2^\top E_t^\top \right\|_2 \le \frac{\sigma \sqrt{8Td_1}}{T-h} \|\Theta_{1,0}^* \|_2^{1/2}.
\end{align*}
Elementary calculation shows that
\begin{eqnarray*}
&&\left| \left\| \sum_{t=h+1}^T A_1 F_{t-h} A_2^\top E_t^\top \right\|_{2} - \left\| \sum_{t=h+1}^T A_1 F_{t-h} A_2^\top E_t^{*\top} \right\|_{2} \right| \\
&\le& \left\| (A_1F_1A_2^\top,...,A_1 F_{T-h}A_2^\top) \begin{pmatrix}
E_{h+1}^\top-E_{h+1}^{*\top} \\
\vdots\\
E_{T}^\top-E_{T}^{*\top}
\end{pmatrix} \right\|_{2} \\
&\le& \left\| 
(A_1F_1A_2^\top,...,A_1 F_{T-h}A_2^\top)
\right\|_{2}^{1/2}
\left\| 
\begin{pmatrix} E_{h+1}^\top-E_{h+1}^{*\top} \\
\vdots\\
E_{T}^\top-E_{T}^{*\top}
\end{pmatrix}\right\|_{2} \\
&\le& \sqrt{T} \|\Theta_{1,0}^*\|_{2}^{1/2} \left\|\begin{pmatrix}
E_{h+1}^\top-E_{h+1}^{*\top} \\
\vdots\\
E_{T}^\top-E_{T}^{*\top}
\end{pmatrix} \right\|_{\rm F}.
\end{eqnarray*}
That is, $\left\| \sum_{t=h+1}^T A_1 F_{t-h} A_2^\top  E_t^\top \right\|_{2}$ is a $\sigma\sqrt{T}\|\Theta_{1,0}^*\|_{2}^{1/2}$ Lipschitz function in $(E_1,\ldots,E_T)$. Then, by Gaussian concentration inequalities for Lipschitz functions,
\begin{eqnarray*}
\P\left( \left\| \sum_{t=h+1}^T \frac{A_1 F_{t-h} A_2^\top E_t^\top} {T-h} \right\|_{2}
- \frac{\sigma(8Td_1)^{1/2}}{T-h} \|\Theta_{1,0}^*\|_{2}^{1/2}
\ge \frac{\sigma\sqrt{T} }{T-h} \|\Theta_{1,0}^*\|_{2}^{1/2} x\right)\le 2e^{-\frac{x^2}{2 } }.
\end{eqnarray*}
As $T\ge 4h_0$ and $K=2$, this implies with $x\asymp \sqrt{d_1}$ that in an event $\Omega_a$
with at least probability $1-e^{-d_1}/6$,
\begin{align}
\left\| \frac{1}{T-h} \sum_{t=h+1}^T A_1 F_{t-h} A_2^\top E_t^\top \right\|_2 \le \frac{C_0 \sqrt{Td_1}}{T-h} \|\Theta_{1,0}^* \|_2^{1/2},
\end{align}
with a constant $C_0$ depending on $K$ only. Then, by Proposition \ref{prop:zt}, in the event $\Omega_a\cap\Omega_0$, \eqref{eq:tipup:spectral:1} follows. Similar arguments yield \eqref{eq:tipup:spectral:2}, \eqref{eq:tipup:spectral:3} and \eqref{eq:tipup:spectral:4} in the event $\Omega_a\cap\Omega_0$.

We split the sum into two terms over the index sets, $S_1=\{(h,2h]\cup(3h,4h]\cup\cdots\} \cap(h,T]$ and its complement $S_2$ in $(h,T]$, so that $\{E_{t-h},t\in S_a\}$ is independent of $\{E_t, t\in S_a\}$ for each $a=1,2$. Let $n_a=|S_a|d_2$. Define $G_a=(E_{t-h},t\in S_a)\in \R^{d_1\times n_a}$ and
$H_a=(E_{t},t\in S_a)\in \R^{d_1\times n_a}$. Then, $G_a$, $H_a$ are two independent Gaussian matrices. Note that
\begin{equation*}
\left\| \sum_{t\in S_a} \frac{E_{t-h} E_t^\top} {T-h}\right\|_{2} = \left\|\frac{G_a H_a^\top}{T-h} \right\|_{2}.
\end{equation*}
Moreover, by Assumption \ref{asmp:error}, $\Var(u^\top \text{vec}(G_a))\le \sigma^2$ and
$\Var(u^\top \text{vec}(H_a))\le \sigma^2$ for all unit vectors $u\in\R^{d_1n_a}$,
so that by Lemme \ref{lm-GH}(i)
\begin{eqnarray*}
\P\Big\{ \|G_aH_a^\top\|_{2}/\sigma^2\ge
d_1+2\sqrt{d_1n_a}
+ x(x+2\sqrt{n_a}+2\sqrt{d_1})
\Big\}
\le 2e^{-x^2/2},\quad x>0.
\end{eqnarray*}
Thus, as $\sum_{a=1}^2 n_a=d_2(T-h)$, it follows from the above inequality that with $h_0\le T/4$, $x\asymp \sqrt{d_1}$ and some constant $C_{0}'$ depending on $K$ only,
\begin{align}
\left\| \sum_{t=h+1}^T \frac{E_{t-h} E_t^\top} {T-h}\right\|_{2} \le \frac{C_0'(d_1+\sqrt{d_1d_2T})}{T-h} ,
\end{align}
in an event $\Omega_c$ with at least probability $1-e^{- d_1}/6$. Then, as in the derivation of $\|\Delta_3^*\|_2$ in the proof of Theorem 2 in \citet{chen2021factor}, in the event $\Omega_c$, \eqref{eq:tipup:spectral:5} follows. Similar arguments yield \eqref{eq:tipup:spectral:6} in the event $\Omega_c$. Set $\Omega_{32}=\Omega_a\cap\Omega_c$, then $\P(\Omega_{32})\ge 1-e^{-d_1}/2$.

\end{proof}

\begin{lemma}\label{lemma:tipup:upf}
Suppose Assumptions \ref{asmp:error}, \ref{asmp:mixing}, \ref{asmp:factor}, \ref{asmp:strength} and \ref{asmp:rank}(b) hold. In an event $\Omega_3$ with $\P(\Omega_3)\ge 1-e^{-d_1}-e^{-d_2}-T\exp(-C_1 T^\vartheta)-\exp(-C_2T)$ with $C_1,C_2>0, 1/\vartheta=1/\theta_1+2/\theta_2$, for any fixed $m$ with $m > r_1$, $\cM^*(r_1,\hat U_1)-\cM^*(m,\hat U_{1,m}) \le C \beta_{d,T}^*$, where $C>0$,
\begin{equation*}
\beta_{d,T}^*= \frac{1}{Td}+ \frac{d_1 }{Td^{1+\delta_0}} + \frac{d_1^{1/2} \eta_d^*}{Td^{1/2+3\delta_0/2}}  + \frac{\eta_d^*}{Td^{1/2+\delta_0}} ,
\end{equation*}
and $\eta_d^*=d_k^{1/2}d^{\delta_1-\delta_0/2-1/2} + d^{\delta_1-1/2}$.
\end{lemma}

\begin{proof}
Let $\Omega_3=\Omega_{31}\cap\Omega_{32}\cap\Omega_0$, where $\Omega_{31}, \Omega_{32}$ are defined in Lemma \ref{lemma:tipup:rate} and \ref{lemma:tipup:spectral}, respectively, and $\Omega_0$ is the event in Proposition \ref{prop:zt}. 
Let $U_{1,r_1}=U_{1}$ and $U_{1,m}=(U_{11},U_{12},...,U_{1m})$.
\begin{align*}
\left| \cM^*(m,\hat U_{1,m})-\cM^*(r_1,\hat U_1) \right| &\le \left| \cM^*(m,\hat U_{1,m})-\cM^*(r_1, U_1) \right| + \left| \cM^*(r_1,\hat U_1)-\cM^*(r_1, U_1) \right| \\
&\le 2 \max_{r_1< m\le m_1} \left| \cM^*(m,\hat U_{1,m})-\cM^*(r_1, U_1) \right|.
\end{align*}
As $m_1$ is fixed, it is sufficient to prove for each $m$ with $m > r_1$,
\begin{equation}
\left| \cM^*(m,\hat U_{1,m})-\cM^*(r_1, U_1) \right| \le C \beta_{d,T}^*.
\end{equation}
In the following, we shall only work on $\Omega_3$.

Elementary calculation shows that
\begin{align*}
&d^2\cdot\cM^*(m,\hat U_{1,m})-d^2\cdot\cM^*(r_1, U_1) \\
=& \sum_{h=1}^{h_0} \frac{1}{(T-h)^2} \sum_{s,t=h+1}^T \tr\left\{ \left(U_1 U_1^\top -\hat U_{1,m} \hat U_{1,m}^\top \right) X_{t-h} X_t^\top X_s X_{s-h}^\top \right\} \\
=& \sum_{h=1}^{h_0} \frac{1}{(T-h)^2} \sum_{s,t=h+1}^T \tr\left\{X_s  A_2 F_{s-h}^\top A_1^\top \left(U_1 U_1^\top -\hat U_{1,m} \hat U_{1,m}^\top \right) A_1 F_{t-h} A_2^\top X_t^\top \right\} \\
&+ \sum_{h=1}^{h_0} \frac{1}{(T-h)^2} \sum_{s,t=h+1}^T \tr\left\{X_s  E_{s-h}^\top \left(U_1 U_1^\top -\hat U_{1,m} \hat U_{1,m}^\top \right) E_{t-h} X_t^\top \right\} \\
&+ \sum_{h=1}^{h_0} \frac{1}{(T-h)^2} \sum_{s,t=h+1}^T \tr\left\{X_s  A_2 F_{s-h}^\top A_1^\top \left(U_1 U_1^\top -\hat U_{1,m} \hat U_{1,m}^\top \right) E_{t-h} X_t^\top \right\} \\
&+ \sum_{h=1}^{h_0} \frac{1}{(T-h)^2} \sum_{s,t=h+1}^T \tr\left\{X_s  E_{s-h}^\top \left(U_1 U_1^\top -\hat U_{1,m} \hat U_{1,m}^\top \right) A_1 F_{t-h} A_2^\top X_t^\top \right\} \\
:=&\I+\II+\III+\IV.
\end{align*}
Note that $\III=\IV$.

We first consider $\II$.
\begin{align*}
\II=&\sum_{h=1}^{h_0} \frac{1}{(T-h)^2} \sum_{s,t=h+1}^T \tr\left\{X_s  E_{s-h}^\top \left(U_1 U_1^\top -\hat U_{1,m} \hat U_{1,m}^\top \right) E_{t-h} X_t^\top \right\} \\
=&  \sum_{h=1}^{h_0} \frac{1}{(T-h)^2} \sum_{s,t=h+1}^T \tr\left\{E_s  E_{s-h}^\top \left(U_1 U_1^\top -\hat U_{1,m} \hat U_{1,m}^\top \right) E_{t-h} E_t^\top \right\} \\
&+ \sum_{h=1}^{h_0} \frac{1}{(T-h)^2} \sum_{s,t=h+1}^T \tr\left\{A_1 F_s A_2^\top  E_{s-h}^\top \left(U_1 U_1^\top -\hat U_{1,m} \hat U_{1,m}^\top \right) E_{t-h} A_2 F_t^\top A_1^\top \right\} \\
&+ 2\sum_{h=1}^{h_0} \frac{1}{(T-h)^2} \sum_{s,t=h+1}^T \tr\left\{A_1 F_s A_2^\top  E_{s-h}^\top \left(U_1 U_1^\top -\hat U_{1,m} \hat U_{1,m}^\top \right) E_{t-h} E_t^\top \right\}\\
:=& \II_1+\II_2+2\II_3.
\end{align*}
By \eqref{eq:split:u1}, Lemma \ref{lemma:tipup:rate} and \ref{lemma:tipup:spectral},
\begin{align*}
\II_1 \le& r_1\sum_{h=1}^{h_0}\left\| \frac{1}{T-h}\sum_{t=h+1}^T E_{t-h} E_t^\top      \right\|_2^2     \left( \left\| \hat U_1 -U_1\tilde U_1^\top \hat U_1 \right\|_2^2 + \left\| \hat U_1 -U_1\tilde U_1^\top \hat U_1 \right\|_2 \cdot \left\| U_1\tilde U_1^\top \hat U_1 \right\|_2   \right) \\
&+(m-r_1) \sum_{h=1}^{h_0}\left\| \frac{1}{T-h}\sum_{t=h+1}^T E_{t-h} E_t^\top      \right\|_2^2 \\
=&O \left( \frac{d}{T} \right),
\end{align*}
using the fact $\eta_d^*=o(\sqrt{T})$.
\begin{align*}
\II_2\le& r_1\sum_{h=1}^{h_0}\left\| \frac{1}{T-h}\sum_{t=h+1}^T E_{t-h} A_2 F_t^\top A_1^\top  \right\|_2^2      \left\| \hat U_1 -U_1\tilde U_1^\top \hat U_1 \right\|_2^2  \\
&+ 2r_1\sum_{h=1}^{h_0}\left\| \frac{1}{T-h}\sum_{t=h+1}^T E_{t-h} A_2 F_t^\top A_1^\top  \right\|_2 \left\| \frac{1}{T-h}\sum_{s=h+1}^T U_1^\top E_{s-h} A_2 F_s^\top A_1^\top  \right\|_2     \left\| \hat U_1 -U_1\tilde U_1^\top \hat U_1 \right\|_2     \\
&+(m-r_1) \sum_{h=1}^{h_0}\left\| \frac{1}{T-h}\sum_{t=h+1}^T E_{t-h} A_2 F_t^\top A_1^\top     \right\|_2^2 \\
=&O \left( \frac{d_1 d^{1-\delta_0}}{T} \right).
\end{align*}
\begin{align*}
\II_3\le& r_1\sum_{h=1}^{h_0}\left\| \frac{1}{T-h}\sum_{s=h+1}^T E_{s-h} A_2 F_s^\top A_1^\top  \right\|_2 \left\| \frac{1}{T-h}\sum_{t=h+1}^T E_{t-h} E_t^\top   \right\|_2   \left\| \hat U_1 -U_1\tilde U_1^\top \hat U_1 \right\|_2^2  \\
&+ r_1\sum_{h=1}^{h_0}\left\| \frac{1}{T-h}\sum_{s=h+1}^T E_{s-h} A_2 F_s^\top A_1^\top  \right\|_2 \left\| \frac{1}{T-h}\sum_{t=h+1}^T U_1^\top E_{t-h} E_t^\top  \right\|_2     \left\| \hat U_1 -U_1\tilde U_1^\top \hat U_1 \right\|_2     \\
&+ r_1\sum_{h=1}^{h_0}\left\| \frac{1}{T-h}\sum_{t=h+1}^T E_{t-h} E_t^\top   \right\|_2 \left\| \frac{1}{T-h}\sum_{s=h+1}^T U_1^\top E_{s-h} A_2 F_s^\top A_1^\top  \right\|_2     \left\| \hat U_1 -U_1\tilde U_1^\top \hat U_1 \right\|_2     \\
&+(m-r_1) \sum_{h=1}^{h_0}\left\| \frac{1}{T-h}\sum_{s=h+1}^T E_{s-h} A_2 F_s^\top A_1^\top  \right\|_2  \left\| \frac{1}{T-h}\sum_{t=h+1}^T E_{t-h} E_t^\top   \right\|_2 \\
=& O \left( \frac{d_1^{1/2} d^{1-\delta_0/2}}{T} \right).
\end{align*}
Combing the bounds of $\II_1,\II_2$ and $\II_3$, we have
\begin{align} \label{eq:lemma:tipup:upf:2}
\II=O \left(\frac{d}{T}+ \frac{d_1 d^{1-\delta_0}}{T} \right).
\end{align}

Next, we consider $\III$.
\begin{align*}
\III=& \sum_{h=1}^{h_0} \frac{1}{(T-h)^2} \sum_{s,t=h+1}^T \tr\left\{X_s  A_2 F_{s-h}^\top A_1^\top \left(U_1 U_1^\top -\hat U_{1,m} \hat U_{1,m}1^\top \right) E_{t-h} X_t^\top \right\} \\
=& \sum_{h=1}^{h_0} \frac{1}{(T-h)^2} \sum_{s,t=h+1}^T \tr\left\{A_1 F_s A_2^\top A_2 F_{s-h}^\top A_1^\top \left(U_1 U_1^\top -\hat U_{1,m} \hat U_{1,m}^\top \right) E_{t-h} A_2 F_t^\top A_1^\top \right\} \\
&+ \sum_{h=1}^{h_0} \frac{1}{(T-h)^2} \sum_{s,t=h+1}^T \tr\left\{E_s  A_2 F_{s-h}^\top A_1^\top \left(U_1 U_1^\top -\hat U_{1,m} \hat U_{1,m}^\top \right) E_{t-h} E_t^\top \right\} \\
&+ \sum_{h=1}^{h_0} \frac{1}{(T-h)^2} \sum_{s,t=h+1}^T \tr\left\{E_s  A_2 F_{s-h}^\top A_1^\top \left(U_1 U_1^\top -\hat U_{1,m} \hat U_{1,m}^\top \right) E_{t-h} A_2 F_t^\top A_1^\top \right\}\\
&+ \sum_{h=1}^{h_0} \frac{1}{(T-h)^2} \sum_{s,t=h+1}^T \tr\left\{A_1 F_s A_2^\top A_2 F_{s-h}^\top A_1^\top \left(U_1 U_1^\top -\hat U_{1,m} \hat U_{1,m}^\top \right) E_{t-h} E_t^\top \right\}\\
:=&\III_1+\III_2+\III_3+\III_4.
\end{align*}
Note that $A_1^\top U_1 U_1^\top =A_1^\top$. Similarly, by \eqref{eq:split:u1}, Lemma \ref{lemma:tipup:rate} and \ref{lemma:tipup:spectral},
\begin{align*}
\III_1\le&  3r_1\sum_{h=1}^{h_0} \left\|\frac{1}{T-h} \sum_{s=h+1}^T A_1 F_s A_2^\top A_2 F_{s-h}^\top A_1^\top \right\|_2 \left\|\frac{1}{T-h} \sum_{t=h+1}^T E_{t-h} A_2 F_t^\top A_1^\top \right\|_2 \left\| \hat U_1 -U_1\tilde U_1^\top \hat U_1 \right\|_2\\
&+   (m-r_1)\sum_{h=1}^{h_0} \left\|\frac{1}{T-h} \sum_{s=h+1}^T A_1 F_s A_2^\top A_2 F_{s-h}^\top A_1^\top \right\|_2 \left\|\frac{1}{T-h} \sum_{t=h+1}^T E_{t-h} A_2 F_t^\top A_1^\top \right\|_2 \left\| U_1 U_1^\top \hat U_{1j}  \right\|_2\\
=& O \left( \frac{d_1^{1/2} d^{3/2-3\delta_0/2}\eta_d^*}{T} \right).
\end{align*}
Moreover,
\begin{align*}
\III_2\le& 3r_1\sum_{h=1}^{h_0} \left\|\frac{1}{T-h} \sum_{s=h+1}^T E_s  A_2 F_{s-h}^\top A_1^\top \right\|_2 \left\|\frac{1}{T-h} \sum_{t=h+1}^T E_{t-h} E_t^\top \right\|_2 \left\| \hat U_1 -U_1\tilde U_1^\top \hat U_1 \right\|_2\\
&+ r_1\sum_{h=1}^{h_0} \left\|\frac{1}{T-h} \sum_{s=h+1}^T E_s  A_2 F_{s-h}^\top A_1^\top \right\|_2 \left\|\frac{1}{T-h} \sum_{t=h+1}^T E_{t-h} E_t^\top \right\|_2 \left\| U_1 U_1^\top \hat U_{1j}  \right\|_2\\
=&O \left( \frac{d_1^{1/2} d^{1-\delta_0/2}\eta_d^*}{T^{3/2}} \right).
\end{align*}
\begin{align*}
\III_3\le& 3r_1\sum_{h=1}^{h_0} \left\|\frac{1}{T-h} \sum_{s=h+1}^T U_1^\top E_s  A_2 F_{s-h}^\top A_1^\top \right\|_2 \left\|\frac{1}{T-h} \sum_{t=h+1}^T  E_{t-h} A_2 F_t^\top A_1^\top U_1\right\|_2 \left\| \hat U_1 -U_1\tilde U_1^\top \hat U_1 \right\|_2\\
&+ (m-r_1)\sum_{h=1}^{h_0} \left\|\frac{1}{T-h} \sum_{s=h+1}^T U_1^\top E_s  A_2 F_{s-h}^\top A_1^\top \right\|_2 \left\|\frac{1}{T-h} \sum_{t=h+1}^T  E_{t-h} A_2 F_t^\top A_1^\top U_1\right\|_2 \left\| U_1 U_1^\top \hat U_{1j}  \right\|_2\\
=&O \left( \frac{d_1^{1/2} d^{1-\delta_0}\eta_d^*}{T^{3/2}} \right).
\end{align*}
\begin{align*}
\III_4\le& r_1\sum_{h=1}^{h_0} \left\| \frac{1}{T-h} \sum_{s=h+1}^T U_1^\top A_1 F_s A_2^\top A_2 F_{s-h}^\top A_1^\top \right\|_2 \left\|\frac{1}{T-h} \sum_{t=h+1}^T E_{t-h} E_t^\top U_1 \right\|_2 \left\| \hat U_1 -U_1\tilde U_1^\top \hat U_1 \right\|_2 \\
&+  (m-r_1)\sum_{h=1}^{h_0} \left\| \frac{1}{T-h} \sum_{s=h+1}^T A_1 F_s A_2^\top A_2 F_{s-h}^\top A_1^\top \right\|_2 \left\|\frac{1}{T-h} \sum_{t=h+1}^T E_{t-h} E_t^\top \right\|_2 \left\| U_1 U_1^\top \hat U_{1j}  \right\|_2\\
=&O \left( \frac{d^{3/2-\delta_0}\eta_d^*}{T} \right).
\end{align*}
Combing $\III_1,\III_2,\III_3$ and $\III_4$, we have
\begin{align} \label{eq:lemma:tipup:upf:3}
\III=O \left( \frac{d_1^{1/2} d^{3/2-3\delta_0/2}\eta_d^*}{T}+  \frac{d_1^{1/2} d^{1-\delta_0/2}\eta_d^*}{T^{3/2}}  + \frac{d^{3/2-\delta_0}\eta_d^*}{T} \right).
\end{align}
In view of \eqref{eq:lemma:tipup:upf:2} and \eqref{eq:lemma:tipup:upf:3},
\begin{align} \label{eq:lemma:tipup:upf:23}
\II+\III+\IV=O \left( \frac{d}{T}+ \frac{d_1 d^{1-\delta_0}}{T} + \frac{d_1^{1/2} d^{3/2-3\delta_0/2}\eta_d^*}{T}  + \frac{d^{3/2-\delta_0}\eta_d^*}{T} \right).
\end{align}

Next, we consider $\I$.
\begin{align*}
\I=&\sum_{h=1}^{h_0} \frac{1}{(T-h)^2} \sum_{s,t=h+1}^T \tr\left\{X_s  A_2 F_{s-h}^\top A_1^\top \left(U_1 U_1^\top -\hat U_{1,m} \hat U_{1,m}^\top \right) A_1 F_{t-h} A_2^\top X_t^\top \right\}  \\
=& \sum_{h=1}^{h_0} \frac{1}{(T-h)^2} \tr\left\{\sum_{s=h+1}^T X_s  A_2 F_{s-h}^\top A_1^\top \left(I-\hat U_{1,m} \hat U_{1,m}^\top \right) \sum_{t=h+1}^T A_1 F_{t-h} A_2^\top X_t^\top \right\} \\
\ge&0,
\end{align*}
using the fact that $I-  \hat U_{1,m} \hat U_{1,m}^\top$ is positive semi-definite, $x^\top \hat U_{1,m} \hat U_{1,m}^\top x \le x^\top x$.

Using the definition of our optimization target, we have $\cM^*(m,\hat U_{1,m}) < \cM^*(r_1,\hat U_{1})\le \cM^*(r_1, U_{1})$. It follows that $\I+\II+\III+\IV\le 0$. As $\I\ge0$, by \eqref{eq:lemma:tipup:upf:23}, $\I=O (d^2\beta_{d,T}^*)$. In summary,
\begin{equation*}
\left| \cM^*(m,\hat U_{1,m})- \cM^*(r_1, U_1) \right| \le C \beta_{d,T}^*.
\end{equation*}

\end{proof}

\begin{lemma}\label{lemma:tipup:lessf}
Suppose Assumptions \ref{asmp:error}, \ref{asmp:mixing}, \ref{asmp:factor}, \ref{asmp:strength} and \ref{asmp:rank}(b) hold. In an event $\Omega_3$ with $\P(\Omega_3)\ge 1-e^{-d_1}-e^{-d_2}-T\exp(-C_1 T^\vartheta)-\exp(-C_2T)$ with $C_1,C_2>0, 1/\vartheta=1/\theta_1+2/\theta_2$, for any $m$ with $m< r_1$, there exist constants $c_m>0$ and $C>0$, such that 
$\cM^*(m,\hat U_{1,m})-\cM^*(r_1,\hat U_1)\ge (c_m+o(1)) d^{-2\delta_1}+C \gamma_{d,T}^*$, where
\begin{equation*}
\gamma_{d,T}^*= \frac{d_1}{Td^{1+\delta_0}} + \frac{1}{Td} + \frac{1}{T^{1/2}d^{1/2+3\delta_0/2}} + \frac{1}{T^{1/2}d^{1/2+\delta_0}d_1^{1/2}} + \frac{d_1^{1/2}\eta_d^*}{Td^{1/2+3\delta_0/2}}+ \frac{\eta_d^*}{Td^{1/2+\delta_0}} ,
\end{equation*}
and $\eta_d^*=d_k^{1/2}d^{\delta_1-\delta_0/2-1/2} + d^{\delta_1-1/2}$.
\end{lemma}

\begin{proof}
Let $\Omega_3=\Omega_{31}\cap\Omega_{32}\cap\Omega_0$, where $\Omega_{31}, \Omega_{32}$ are defined in Lemma \ref{lemma:tipup:rate} and \ref{lemma:tipup:spectral}, respectively, and $\Omega_0$ is the event in Proposition \ref{prop:zt}. In the following, we shall only work on $\Omega_3$.

Let $U_{1,r_1}=U_{1}$ and $U_{1,m}=(U_{11},U_{12},...,U_{1m})$.
\begin{align*}
&d^2\cdot\cM^*(m,\hat U_{1,m})-d^2\cdot\cM^*(r_1, \hat U_1) \\
=& \sum_{h=1}^{h_0}\frac{1}{(T-h)^2}\sum_{s,t=h+1}^T \tr\left\{ \left(\hat U_1 \hat U_1^\top - \hat U_{1,m} \hat U_{1,m}^\top \right) X_{t-h} X_t^\top X_{s-h} X_s^\top  \right\} \\
=& \sum_{h=1}^{h_0}\sum_{j=m+1}^{r_1}\frac{1}{(T-h)^2}\sum_{s,t=h+1}^T \tr\left\{  X_s A_2 F_{s-h}^\top A_1^\top \hat U_{1j} \hat U_{1j}^\top  A_1 F_{t-h} A_2^\top  X_t^\top  \right\} \\
&+ \sum_{h=1}^{h_0}\sum_{j=m+1}^{r_1}\frac{1}{(T-h)^2}\sum_{s,t=h+1}^T \tr\left\{  X_s E_{s-h}^\top \hat U_{1j} \hat U_{1j}^\top E_{t-h} X_t^\top  \right\} \\
&+ \sum_{h=1}^{h_0}\sum_{j=m+1}^{r_1}\frac{1}{(T-h)^2}\sum_{s,t=h+1}^T \tr\left\{  X_s A_2 F_{s-h}^\top A_1^\top \hat U_{1j} \hat U_{1j}^\top E_{t-h} X_t^\top  \right\} \\
&+ \sum_{h=1}^{h_0}\sum_{j=m+1}^{r_1}\frac{1}{(T-h)^2}\sum_{s,t=h+1}^T \tr\left\{  X_s E_{s-h}^\top \hat U_{1j} \hat U_{1j}^\top A_1 F_{t-h} A_2^\top X_t^\top  \right\} \\
:=&\I+\II+\III+\IV.
\end{align*}
Note that $\III=\IV$.

We first consider $\I$.
\begin{align*}
\I=& \sum_{h=1}^{h_0}\sum_{j=m+1}^{r_1}\frac{1}{(T-h)^2}\sum_{s,t=h+1}^T \tr\left\{  A_1 F_s A_2^\top A_2 F_{s-h}^\top A_1^\top \hat U_{1j} \hat U_{1j}^\top  A_1 F_{t-h} A_2^\top  A_2 F_t^\top A_1^\top  \right\} \\
&+ \sum_{h=1}^{h_0}\sum_{j=m+1}^{r_1}\frac{1}{(T-h)^2}\sum_{s,t=h+1}^T \tr\left\{  E_s A_2 F_{s-h}^\top A_1^\top \hat U_{1j} \hat U_{1j}^\top  A_1 F_{t-h} A_2^\top  E_t^\top  \right\} \\
&+ 2\sum_{h=1}^{h_0}\sum_{j=m+1}^{r_1}\frac{1}{(T-h)^2}\sum_{s,t=h+1}^T \tr\left\{  E_s A_2 F_{s-h}^\top A_1^\top \hat U_{1j} \hat U_{1j}^\top  A_1 F_{t-h} A_2^\top  A_2 F_t^\top A_1^\top  \right\} \\
:=& \I_1+\I_2+\I_3.
\end{align*}
For $\I_1$,
\begin{align*}
\I_1=&\sum_{h=1}^{h_0}\sum_{j=m+1}^{r_1}\frac{1}{(T-h)^2}\sum_{s,t=h+1}^T \tr\left\{ \hat U_{1j}^\top  A_1 F_{t-h} A_2^\top  A_2 F_t^\top A_1^\top A_1 F_s A_2^\top A_2 F_{s-h}^\top A_1^\top \hat U_{1j}   \right\}    \\
=& \sum_{h=1}^{h_0}\sum_{j=m+1}^{r_1}  \tr\left\{ \hat U_{1j}^\top U_1 U_1^\top \Theta_{1,h}^* \Theta_{1,h}^{*\top}U_1 U_1^\top \hat U_{1j}   \right\}  \\
\ge& \lambda_{\min}\left( \sum_{h=1}^{h_0} U_1^\top \Theta_{1,h}^* \Theta_{1,h}^{*\top}U_1 \right)   \tr\left\{\sum_{j=m+1}^{r_1}  \hat U_{1j}^\top U_1 U_1^\top \hat U_{1j}   \right\} \\
=& \sigma_{r_1}^2 \left( U_1^\top \overline\E(\text{TIPUP}_1) \right)   \sum_{j=m+1}^{r_1}  \hat U_{1j}^\top U_1 U_1^\top \hat U_{1j} \\
\ge& \sigma_{r_1}^2 \left( U_1^\top \overline\E(\text{TIPUP}_1) \right)  (r_1-m) ( 1- \|\hat U_1 \hat U_1^\top -U_1 U_1^\top \|_2^2) ,
\end{align*}
where the last step follows from \eqref{eq:topup:lessf:I1}. By Proposition \ref{prop:zt}, in the event $\Omega_0$, $\sigma_{r_1} \left( U_1^\top \overline\E (\text{TIPUP}_1)\right)=\tau_{1,r_1}^*\asymp d^{1-\delta_1}$, with $0\le \delta_0\le \delta_1 \le 1$. Hence, there exists a constant $c_m>0$ such that in the event $\Omega_3$,
\begin{align}\label{eq:tipup:plim}
\I_1\ge (c_m+o(1)) d^{2-2\delta_1}.
\end{align}
By Lemma \ref{lemma:tipup:spectral},
\begin{align*}
\I_2=&\sum_{h=1}^{h_0}\sum_{j=m+1}^{r_1} \left\| \frac{1}{T-h}\sum_{s=h+1}^T   E_s A_2 F_{s-h}^\top A_1^\top \right\|_2^2 = O \left( \frac{d_1d^{1-\delta_0}}{T}  \right).    
\end{align*}
\begin{align*}
\I_3=&\sum_{h=1}^{h_0}\sum_{j=m+1}^{r_1}\left\| \frac{1}{T-h}\sum_{s=h+1}^T U_1^\top E_s A_2 F_{s-h}^\top A_1^\top \right\|_2 \left\| \frac{1}{T-h}\sum_{t=h+1}^T A_1 F_{t-h} A_2^\top  A_2 F_t^\top A_1^\top U_1  \right\|_2 = O \left( \frac{d^{3/2-3\delta_0/2}}{T^{1/2}}  \right).    
\end{align*}
This implies that
\begin{align}\label{eq:tipup:lessf:i23}
\I_2+\I_3=& O \left( \frac{d_1d^{1-\delta_0}}{T} + \frac{d^{3/2-3\delta_0/2}}{T^{1/2}} \right).
\end{align}

Next, we consider $\II$.
\begin{align*}
\II=&\sum_{h=1}^{h_0}\sum_{j=m+1}^{r_1}\frac{1}{(T-h)^2}\sum_{s,t=h+1}^T \tr\left\{  E_s E_{s-h}^\top \hat U_{1j} \hat U_{1j}^\top E_{t-h} E_t^\top  \right\} \\
&+   \sum_{h=1}^{h_0}\sum_{j=m+1}^{r_1}\frac{1}{(T-h)^2}\sum_{s,t=h+1}^T \tr\left\{  A_1 F_s A_2^\top E_{s-h}^\top \hat U_{1j} \hat U_{1j}^\top E_{t-h} A_2 F_t^\top A_1^\top \right\} \\
&+ 2\sum_{h=1}^{h_0}\sum_{j=m+1}^{r_1}\frac{1}{(T-h)^2}\sum_{s,t=h+1}^T \tr\left\{  A_1 F_s A_2^\top E_{s-h}^\top \hat U_{1j} \hat U_{1j}^\top E_{t-h} E_t^\top  \right\} \\
:=&\II_1+\II_2+2\II_3.
\end{align*}
By Lemma \ref{lemma:tipup:spectral},
\begin{align*}
\II_1=\sum_{h=1}^{h_0}\sum_{j=m+1}^{r_1}\frac{1}{(T-h)^2}\sum_{s,t=h+1}^T \tr\left\{  E_s E_{s-h}^\top \hat U_{1j} \hat U_{1j}^\top E_{t-h} E_t^\top  \right\}=O \left( \frac{d}{T} \right).
\end{align*}
Applying \eqref{eq:split:u1},
\begin{align*}
\II_2=&\sum_{h=1}^{h_0}\sum_{j=m+1}^{r_1}\frac{1}{(T-h)^2}\sum_{s,t=h+1}^T \tr\left\{  A_1 F_s A_2^\top E_{s-h}^\top (\hat U_{1j}-U_1\tilde U_1^\top \hat U_{1j}) (\hat U_{1j}-U_1\tilde U_1^\top \hat U_{1j})^\top E_{t-h} A_2 F_t^\top A_1^\top \right\} \\
&+ 2 \sum_{h=1}^{h_0}\sum_{j=m+1}^{r_1}\frac{1}{(T-h)^2}\sum_{s,t=h+1}^T \tr\left\{  A_1 F_s A_2^\top E_{s-h}^\top (\hat U_{1j}-U_1\tilde U_1^\top \hat U_{1j}) (U_1\tilde U_1^\top \hat U_{1j})^\top E_{t-h} A_2 F_t^\top A_1^\top \right\} \\
&+ \sum_{h=1}^{h_0}\sum_{j=m+1}^{r_1}\frac{1}{(T-h)^2}\sum_{s,t=h+1}^T \tr\left\{  A_1 F_s A_2^\top E_{s-h}^\top (U_1\tilde U_1^\top \hat U_{1j}) (U_1\tilde U_1^\top \hat U_{1j})^\top E_{t-h} A_2 F_t^\top A_1^\top \right\} \\
=& O \left( \frac{d_1d^{1-\delta_0}\eta_d^{*2}}{T^2} + \frac{d^{1-\delta_0}}{T} \right).
\end{align*}
\begin{align*}
\II_3=& \sum_{h=1}^{h_0}\sum_{j=m+1}^{r_1}\frac{1}{(T-h)^2}\sum_{s,t=h+1}^T \tr\left\{  A_1 F_s A_2^\top E_{s-h}^\top (\hat U_{1j}-U_1\tilde U_1^\top \hat U_{1j}) (\hat U_{1j}-U_1\tilde U_1^\top \hat U_{1j})^\top E_{t-h} E_t^\top  \right\} \\
&+ \sum_{h=1}^{h_0}\sum_{j=m+1}^{r_1}\frac{1}{(T-h)^2}\sum_{s,t=h+1}^T \tr\left\{  A_1 F_s A_2^\top E_{s-h}^\top (U_1\tilde U_1^\top \hat U_{1j}) (\hat U_{1j}-U_1\tilde U_1^\top \hat U_{1j})^\top E_{t-h} E_t^\top  \right\} \\
&+ \sum_{h=1}^{h_0}\sum_{j=m+1}^{r_1}\frac{1}{(T-h)^2}\sum_{s,t=h+1}^T \tr\left\{  U_1^\top A_1 F_s A_2^\top E_{s-h}^\top (\hat U_{1j}-U_1\tilde U_1^\top \hat U_{1j}) (U_1\tilde U_1^\top \hat U_{1j})^\top E_{t-h} E_t^\top U_1  \right\} \\
&+ \sum_{h=1}^{h_0}\sum_{j=m+1}^{r_1}\frac{1}{(T-h)^2}\sum_{s,t=h+1}^T \tr\left\{  U_1^\top A_1 F_s A_2^\top E_{s-h}^\top (U_1\tilde U_1^\top \hat U_{1j}) (U_1\tilde U_1^\top \hat U_{1j})^\top E_{t-h} E_t^\top U_1 \right\} \\
=& O \left( \frac{d_1^{1/2}d^{1-\delta_0/2}\eta_d^{*2}}{T^2} + \frac{d^{1-\delta_0/2}\eta_d^*}{T^{3/2}} + \frac{d^{1-\delta_0/2}}{Td_1^{1/2}} \right).
\end{align*}
Combing $\II_1,\II_2$ and $\II_3$,
\begin{align}\label{eq:tipup:lessf:ii}
\II=& O \left( \frac{d}{T} + \frac{d_1d^{1-\delta_0}\eta_d^{*2}}{T^2} \right).
\end{align}

Next, we consider $\III$.
\begin{align*}
\III=&\sum_{h=1}^{h_0}\sum_{j=m+1}^{r_1}\frac{1}{(T-h)^2}\sum_{s,t=h+1}^T \tr\left\{  E_s A_2 F_{s-h}^\top A_1^\top \hat U_{1j} \hat U_{1j}^\top E_{t-h} E_t^\top  \right\} \\
&+ \sum_{h=1}^{h_0}\sum_{j=m+1}^{r_1}\frac{1}{(T-h)^2}\sum_{s,t=h+1}^T \tr\left\{  A_1 F_s A_2^\top A_2 F_{s-h}^\top A_1^\top \hat U_{1j} \hat U_{1j}^\top E_{t-h} A_2 F_t^\top A_1^\top \right\} \\
&+ \sum_{h=1}^{h_0}\sum_{j=m+1}^{r_1}\frac{1}{(T-h)^2}\sum_{s,t=h+1}^T \tr\left\{  E_s A_2 F_{s-h}^\top A_1^\top \hat U_{1j} \hat U_{1j}^\top E_{t-h} A_2 F_t^\top A_1^\top \right\} \\
&+ \sum_{h=1}^{h_0}\sum_{j=m+1}^{r_1}\frac{1}{(T-h)^2}\sum_{s,t=h+1}^T \tr\left\{  A_1 F_s A_2^\top A_2 F_{s-h}^\top A_1^\top \hat U_{1j} \hat U_{1j}^\top E_{t-h} E_t^\top  \right\} \\
:=&\III_1+\III_2+\III_3+\III_4.
\end{align*}
We bound each term in turn. Applying \eqref{eq:split:u1}, using same arguments in the proof of $\II$, we have,
\begin{align*}
\III_1&=O \left( \frac{d_1^{1/2}d^{1-\delta_0/2}}{T} \right),\\
\III_2&=O \left( \frac{d_1^{1/2}d^{3/2-3\delta_0/2}\eta_d^*}{T}+ \frac{d^{3/2-3\delta_0/2}}{T^{1/2}} \right), \\
\III_3&=O \left( \frac{d_1^{1/2}d^{1-\delta_0}\eta_d^*}{T^{3/2}}+ \frac{d^{1-\delta_0}}{T} \right), \\
\III_4&=O \left( \frac{d^{3/2-\delta_0}\eta_d^*}{T}+ \frac{d^{3/2-\delta_0}}{T^{1/2}d_1^{1/2}} \right). \\
\end{align*}
This implies that
\begin{align}\label{eq:tipup:lessf:iii}
\III=& O \left( \frac{d_1^{1/2}d^{1-\delta_0/2}}{T} + \frac{d_1^{1/2}d^{3/2-3\delta_0/2}\eta_d^*}{T}+ \frac{d^{3/2-3\delta_0/2}}{T^{1/2}} + \frac{d^{1-\delta_0}}{T}+ \frac{d^{3/2-\delta_0}\eta_d^*}{T}+ \frac{d^{3/2-\delta_0}}{T^{1/2}d_1^{1/2}} \right).
\end{align}
Employing \eqref{eq:tipup:plim}, \eqref{eq:tipup:lessf:i23}, \eqref{eq:tipup:lessf:ii} and \eqref{eq:tipup:lessf:iii}, in the event $\Omega_3$, there exist constants $c_m>0$ and $C>0$, such that
\begin{align}
&\cM^*(m,\hat U_{1,m})-\cM^*(r_1, \hat U_1) \notag\\
\ge& (c_m+o(1)) d^{-2\delta_1}+ C\left( \frac{d_1}{Td^{1+\delta_0}} + \frac{1}{Td} + \frac{1}{T^{1/2}d^{1/2+3\delta_0/2}} + \frac{1}{T^{1/2}d^{1/2+\delta_0}d_1^{1/2}} + \frac{d_1^{1/2}\eta_d^*}{Td^{1/2+3\delta_0/2}}+ \frac{\eta_d^*}{Td^{1/2+\delta_0}} \right).
\end{align}

\end{proof}

\begin{proof}[Proof of Theorem \ref{thm:all} for non-iterative TIPUP]
The proof is similar to Theorem \ref{thm:all} for non-iterative TOPUP. For the IC estimators base on non-iterative TIPUP, it is followed by Lemma \ref{lemma:tipup:upf}, \ref{lemma:tipup:lessf} and Assumption \ref{asmp:penalty}(a). 

Applying similar arguments in the proof of Theorem \ref{thm:all} for non-iterative TOPUP, the IC estimators base on non-iterative TIPUP in Theorem \ref{thm:all} follow from Lemma \ref{lemma:tipup:upf}, \ref{lemma:tipup:lessf} and Assumption \ref{asmp:penalty}(b). 
\end{proof}

\subsection{Proof of Theorem \ref{thm:all} for iTIPUP }

\begin{lemma}\label{lemma:tipup:spectral:iterative}
Suppose Assumptions \ref{asmp:error}, \ref{asmp:mixing}, \ref{asmp:factor}, \ref{asmp:strength} and \ref{asmp:rank}(b) hold. Assume that $\|\hat U_{2} \hat U_{2}^\top - U_2 U_2^\top \|_2\le c_0\sqrt{d_1/d_2}$ for some positive constant $c_0$. Then, in an event $\Omega_{42}\cap \Omega_0$ with $\P(\Omega_{42})\ge 1-e^{-d_1}/2$ and $\P(\Omega_0)\ge 1-T\exp(-C_1 T^\vartheta)-\exp(-C_2T)$ with $C_1,C_2>0, 1/\vartheta=1/\theta_1+2/\theta_2$, we have
\begin{align}
&\left\| \frac{1}{T-h} \sum_{t=h+1}^T A_1 F_{t-h} A_2^\top \hat U_{2} \hat U_{2}^\top E_t^\top   \right\|_2 \le C \left( \frac{d_1^{1/2}d^{1/2-\delta_0/2} }{\sqrt{T}} \right), \label{eq:tipup:spectral:iter1} \\
&\left\| \frac{1}{T-h} \sum_{t=h+1}^T E_{t-h} \hat U_{2} \hat U_{2}^\top A_2 F_t^\top A_1^\top   \right\|_2 \le C \left( \frac{d_1^{1/2}d^{1/2-\delta_0/2} }{\sqrt{T}} \right), \label{eq:tipup:spectral:iter2} \\
&\left\| \frac{1}{T-h} \sum_{t=h+1}^T A_1 F_{t-h} A_2^\top \hat U_{2} \hat U_{2}^\top E_t^\top U_1  \right\|_2 \le C \left( \frac{d^{1/2-\delta_0/2} }{\sqrt{T}} \right), \label{eq:tipup:spectral:iter3} \\
&\left\| \frac{1}{T-h} \sum_{t=h+1}^T U_1^\top E_{t-h} \hat U_{2} \hat U_{2}^\top A_2^\top F_t^\top A_1^\top  \right\|_2 \le C \left( \frac{d^{1/2-\delta_0/2} }{\sqrt{T}} \right), \label{eq:tipup:spectral:iter4} \\
&\left\| \frac{1}{T-h} \sum_{t=h+1}^T E_{t-h} \hat U_{2} \hat U_{2}^\top E_t^\top  \right\|_2 \le C \left( \frac{d_1^{1/2} }{\sqrt{T}} \right), \label{eq:tipup:spectral:iter5} \\
&\left\| \frac{1}{T-h} \sum_{t=h+1}^T U_1^\top E_{t-h} \hat U_{2} \hat U_{2}^\top E_t^\top U_1  \right\|_2 \le C \left( \frac{1 }{\sqrt{T}} \right), \label{eq:tipup:spectral:iter6} 
\end{align}
for some constant positive $C$ depending on $K$ only.
\end{lemma}

\begin{lemma}\label{lemma:tipup:upf:iterative}
Suppose Assumptions \ref{asmp:error}, \ref{asmp:mixing}, \ref{asmp:factor}, \ref{asmp:strength} and \ref{asmp:rank}(b) hold. Let $r_1\le r_1^{(j)}\le m_1<d_1$ for all $1\le j\le i-1$, and $m_1=O(r_1)$. There exists an event $\Omega_4$ such that $\P(\Omega_4)\ge 1 -e^{-d_1}-e^{-d_2}-T\exp(-C_1 T^\vartheta)-\exp(-C_2T)$, $C_1,C_2>0, 1/\vartheta=1/\theta_1+2/\theta_2$ and $\Omega_4$ is independent of iteration number $i$. Then, in the event $\Omega_4$, at $i$-th iteration, for any fixed $m$ with $m> r_1$, $\cM^{*(i)}(r_1,\hat U_1) - \cM^{*(i)}(m,\hat U_{1,m}) \le C \beta_{d,T}^*$, where $C>0$, 
\begin{equation*}
\beta_{d,T}^*= \frac{d_1 }{Td^{1+\delta_0}} + \frac{d_1^{1/2} \eta_k^{*(i)}}{Td^{1/2+3\delta_0/2}}   ,
\end{equation*}
and $\eta_k^{*(i)}=d_k^{1/2}d^{\delta_1-\delta_0/2-1/2} + d_k^{1/2}d^{\delta_1-1}$.
\end{lemma}

\begin{lemma}\label{lemma:tipup:lessf:iterative}
Suppose Assumptions \ref{asmp:error}, \ref{asmp:mixing}, \ref{asmp:factor}, \ref{asmp:strength} and \ref{asmp:rank}(b) hold. Let $r_1\le r_1^{(j)}\le m_1<d_1$ for all $1\le j\le i-1$, and $m_1=O(r_1)$. There exists an event $\Omega_4$ such that $\P(\Omega_4)\ge 1 -e^{-d_1}-e^{-d_2}-T\exp(-C_1 T^\vartheta)-\exp(-C_2T)$, $C_1,C_2>0, 1/\vartheta=1/\theta_1+2/\theta_2$ and $\Omega_4$ is independent of iteration number $i$. Then, in the event $\Omega_4$, at $i$-th iteration, for any $m$ with $m< r_1$, there exist constants $c_m>0$ and $C>0$, such that  
$\cM^{*(i)}(m,\hat U_{1,m})-\cM^{*(i)}(r_1,\hat U_1)\ge (c_m+o(1)) d^{-2\delta_1}+C\gamma_{d,T}^*$, where
\begin{equation*}
\gamma_{d,T}^*= \frac{d_1}{Td^{1+\delta_0}}  + \frac{1}{T^{1/2}d^{1/2+3\delta_0/2}} +  \frac{d_1^{1/2}\eta_k^{*(i)}}{Td^{1/2+3\delta_0/2}} ,
\end{equation*}
and $\eta_k^{*(i)}=d_k^{1/2}d^{\delta_1-\delta_0/2-1/2} + d_k^{1/2}d^{\delta_1-1}$.
\end{lemma}

Note that by \cite{han2020}, in the event $\Omega_4$, at $i$-th iteration ($i\ge1$), for the iTIPUP estimator, assume that $\|\hat U_{k,r_k}^{(i)} \hat U_{k,r_k}^{(i)\top} - U_k U_k^\top \|_2 \le C_0 (\eta_k^{*(i)}T^{-1/2})$, $1\le k\le K$, for some $C_0>0$.
The proofs of Lemma \ref{lemma:tipup:spectral:iterative}, \ref{lemma:tipup:upf:iterative} and \ref{lemma:tipup:lessf:iterative} are omitted, since they are similar to Lemma \ref{lemma:topup:spectral:iterative}, \ref{lemma:topup:upf:iterative} and \ref{lemma:topup:lessf:iterative}.

\begin{proof}[Proof of Theorem \ref{thm:all} for iTIPUP ]
Theorem \ref{thm:all} for iTIPUP part follows by similar arguments in the proofs of Theorem \ref{thm:all} for non-iterative iTIPUP.
\end{proof}

\subsection{Proof of Propositions and Theorem \ref{thm:eigen} }
\begin{proof}[Proof of Proposition \ref{prop:zt}]
Let $A_k$ has SVD $A_k=U_k\Lambda_k V_k^\top$. As $|\P(A\cap B) - \P(A)\P(B)|\le 1/4 \le e^{-1}$ for all event pairs $\{A,B\}$,
\begin{align*}
& \sup_{t}\Big\{\Big|\P(A\cap B) - \P(A)\P(B)\Big|:
A\in \sigma(F_{s+\tau}\otimes F_s, s\le t, \tau\le \tau_0),
B\in \sigma(F_s, s\ge t+h)\Big\} \\
\le & \sup_{t}\Big\{\Big|\P(A\cap B) - \P(A)\P(B)\Big|:
A\in \sigma(F_s, s\le t+\tau_0),
B\in \sigma(F_s, s\ge t+h)\Big\} \\
\le & \exp\Big\{- \max\{1,c_0(h-\tau_0)^{\theta_1}\}\Big\} \\
\le & \exp\Big\{ - c_0(1+c_0^{1/\theta_1}\tau_0)^{-\theta_1} h^{\theta_1}\Big\} = \exp\Big\{ - c_0' h^{\theta_1}\Big\}.
\end{align*}
where $c_0'=c_0(1+c_0^{1/\theta_1}\tau_0)^{-\theta_1}$.

Since $r_k$ is fixed, $\E( F_{i,k,t-h} F_{j,l,t}) =O(1)$ for any $0\le h\le T/4$, $1\le i,j\le r_1$, $1\le k,l\le r_2$. Pick $a>0$, $x> \E( F_{i,k,t-h} F_{j,l,t} )$, then
\begin{align*}
&\P\left(  \left| F_{i,k,t-h} F_{j,l,t} -\E F_{i,k,t-h} F_{j,l,t} \right| >x \right)\\
=&\P\left(  \left| F_{i,k,t-h} F_{j,l,t} -\E F_{i,k,t-h} F_{j,l,t} \right| >x, \left| F_{i,k,t-h} \right|> x^a \right) \\
&\quad + \P\left(  \left| F_{i,k,t-h} F_{j,l,t} -\E F_{i,k,t-h} F_{j,l,t} \right| >x, \left| F_{i,k,t-h} \right|\le x^a  \right)\\
\le& \P\left( \left| F_{i,k,t-h} \right|>x^a \right) + \P\left( \left| F_{j,l,t} \right| > x^{1-a} \right) \\
\le& c_1 \exp\left( -c_2 x^{a\theta_2} \right) + c_1 \exp\left( -c_2 x^{(1-a)\theta_2} \right) \\
\le& 2c_1 \exp\left( -c_2 x^{\theta_2/2} \right). 
\end{align*}
Let $1/\vartheta=1/\theta_1+2/\theta_2$. Hence, by Theorem 1 in \citet{merlevede2011}, for any $\|u\|_2=\|v\|_2=1$,
\begin{align*}
& \P\left( \left| \frac{1}{T-h}\sum_{t=h+1}^T u^\top \left( F_{t-h} F_t^{\top} - \E F_{t-h} F_t^{\top} \right) v \right| > x  \right) \\
\le& (T-h)\exp\left(-c_3(T-h)^{\vartheta} x^{\vartheta}\right) +\exp\left( -c_4(T-h) x^2 \right).
\end{align*}
By Lemma \ref{lemma:epsilonnet}(i),
\begin{align*}
&\P\left( \left\| \frac{1}{T-h}\sum_{t=h+1}^T \left( F_{t-h} F_t^{\top} - \E F_{t-h} F_t^{\top} \right)  \right\|_2 > x  \right) \\
\le& 5^{r_1} \cdot\P\left( \left| \frac{1}{T-h}\sum_{t=h+1}^T u^\top \left( F_{t-h} F_t^{\top} - \E F_{t-h} F_t^{\top} \right) v \right| > x/2  \right) \\
\le& 5^{r_1}(T-h)\exp\left(-c_3(T-h)^{\vartheta} (x/2)^{\vartheta}\right) +5^{r_1}\exp\left( -c_4(T-h) x^2/4 \right).
\end{align*}
As $h\le T/4$, choosing $x=2^{-1}\sigma_{r_1} \left( (T-h)^{-1}\sum_{t=h+1}^T \E F_{t-h} F_t^{\top}  \right) \vee 1$, in an event $\Omega_0$ with probability at least $1-T\exp(-C_1 T^\vartheta)-\exp(-C_2T)$,
\begin{align}\label{prop2:eq1}
\left\| \frac{1}{T-h}\sum_{t=h+1}^T \left( F_{t-h} F_t^{\top} - \E F_{t-h} F_t^{\top} \right)  \right\|_2  \le \frac12 \sigma_{r_1} \left( \frac{1}{T-h}\sum_{t=h+1}^T \E F_{t-h} F_t^{\top}  \right)\vee 1.
\end{align}
By triangle inequality,
\begin{align*}
&\left\| \frac{1}{T-h}\sum_{t=h+1}^T \E M_{t-h} M_t^\top \right\|_2 -\left\| \frac{1}{T-h}\sum_{t=h+1}^T \left( M_{t-h} M_t^\top - \E M_{t-h} M_t^\top \right)  \right\|_2 \le \left\| \frac{1}{T-h}\sum_{t=h+1}^T M_{t-h} M_t^\top \right\|_2 \\
&\le \left\| \frac{1}{T-h}\sum_{t=h+1}^T \E M_{t-h} M_t^\top \right\|_2 + \left\| \frac{1}{T-h}\sum_{t=h+1}^T \left( M_{t-h} M_t^\top - \E M_{t-h} M_t^\top \right)  \right\|_2.
\end{align*}
In the event $\Omega_0$, by \eqref{prop2:eq1} and Assumption \ref{asmp:rank}(b),
\begin{align*}
&\left\| \frac{1}{T-h}\sum_{t=h+1}^T \left( M_{t-h} M_t^\top - \E M_{t-h} M_t^\top \right)  \right\|_2 \le d^{1-\delta_0}\left\| \frac{1}{T-h}\sum_{t=h+1}^T \left( F_{t-h} F_t^{\top} - \E F_{t-h} F_t^{\top} \right)  \right\|_2  \\
\le& \frac12 \sigma_{r_1} \left( \frac{1}{T-h}\sum_{t=h+1}^T \E M_{t-h} M_t^\top  \right).
\end{align*}
It follows that
\begin{align*}
\|\Theta_{1,h}^*\|_2 = \left\| \frac{1}{T-h}\sum_{t=h+1}^T M_{t-h} M_t^\top \right\|_2 \asymp \left\| \frac{1}{T-h}\sum_{t=h+1}^T \E M_{t-h} M_t^\top \right\|_2 \asymp d^{1-\delta_0},
\end{align*}
for $h=0$ and some $1\le h\le h_0 \le T/4$. 
Similarly, by \eqref{prop2:eq1} and Assumption \ref{asmp:rank}(b), for some $1\le h\le h_0 \le T/4$, 
\begin{align*}
\sigma_{r_1}\left( \frac{1}{T-h}\sum_{t=h+1}^T M_{t-h} M_t^\top \right) \asymp d^{1-\delta_1},
\end{align*}
Then,
\begin{align*}
\tau_{1,r_1}^* \asymp d^{1-\delta_1}.
\end{align*}
Hence, in the event $\Omega_0$ with probability at least $1-C_1 \exp(-C_2 T^\vartheta)-C_3\exp(-C_4T)$,
\begin{align}
\|\Theta_{1,0}^*\|_2\asymp d^{1-\delta_0}, \ \ \tau_{1,r_1}^* \asymp d^{1-\delta_1}.
\end{align}
Similarly, applying arguments of proving Lemma 5 in \citet{wang2019}, we can show, in an event with probability at least $1- T\exp(-C_1 T^\vartheta)-\exp(-C_2T)$,
\begin{align}
\|\Theta_{1,0}\|_{\rm{op}}\asymp d^{1-\delta_0}, \ \ \tau_{1,r_1} \asymp d^{1-\delta_1}.
\end{align}
\end{proof}

\begin{proof}[Proof of Theorem \ref{thm:eigen}]
We only consider the case of non-iterative TOPUP, as the other cases will be similar. Let $\eta_d=d^{\delta_1-\delta_0/2} + d^{\delta_1}d_1^{-1/2}$. For any $0\le m< n<d_1$,
\begin{align}\label{eq:cor1}
\sum_{j=m+1}^n\hat\lambda_{1,j}=d^2\cM(m,\hat U_{1,m})-d^2\cM(n,\hat U_{n}).
\end{align}
By Lemma \ref{lemma:topup:upf}, in the same event $\Omega_1$, $\hat\lambda_{r_1+1}=O(d^2\beta_{d,T})$. As $\hat\lambda_{r_1+1}\ge \hat\lambda_{r_1+2}\ge ... \ge \hat\lambda_{d_1}$, part (i) follows. 

Set $n=m+1\le r_1$ in \eqref{eq:cor1}. Employing the same arguments in the proof of Lemma \ref{lemma:topup:lessf}, we can show $\I_2+\I_3+\II+\III=O(d^2\gamma_{d,T})$
in the same event $\Omega_1$. For $\I_1$,
\begin{align*}
&\I_1\\
=&\sum_{h=1}^{h_0}\frac{1}{(T-h)^2}\sum_{s,t=h+1}^T \tr\left\{  {\mat1}^\top (A_1 F_{s-h} A_2^\top\otimes A_1 F_s A_2^\top) \hat U_{1m} \hat U_{1m}^\top  \mat1(A_1 F_{t-h} A_2^\top\otimes A_1 F_t A_2^\top)  \right\} \\
=&\sum_{h=1}^{h_0}\frac{1}{(T-h)^2}\sum_{s,t=h+1}^T \tr\left\{  {\mat1}^\top (A_1 F_{s-h} A_2^\top\otimes A_1 F_s A_2^\top) (U_{1m} U_{1m}^\top -\hat U_{1m} \hat U_{1m}^\top)  \mat1(A_1 F_{t-h} A_2^\top\otimes A_1 F_t A_2^\top)  \right\} \\
&+\sum_{h=1}^{h_0}\frac{1}{(T-h)^2}\sum_{s,t=h+1}^T \tr\left\{  {\mat1}^\top (A_1 F_{s-h} A_2^\top\otimes A_1 F_s A_2^\top) U_{1m} U_{1m}^\top  \mat1(A_1 F_{t-h} A_2^\top\otimes A_1 F_t A_2^\top)  \right\}  \\
:=&\I_{11}+\I_{12}. 
\end{align*}
By \eqref{eq:split:u1},
\begin{align*}
\I_{11}&\le \| U_{1m} U_{1m}^\top -\hat U_{1m} \hat U_{1m}^\top \|_2 \left\|\sum_{h=1}^{h_0} U_1^\top \mat1 \left(\Theta_{1,h} \right) {\mat1}^\top \left(\Theta_{1,h} \right) U_1 \right\|_2 \\
&= \| U_{1m} U_{1m}^\top -\hat U_{1m} \hat U_{1m}^\top \|_2 \cdot \sigma_{1}^2 \left( U_1^\top \overline\E \mat1(\text{TOPUP}_1)\right)
\end{align*}
Followed by Lemma \ref{lemma:topup:rate} and Proposition \ref{prop:zt}, $\I_{11}=O(d^{2-2\delta_0}T^{-1/2}\eta_{d})$ in the event $\Omega_1$. For $\I_{12}$
\begin{align*}
\I_{12}&= \sum_{h=1}^{h_0}\frac{1}{(T-h)^2}\sum_{s,t=h+1}^T \tr\left\{   U_{1n}^\top  \mat1(A_1 F_{t-h} A_2^\top\otimes A_1 F_t A_2^\top)  {\mat1}^\top (A_1 F_{s-h} A_2^\top\otimes A_1 F_s A_2^\top) U_{1n} \right\} \\
&= \sum_{h=1}^{h_0}   U_{1n}^\top U_1 U_1^\top \mat1 \left(\Theta_{1,h} \right) {\mat1}^\top \left(\Theta_{1,h} \right) U_1  U_1^\top U_{1n}  \\
&=\sigma_{n}^2 \left( \overline\E \mat1(\text{TOPUP}_1)\right)=\tau_{1,n}^2 .
\end{align*}
Applying similar arguments in the proof of Proposition \ref{prop:zt}, we can show, in an event with probability at least $1- T\exp(-C_1 T^{\vartheta/2}\eta_{d}^\vartheta)-\exp(-C_2\eta_{d}^2)$ with $1/\vartheta=1/\theta_1+2/\theta_2$ and $C_1,C_2>0$, $$\I_{12}=\tau_{1,n}^2=\lambda_{1,n}+O(T^{-1/2}d^{2-2\delta_0}\eta_{d}).$$
Then part (ii) follows by combing the bounds for $\I_2+\I_3+\II+\III$, $\I_{11}$ and $\I_{12}$.

\end{proof}

\section{Techinical Lemmas}

\begin{lemma}\label{lemma:sintheta}
Let $\hat V, V\in\mathbb O_{d,r}$, and $V_{\perp}$ be complement part of $V$, namely $[V,V_{\perp} ]\in\mathbb O_{d}$. Then, the following equivalent forms hold
\begin{equation*}
\|\hat V\hat V^\top -V V^\top\|_2 =\sqrt{1-\sigma_{\min}(\hat V^\top V_{\perp}) }
= \| \hat V^\top V_{\perp} \|_{2} .
\end{equation*}
\end{lemma}
\begin{proof}
The singular values $\sigma_j$ of $\hat V^\top V$ correspond to eigenvalues $\pm\sqrt{1-\sigma_j^2}$
for $\hat V\hat V^\top -V V^\top$.
\end{proof}

\begin{lemma}\label{lemma:perturbation}
Suppose $A$ and $A+E$ are $n\times n$ symmetric matrices and that
$$ Q=[Q_1\ Q_2] $$
is an orthogonal matrix such that span$(Q_1)$ is an invariant subspace for $A$, where $Q_1\in\RR^{n\times r}$ and $Q_2\in\RR^{n\times (n-r)}$. Partition the matrices $Q^\top A Q$ and $Q^\top E Q$ as follows:
\begin{align*}
Q^\top A Q=\left(\begin{matrix}D_1& 0\\ 0&D_2 \end{matrix} \right) \quad\quad Q^\top E Q= \left(\begin{matrix}E_{11} & E_{21}^\top\\ E_{21}& E_{22} \end{matrix} \right).
\end{align*}
If sep$(D_1,D_2)=\min_{\lambda\in\lambda(D_1),\mu\in\lambda(D_2)}|\lambda-\mu|>0$, where $\lambda(M)$ denotes the set of eigenvalues of the matrix $M$, and $\|E\|_2\le $sep$(D_1,D_2)/5$, then there exists a matrix $P\in \RR^{(n-r)\times r}$ with
\begin{align*}
\| P\|_2 \le \frac{4}{\text{sep}(D_1,D_2)} \|E_{21} \|_2
\end{align*}
such that the columns of $\hat Q_1=(Q_1+Q_2 P)(I+P^\top P)^{-1/2}$ define an orthonormal basis for a subspace that is invariant for $A+E$.

\end{lemma}
\begin{proof}
See Theorem 8.1.10 in \citet{golub2012}.
\end{proof}

\begin{lemma}\label{lemma:epsilonnet}
Let $d, d_j, d_*, r\le d\wedge d_j$ be positive integers, $\epsilon>0$ and
$N_{d,\epsilon} = \lfloor(1+2/\epsilon)^d\rfloor$. \\
(i) For any norm $\|\cdot\|$ in $\R^d$, there exist
$M_j\in \R^d$ with $\|M_j\|\le 1$, $j=1,\ldots,N_{d,\epsilon}$,
such that $\max_{\|M\|\le 1}\min_{1\le j\le N_{d,\epsilon}}\|M - M_j\|\le \epsilon$.
Consequently, for any linear mapping $f$ and norm $\|\cdot\|_*$,
$$
\sup_{M\in \R^d,\|M\|\le 1}\|f(M)\|_* \le 2\max_{1\le j\le N_{d,1/2}}\|f(M_j)\|_*.
$$
(ii) Given $\epsilon >0$, there exist $U_j\in \R^{d\times r}$
and $V_{j'}\in \R^{d'\times r}$ with $\|U_j\|_{2}\vee\|V_{j'}\|_{2}\le 1$ such that
$$
\max_{M\in \R^{d\times d'},\|M\|_{2}\le 1,\text{rank}(M)\le r}\
\min_{j\le N_{dr,\epsilon/2}, j'\le N_{d'r,\epsilon/2}}\|M - U_jV_{j'}^\top\|_{2}\le \epsilon.
$$
Consequently, for any linear mapping $f$ and norm $\|\cdot\|_*$ in the range of $f$,
\begin{equation}\label{lm-3-2}
\sup_{M, \widetilde M\in \R^{d\times d'}, \|M-\widetilde M\|_{2}\le \epsilon
\atop{\|M\|_{2}\vee\|\widetilde M\|_{2}\le 1\atop
\text{rank}(M)\vee\text{rank}(\widetilde M)\le r}}
\frac{\|f(M-\widetilde M)\|_*}{\epsilon 2^{I_{r<d\wedge d'}}}
\le \sup_{\|M\|_{2}\le 1\atop \text{rank}(M)\le r}\|f(M)\|_*
\le 2\max_{1\le j \le N_{dr,1/8}\atop 1\le j' \le N_{d'r,1/8}}\|f(U_jV_{j'}^\top)\|_*.
\end{equation}
(iii) Given $\epsilon >0$, there exist $U_{j,k}\in \R^{d_k\times r_k}$
and $V_{j',k}\in \R^{d'_k\times r_k}$ with $\|U_{j,k}\|_{2}\vee\|V_{j',k}\|_{2}\le 1$ such that
$$
\max_{M_k\in \R^{d_k\times d_k'},\|M_k\|_{2}\le 1\atop \text{rank}(M_k)\le r_k, \forall k\le K}\
\min_{j_k\le N_{d_kr_k,\epsilon/2} \atop j'_k\le N_{d_k'r_k,\epsilon/2}, \forall k\le K}
\Big\|\odot_{k=2}^K M_k - \odot_{k=2}^K(U_{j_k,k}V_{j_k',k}^\top)\Big\|_{2}\le \epsilon (K-1).
$$
For any linear mapping $f$ and norm $\|\cdot\|_*$ in the range of $f$,
\begin{equation}\label{lm-3-3}
\sup_{M_k, \widetilde M_k\in \R^{d_k\times d_k'},\|M_k-\widetilde M_k\|_{2}\le\epsilon\atop
{\text{rank}(M_k)\vee\text{rank}(\widetilde M_k)\le r_k \atop \|M_k\|_{2}\vee\|\widetilde M_k\|_{2}\le 1\ \forall k\le K}}
\frac{\|f(\odot_{k=2}^KM_k-\odot_{k=2}^K\widetilde M_k)\|_*}{\epsilon(2K-2)}
\le \sup_{M_k\in \R^{d_k\times d_k'}\atop {\text{rank}(M_k)\le r_k \atop \|M_k\|_{2}\le 1, \forall k}}
\Big\|f\big(\odot_{k=2}^K M_k\big)\Big\|_*
\end{equation}
and
\begin{equation}\label{lm-3-4}
\sup_{M_k\in \R^{d_k\times d_k'},\|M_k\|_{2}\le 1\atop \text{rank}(M_k)\le r_k\ \forall k\le K}
\Big\|f\big(\odot_{k=2}^K M_k\big)\Big\|_*
\le 2\max_{1\le j_k \le N_{d_kr_k,1/(8K-8)}\atop 1\le j_k' \le N_{d_k'r_k,1/(8K-8)}}
\Big\|f\big(\odot_{k=2}^K U_{j_k,k}V_{j_k',k}^\top\big)\Big\|_*.
\end{equation}
\end{lemma}

\begin{proof}
See Lemma 1 in \citet{han2020}.
\end{proof}

\begin{lemma}\label{lm-GH}
(i) Let $G\in \R^{d_1\times n}$ and $H\in \R^{d_2\times n}$ be two centered independent
Gaussian matrices such that $\E(u^\top \text{vec}(G))^2 \le \sigma^2\ \forall\ u\in \R^{d_1n}$ and
$\E(v^\top \text{vec}(H))^2\le \sigma^2\ \forall\ v\in \R^{d_2n}$. Then,
\begin{align}\label{lm-GH-1}
\|GH^\top\|_{\rm S} \le \sigma^2\big(\sqrt{d_1d_2}+\sqrt{d_1n} + \sqrt{d_2n}\big)
+ \sigma^2x(x+2\sqrt{n}+\sqrt{d_1}+\sqrt{d_2})
\end{align}
with at least probability $1 - 2e^{-x^2/2}$ for all $x\ge 0$. \\
(ii) Let $G_i\in \R^{d_1\times d_2}, H_i\in \R^{d_3\times d_4}, i=1,\ldots, n$,
be independent centered Gaussian matrices
such that $\E(u^\top \text{vec}(G_i))^2 \le \sigma^2\ \forall\ u\in \R^{d_1d_2}$ and
$\E(v^\top \text{vec}(H_i))^2\le \sigma^2\ \forall\ v\in \R^{d_3d_4}$. Then,
\begin{align}
\bigg\|\text{mat}_1\bigg(\sum_{i=1}^n G_i\otimes H_i\bigg)\bigg\|_{2}
\le& \sigma^2\big(\sqrt{d_1n}+\sqrt{d_1d_3d_4} + \sqrt{nd_2d_3d_4}\big) \notag\\
& + \sigma^2 x\big(x + \sqrt{n} + \sqrt{d_1} + \sqrt{d_2}+\sqrt{d_3d_4}\big)
\end{align}
with at least probability $1 - 2e^{-x^2/2}$ for all $x\ge 0$.
\end{lemma}

\begin{proof}
See Lemma 2 in \citet{han2020}.
\end{proof}

\section{Additional Simulation Results}
In this section, we show detailed comparison among IC1-IC5, and ER1-ER5 for the first part and third part simulation in Section \ref{section:simulation}. We also study a strong factor model with $r_1=r_2=2$.
\begin{enumerate}
\item[(M0).] Set $r_1=r_2=2$. The univariate $f_{ijt}$ follows AR(1) with AR coefficient $\phi_{11}=\phi_{22}=0.8$ and $\phi_{12}=\phi_{21}=0.3$; All elements of $A_1$ and $A_2$ are i.i.d N(0,1). 
\end{enumerate}

\begin{table}[htbp]
\centering
\caption{Proportion of correct identification of rank $r$ 
using IC estimators based on TOPUP procedures for Model M1, over 1000 replications}
\label{tab:topup.bic.r5.strong}
\resizebox{\textwidth}{!}
{\begin{tabular}{cccccc|ccccc|cccccc}\hline \hline
 &\multicolumn{5}{c}{initial estimator} & \multicolumn{5}{c}{one step estimator} & \multicolumn{5}{c}{final estimator}\\ \hline
$T$ & IC1 & IC2 & IC3 & IC4 & IC5 & IC1 & IC2 & IC3 & IC4 & IC5 & IC1 & IC2 & IC3 & IC4 & IC5\\ \hline
&\multicolumn{5}{l}{ $d_1=d_2=20$}\\ \hline
100 &  0 & 0.096 & 0.005 & 0.203 & 0 & 0.313 & 0.753 & 0.421 & 0.841 & 0.094 & 0.437 & 0.853 & 0.544 & 0.903 & 0.154\\
200 & 0.009 & 0.051 & 0.881 & 0.957 & 0.001 & 0.757 & 0.863 & 0.994 & 0.999 & 0.561 & 0.997 & 0.997 & 1 & 1 & 0.997 \\
300 & 0.022 & 0.995 & 0.155 & 1 & 0.033 & 0.853 & 1 & 0.943 & 1 & 0.872 & 1 & 1 & 1 &  1 & 1 \\
500 & 0.064 & 1 & 0.292 & 1 & 0.673 & 0.917 & 1 & 0.969 & 1 & 0.995 & 1 & 1 & 1 &  1 & 1 \\
1000 & 0.127 & 1 & 0.268 & 1 & 1 & 0.949 & 1 & 0.976 & 1 & 1 & 1 & 1 & 1 &  1 & 1 \\
\hline
&\multicolumn{5}{l}{ $d_1=d_2=40$}\\ \hline
100 & 0 & 0 & 0 & 0 & 0 & 0.314 & 0.467 & 0.339 & 0.491 & 0.073 & 0.337 & 0.482 & 0.366 & 0.510 & 0.080\\
200 & 0 & 0 & 0 & 0.002 & 0 & 0.684 & 0.736 & 0.900 & 0.916 & 0.170 & 0.989 & 0.991 & 0.998 & 1.000 & 0.936 \\
300 & 0 & 0.025 & 0 & 0.067 & 0 & 0.858 & 0.979 & 0.891 & 0.988 & 0.302 & 0.994 & 0.999 & 0.995 & 0.999 & 0.981 \\
500 & 0 & 0.457 & 0.001 & 0.654 & 0 & 0.939 & 0.999 & 0.967 & 0.999 & 0.689 & 0.999 & 1 & 0.999 & 1 & 0.999 \\
1000 & 0.004 & 1 & 0.048 & 1 & 0 & 0.996 & 1 & 1 & 1 & 0.988 & 1 & 1 & 1 &  1 & 1 \\
\hline
&\multicolumn{5}{l}{ $d_1=d_2=80$}\\ \hline
100 & 0 & 0 & 0 & 0 & 0 & 0.626 & 0.659 & 0.628 & 0.667 & 0.567 & 0.973 & 0.977 & 0.974 & 0.977 & 0.967 \\
200 & 0 & 0 & 0 & 0 & 0 & 0.913 & 0.918 & 0.938 & 0.939 & 0.681 & 0.998 & 0.998 & 0.999 & 0.999 & 0.992 \\
300 & 0 & 0 & 0 & 0.001 & 0 & 0.969 & 0.982 & 0.970 & 0.983 & 0.787 & 1 & 1 & 1 & 1 & 0.997 \\
500 & 0.001 & 0.025 & 0.001 & 0.038 & 0 & 0.993 & 0.996 & 0.993 & 0.998 & 0.883 & 1 & 1 & 1 & 1 & 0.999 \\
1000 & 0.058 & 0.698 & 0.100 & 0.780 & 0 & 1 & 1 & 1 & 1 & 0.954 & 1 & 1 & 1 & 1 & 1 \\
\hline
\end{tabular}}
\end{table}


For Model M0 with small $r_1=r_2=2$ and all strong factors, all versions of our methods show perfect
accuracy for all sample sizes except $T=100$ when some versions show error rates less than 1\%.
For Model M1, the results in Table \ref{tab:topup.bic.r5.strong} show clearly that, among all five penalty functions, IC2 and IC4 seem to be
slightly better than the others. With the same settings, the determination of the ranks are perfect using IC and TIPUP, or using ER estimators. Thus, we omitted the tables.

\begin{table}[htbp]
\centering
\caption{Proportion of correct identification of rank $r$ 
using IC estimators based on TOPUP procedures for Model M2, over 1000 replications}
\label{tab:topup.bic.r5.strongweak}
\resizebox{\textwidth}{!}
{\begin{tabular}{cccccc|ccccc|cccccc}\hline \hline
 &\multicolumn{5}{c}{initial estimator} & \multicolumn{5}{c}{one step estimator} & \multicolumn{5}{c}{final estimator}\\ \hline
$T$ & IC1 & IC2 & IC3 & IC4 & IC5 & IC1 & IC2 & IC3 & IC4 & IC5 & IC1 & IC2 & IC3 & IC4 & IC5\\ \hline
&\multicolumn{5}{l}{ $d_1=d_2=20$}\\ \hline
100 & 0.956 & 0.800 & 0.939 & 0.748 & 0.983 & 0.911 & 0.727 & 0.885 & 0.668 & 0.970 & 0.758 & 0.595 & 0.730 & 0.528 & 0.875 \\
200 & 0.990 & 0.979 & 0.746 & 0.641 & 0.997 & 0.979 & 0.962 & 0.675 & 0.557 & 0.990 & 0.883 & 0.822 & 0.546 & 0.453 & 0.934  \\
300 & 1 & 0.734 & 0.998 & 0.565 & 1 & 0.999 & 0.664 & 0.993 & 0.485 & 0.998 & 0.999 & 0.664 & 0.993 & 0.485 & 0.998 \\
500 & 1 & 0.582 & 1 & 0.383 & 1 & 1 & 0.511 & 1 & 0.326 & 1 & 0.999 & 0.378 & 0.996 & 0.241 & 0.982 \\
1000 & 1 & 0.459 & 1 & 0.329 & 1 & 1 & 0.408 & 1 & 0.277 & 1 & 1 & 0.289 & 1 & 0.198 & 0.999  \\
\hline
&\multicolumn{5}{l}{ $d_1=d_2=40$}\\ \hline
100 & 0.970 & 0.946 & 0.963 & 0.945 & 0.974 & 0.936 & 0.901 & 0.931 & 0.895 & 0.981 & 0.912 & 0.873 & 0.903 & 0.862 & 0.970\\
200 & 0.981 & 0.972 & 0.936 & 0.925 & 0.992 & 0.955 & 0.942 & 0.883 & 0.860 & 0.998 & 0.937 & 0.931 & 0.854 & 0.833 & 0.994 \\
300 & 0.998 & 0.960 & 0.995 & 0.943 & 0.998 & 0.991 & 0.931 & 0.985 & 0.892 & 1 & 0.986 & 0.897 & 0.978 & 0.866 & 0.999 \\
500 & 1 & 0.978 & 1 & 0.960 & 0.999 & 1 & 0.955 & 0.999 & 0.927 & 1 & 1 & 0.948 & 0.999 & 0.908 & 1 \\
1000 & 1 & 0.992 & 1 & 0.979 & 1 & 1 & 0.987 & 1 & 0.963 & 1 & 1 & 0.985 & 1 & 0.955 & 1 \\
\hline
&\multicolumn{5}{l}{ $d_1=d_2=80$}\\ \hline
100 & 0.595 & 0.570 & 0.591 & 0.565 & 0.635 & 0.466 & 0.445 & 0.462 & 0.438 & 0.508 & 0.370 & 0.350 & 0.364 & 0.348 & 0.405 \\
200 & 0.673 & 0.663 & 0.623 & 0.612 & 0.889 & 0.553 & 0.546 & 0.489 & 0.478 & 0.794 & 0.458 & 0.447 & 0.415 & 0.405 & 0.598 \\
300 & 0.815 & 0.747 & 0.803 & 0.735 & 0.971 & 0.719 & 0.613 & 0.700 & 0.594 & 0.950 & 0.651 & 0.560 & 0.627 & 0.549 & 0.903 \\
500 & 0.965 & 0.900 & 0.958 & 0.892 & 0.998 & 0.915 & 0.836 & 0.907 & 0.823 & 0.997 & 0.892 & 0.813 & 0.884 & 0.790 & 0.984 \\
1000 & 1 & 0.998 & 1 & 0.998 & 1 & 1 & 0.996 & 1 & 0.995 & 1 & 1 & 0.995 & 1 & 0.993 & 1 \\
\hline
\end{tabular}}
\end{table}

\begin{table}[htbp]
\centering
\caption{Proportion of correct identification of rank $r$ 
using IC estimators based on TIPUP procedures for Model M2, over 1000 replications}
\label{tab:tipup.bic.r5.strongweak}
\resizebox{\textwidth}{!}
{\begin{tabular}{cccccc|ccccc|cccccc}\hline \hline
 &\multicolumn{5}{c}{initial estimator} & \multicolumn{5}{c}{one step estimator} & \multicolumn{5}{c}{final estimator}\\ \hline
$T$ & IC1 & IC2 & IC3 & IC4 & IC5 & IC1 & IC2 & IC3 & IC4 & IC5 & IC1 & IC2 & IC3 & IC4 & IC5\\ \hline
&\multicolumn{5}{l}{ $d_1=d_2=20$}\\ \hline
100 & 0.050 & 0.010 & 0.038 & 0.005 & 0.115 & 0.050 & 0.009 & 0.037 & 0.005 & 0.118 & 0.008 & 0.001 & 0.007 & 0.000 & 0.029 \\
200 & 0.612 & 0.488 & 0.111 & 0.071 & 0.715 & 0.608 & 0.489 & 0.111 & 0.071 & 0.717 & 0.311 & 0.225 & 0.038 & 0.018 & 0.413  \\
300 & 0.980 & 0.347 & 0.940 & 0.193 & 0.974 & 0.981 & 0.344 & 0.940 & 0.195 & 0.974 & 0.891 & 0.200 & 0.789 & 0.104 & 0.879 \\
500 & 1 & 0.792 & 1 & 0.602 & 1 & 1 & 0.792 & 1 & 0.605 & 1 & 1 & 0.705 & 1 & 0.509 & 1 \\
1000 & 1 & 1 & 1 & 0.991 & 1 & 1 & 1 & 1 & 0.991 & 1 & 1 & 0.999 & 1 & 0.990 & 1  \\
\hline
&\multicolumn{5}{l}{ $d_1=d_2=40$}\\ \hline
100 & 0.012 & 0.005 & 0.010 & 0.005 & 0.034 & 0.011 & 0.006 & 0.009 & 0.005 & 0.034 & 0.007 & 0.005 & 0.006 & 0.005 & 0.015\\
200 & 0.965 & 0.961 & 0.914 & 0.898 & 0.991 & 0.964 & 0.961 & 0.912 & 0.896 & 0.991 & 0.843 & 0.824 & 0.731 & 0.708 & 0.952 \\
300 & 0.876 & 0.636 & 0.848 & 0.585 & 0.986 & 0.877 & 0.637 & 0.846 & 0.583 & 0.987 & 0.877 & 0.637 & 0.846 & 0.583 & 0.987 \\
500 & 1 & 1 & 1 & 1 & 1 & 1 & 1 & 1 & 1 & 1 & 1 & 1 & 1 & 1 & 1 \\
1000 & 1 & 1 & 1 & 1 & 1 & 1 & 1 & 1 & 1 & 1 & 1 & 1 & 1 & 1 & 1 \\
\hline
&\multicolumn{5}{l}{ $d_1=d_2=80$}\\ \hline
100 & 0.003 & 0.002 & 0.003 & 0.002 & 0.003 & 0.003 & 0.002 & 0.003 & 0.002 & 0.003 & 0 & 0 & 0 & 0 & 0 \\
200 & 0.193 & 0.192 & 0.166 & 0.157 & 0.381 & 0.192 & 0.191 & 0.166 & 0.158 & 0.381 & 0.163 & 0.163 & 0.139 & 0.133 & 0.309 \\
300 & 0.795 & 0.710 & 0.780 & 0.698 & 0.959 & 0.796 & 0.709 & 0.779 & 0.697 & 0.959 & 0.796 & 0.709 & 0.779 & 0.697 & 0.959 \\
500 & 1 & 1 & 1 & 1 & 1 & 1 & 1 & 1 & 1 & 1 & 1 & 1 & 1 & 1 & 1 \\
1000 & 1 & 1 & 1 & 1 & 1 & 1 & 1 & 1 & 1 & 1 & 1 & 1 & 1 & 1 & 1 \\
\hline
\end{tabular}}
\end{table}

\begin{table}[htbp]
\centering
\caption{Proportion of correct identification of rank $r$ 
using ER estimators based on TOPUP procedures for Model M2, over 1000 replications}
\label{tab:topup.ratio.r5.strongweak}
\resizebox{\textwidth}{!}
{\begin{tabular}{cccccc|ccccc|cccccc}\hline \hline
 &\multicolumn{5}{c}{initial estimator} & \multicolumn{5}{c}{one step estimator} & \multicolumn{5}{c}{final estimator}\\ \hline
$T$ & ER1 & ER2 & ER3 & ER4 & ER5 & ER1 & ER2 & ER3 & ER4 & ER5 & ER1 & ER2 & ER3 & ER4 & ER5\\ \hline
&\multicolumn{5}{l}{ $d_1=d_2=20$}\\ \hline
100 & 0.029 & 0.027 & 0.029 & 0.029 & 0.029 & 0.677 & 0.663 & 0.677 & 0.676 & 0.676 & 0.788 & 0.765 & 0.788 & 0.788 & 0.783 \\
200 & 0.008 & 0.007 & 0.008 & 0.008 & 0.007 & 0.706 & 0.694 & 0.706 & 0.706 & 0.705 & 0.761 & 0.757 & 0.761 & 0.761 & 0.760 \\
300 & 0.006 & 0.006 & 0.006 & 0.006 & 0.006 & 0.743 & 0.739 & 0.743 & 0.743 & 0.743 & 0.781 & 0.779 & 0.781 & 0.781 & 0.781 \\
500 & 0.003 & 0.003 & 0.003 & 0.003 & 0.003 & 0.802 & 0.799 & 0.802 & 0.802 & 0.802 & 0.881 & 0.879 & 0.881 & 0.881 & 0.881 \\
1000 & 0.004 & 0.004 & 0.004 & 0.004 & 0.004 & 0.866 & 0.864 & 0.866 & 0.866 & 0.866 & 0.986 & 0.986 & 0.986 & 0.986 & 0.986 \\
\hline
&\multicolumn{5}{l}{ $d_1=d_2=40$}\\ \hline
100 & 0 & 0 & 0 & 0 & 0 & 0.744 & 0.727 & 0.744 & 0.744 & 0.744 & 0.811 & 0.804 & 0.811 & 0.811 & 0.811 \\
200 & 0 & 0 & 0 & 0 & 0 & 0.554 & 0.543 & 0.554 & 0.554 & 0.554 & 0.588 & 0.578 & 0.588 & 0.588 & 0.587 \\
300 & 0 & 0 & 0 & 0 & 0 & 0.422 & 0.412 & 0.422 & 0.422 & 0.421 & 0.476 & 0.467 & 0.476 & 0.476 & 0.476 \\
500 & 0 & 0 & 0 & 0 & 0 & 0.344 & 0.338 & 0.344 & 0.344 & 0.344 & 0.401 & 0.396 & 0.401 & 0.401 & 0.400\\
1000 & 0 & 0 & 0 & 0 & 0 & 0.367 & 0.362 & 0.367 & 0.367 & 0.366 & 0.434 & 0.432 & 0.434 & 0.434 & 0.434\\
\hline
&\multicolumn{5}{l}{ $d_1=d_2=80$}\\ \hline
100 & 0 & 0 & 0 & 0 & 0 & 0.770 & 0.725 & 0.770 & 0.770 & 0.770 & 0.840 & 0.818 & 0.840 & 0.840 & 0.839 \\
200 & 0 & 0 & 0 & 0 & 0 & 0.599 & 0.565 & 0.599 & 0.599 & 0.598 & 0.662 & 0.639 & 0.662 & 0.662 & 0.662 \\
300 & 0 & 0 & 0 & 0 & 0 & 0.507 & 0.480 & 0.507 & 0.507 & 0.506 & 0.610 & 0.584 & 0.610 & 0.610 & 0.609 \\
500 & 0 & 0 & 0 & 0 & 0 & 0.497 & 0.468 & 0.497 & 0.497 & 0.497 & 0.585 & 0.573 & 0.585 & 0.585 & 0.585 \\
1000 & 0 & 0 & 0 & 0 & 0 & 0.614 & 0.603 & 0.614 & 0.614 & 0.614 & 0.712 & 0.702 & 0.712 & 0.712 & 0.711 \\
\hline
\end{tabular}}
\end{table}

\begin{table}[htbp]
\centering
\caption{Proportion of correct identification of rank $r$ 
using ER estimators based on TIPUP procedures for Model M2, over 1000 replications}
\label{tab:tipup.ratio.r5.strongweak}
\resizebox{\textwidth}{!}
{\begin{tabular}{cccccc|ccccc|cccccc}\hline \hline
 &\multicolumn{5}{c}{initial estimator} & \multicolumn{5}{c}{one step estimator} & \multicolumn{5}{c}{final estimator}\\ \hline
$T$ & ER1 & ER2 & ER3 & ER4 & ER5 & ER1 & ER2 & ER3 & ER4 & ER5 & ER1 & ER2 & ER3 & ER4 & ER5\\ \hline
&\multicolumn{5}{l}{ $d_1=d_2=20$}\\ \hline
100 & 0.999 & 0.994 & 0.999 & 0.999 & 0.998 & 1 & 0.998 & 1 & 1 & 1 & 1 & 0.999 & 1 & 1 & 1 \\
200 & 1 & 1 &  1 & 1 & 1 & 1 & 1 & 1 & 1 & 1 & 1 & 1 & 1 &  1 & 1 \\
300 & 1 & 1 &  1 & 1 & 1 & 1 & 1 & 1 & 1 & 1 & 1 & 1 & 1 &  1 & 1 \\
500 & 1 & 1 &  1 & 1 & 1 & 1 & 1 & 1 & 1 & 1 & 1 & 1 & 1 &  1 & 1 \\
1000 & 1 & 1 &  1 & 1 & 1 & 1 & 1 & 1 & 1 & 1 & 1 & 1 & 1 &  1 & 1 \\
\hline
&\multicolumn{5}{l}{ $d_1=d_2=40$}\\ \hline
100 & 0.986 & 0.958 & 0.986 & 0.986 & 0.985 & 0.992 & 0.979 & 0.992 & 0.992 & 0.992 & 1 & 0.997 & 1 & 1 & 1 \\
200 & 1 & 1 &  1 & 1 & 1 & 1 & 1 & 1 & 1 & 1 & 1 & 1 & 1 &  1 & 1 \\
300 & 1 & 1 &  1 & 1 & 1 & 1 & 1 & 1 & 1 & 1 & 1 & 1 & 1 &  1 & 1 \\
500 & 1 & 1 &  1 & 1 & 1 & 1 & 1 & 1 & 1 & 1 & 1 & 1 & 1 &  1 & 1 \\
1000 & 1 & 1 &  1 & 1 & 1 & 1 & 1 & 1 & 1 & 1 & 1 & 1 & 1 &  1 & 1 \\
\hline
&\multicolumn{5}{l}{ $d_1=d_2=80$}\\ \hline
100 & 0.967 & 0.870 & 0.967 & 0.967 & 0.965 & 0.990 & 0.902 & 0.990 & 0.990 & 0.989 & 1 & 0.979 & 1 & 1 & 1 \\
200 & 0.999 & 0.999 & 0.999 & 0.999 & 0.999 & 1 & 1 & 1 & 1 & 1 & 1 & 1 & 1 &  1 & 1 \\
300 & 1 & 1 &  1 & 1 & 1 & 1 & 1 & 1 & 1 & 1 & 1 & 1 & 1 &  1 & 1 \\
500 & 1 & 1 &  1 & 1 & 1 & 1 & 1 & 1 & 1 & 1 & 1 & 1 & 1 &  1 & 1 \\
1000 & 1 & 1 &  1 & 1 & 1 & 1 & 1 & 1 & 1 & 1 & 1 & 1 & 1 &  1 & 1 \\
\hline
\end{tabular}}
\end{table}


For Model M2, Tables \ref{tab:topup.bic.r5.strongweak} and \ref{tab:tipup.bic.r5.strongweak} report 
the proportion of correct rank identification using IC estimators based on TOPUP and TIPUP procedures, respectively. Tables \ref{tab:topup.ratio.r5.strongweak} and \ref{tab:tipup.ratio.r5.strongweak} report the proportion of correct rank identification using ER estimators based on TOPUP and TIPUP procedures, respectively.

\begin{table}[htbp]
\centering
\caption{Root mean squared errors (RMSEs) of the ER estimators based on TOPUP procedures for Model M3, averaging over 1000 replications}
\label{tab:topup.ratio.r5.weak}
\resizebox{\textwidth}{!}
{\begin{tabular}{cccccc|ccccc|cccccc}\hline \hline
 &\multicolumn{5}{c}{initial estimator} & \multicolumn{5}{c}{one step estimator} & \multicolumn{5}{c}{final estimator}\\ \hline
$T$ & ER1 & ER2 & ER3 & ER4 & ER5 & ER1 & ER2 & ER3 & ER4 & ER5 & ER1 & ER2 & ER3 & ER4 & ER5\\ \hline
&\multicolumn{5}{l}{ $d_1=d_2=20$}\\ \hline
100 & 2.15 & 2.29 & 2.15 & 2.15 & 2.17 & 1.84 & 2.13 & 1.84 & 1.84 & 1.89 & 1.66 & 2.00 & 1.67 & 1.68 & 1.77 \\
200 & 1.68 & 1.76 & 1.68 & 1.68 & 1.69 & 1.23 & 1.33 & 1.23 & 1.23 & 1.24 & 1.13 & 1.29 & 1.13 & 1.13 & 1.16 \\
300 & 1.17 & 1.21 & 1.17 & 1.17 & 1.19 & 1.07 & 1.10 & 1.07 & 1.07 & 1.08 & 0.86 & 0.92 & 0.86 & 0.86 & 0.86 \\
500 & 1.02 & 1.02 & 1.02 & 1.02 & 1.02 & 1.04 & 1.04 & 1.04 & 1.04 & 1.04 & 0.60 & 0.60 & 0.60 & 0.60 & 0.60 \\
1000 & 1.00 & 1.00 & 1.00 & 1.00 & 1.00 & 0.95 & 0.95 & 0.95 & 0.95 & 0.95 & 0.19 & 0.19 & 0.19 & 0.19 & 0.19 \\
\hline
&\multicolumn{5}{l}{ $d_1=d_2=40$}\\ \hline
100 & 1.01 & 1.01 & 1.01 & 1.01 & 1.01 & 1.03 & 1.16 & 1.03 & 1.03 & 1.04 & 1.02 & 1.26 & 1.02 & 1.03 & 1.07 \\
200 & 1.00 & 1.00 & 1.00 & 1.00 & 1.00 & 1.00 & 1.00 & 1.00 & 1.00 & 1.00 & 0.89 & 0.93 & 0.89 & 0.89 & 0.90 \\
300 & 1.00 & 1.00 & 1.00 & 1.00 & 1.00 & 1.00 & 1.00 & 1.00 & 1.00 & 1.00 & 0.76 & 0.74 & 0.76 & 0.76 & 0.75 \\
500 & 1.00 & 1.00 & 1.00 & 1.00 & 1.00 & 1.00 & 1.00 & 1.00 & 1.00 & 1.00 & 0.49 & 0.48 & 0.49 & 0.49 & 0.49 \\
1000 & 1.00 & 1.00 & 1.00 & 1.00 & 1.00 & 0.99 & 0.98 & 0.99 & 0.99 & 0.99 & 0.11 & 0.11 & 0.11 & 0.11 & 0.11 \\
\hline
&\multicolumn{5}{l}{ $d_1=d_2=80$}\\ \hline
100 & 2.71 & 3.38 & 2.71 & 2.71 & 2.74 & 2.06 & 2.51 & 2.06 & 2.06 & 2.09 & 1.90 & 2.78 & 1.92 & 1.93 & 2.29 \\
200 & 1.10 & 1.26 & 1.10 & 1.10 & 1.10 & 1.09 & 1.14 & 1.09 & 1.09 & 1.09 & 1.01 & 1.68 & 1.02 & 1.02 & 1.19\\
300 & 1.00 & 1.00 & 1.00 & 1.00 & 1.00 & 1.01 & 1.02 & 1.01 & 1.01 & 1.01 & 0.95 & 1.36 & 0.95 & 0.95 & 0.97 \\
500 & 1.00 & 1.00 & 1.00 & 1.00 & 1.00 & 1.00 & 1.00 & 1.00 & 1.00 & 1.00 & 0.90 & 1.00 & 0.90 & 0.90 & 0.90 \\
1000 & 1.00 & 1.00 & 1.00 & 1.00 & 1.00 & 1.00 & 1.00 & 1.00 & 1.00 & 1.00 & 0.61 & 0.59 & 0.61 & 0.61 & 0.61 \\
\hline
\end{tabular}}
\end{table}

\begin{table}[htbp]
\centering
\caption{Root mean squared errors (RMSEs) of the ER estimators based on TIPUP procedures for Model M3, averaging over 1000 replications}
\label{tab:tipup.ratio.r5.weak}
\resizebox{\textwidth}{!}
{\begin{tabular}{cccccc|ccccc|cccccc}\hline \hline
 &\multicolumn{5}{c}{initial estimator} & \multicolumn{5}{c}{one step estimator} & \multicolumn{5}{c}{final estimator}\\ \hline
$T$ & ER1 & ER2 & ER3 & ER4 & ER5 & ER1 & ER2 & ER3 & ER4 & ER5 & ER1 & ER2 & ER3 & ER4 & ER5\\ \hline
&\multicolumn{5}{l}{ $d_1=d_2=20$}\\ \hline
100 & 1.50 & 1.76 & 1.50 & 1.50 & 1.53 & 1.14 & 1.65 & 1.15 & 1.15 & 1.20 & 0.45 & 1.19 & 0.45 & 0.45 & 0.52 \\
200 & 0.84 & 1.00 & 0.84 & 0.84 & 0.86 & 0.61 & 0.84 & 0.61 & 0.61 & 0.65 & 0.20 & 0.35 & 0.20 & 0.20 & 0.20 \\
300 & 0.60 & 0.66 & 0.60 & 0.60 & 0.61 & 0.43 & 0.50 & 0.43 & 0.43 & 0.44 & 0.12 & 0.14 & 0.12 & 0.12 & 0.12 \\
500 & 0.30 & 0.33 & 0.30 & 0.30 & 0.30 & 0.17 & 0.18 & 0.17 & 0.18 & 0.17 & 0.04 & 0.04 & 0.04 & 0.04 & 0.04 \\
1000 & 0.07 & 0.07 & 0.07 & 0.07 & 0.07 & 0.04 & 0.04 & 0.04 & 0.04 & 0.04 & 0.03 & 0.03 & 0.03 & 0.03 & 0.03 \\
\hline
&\multicolumn{5}{l}{ $d_1=d_2=40$}\\ \hline
100 & 1.19 & 1.73 & 1.19 & 1.19 & 1.21 & 0.86 & 1.55 & 0.86 & 0.86 & 0.89 & 0.35 & 0.97 & 0.35 & 0.35 & 0.37 \\
200 & 0.50 & 0.56 & 0.50 & 0.50 & 0.50 & 0.37 & 0.38 & 0.37 & 0.37 & 0.36 & 0.19 & 0.15 & 0.19 & 0.19 & 0.18 \\
300 & 0.27 & 0.25 & 0.27 & 0.27 & 0.27 & 0.23 & 0.18 & 0.23 & 0.23 & 0.23 & 0.13 & 0.10 & 0.13 & 0.13 & 0.13 \\
500 & 0.17 & 0.11 & 0.17 & 0.17 & 0.17 & 0.14 & 0.05 & 0.14 & 0.14 & 0.13 & 0.04 & 0.04 & 0.04 & 0.04 & 0.04 \\
1000 & 0.04 & 0.03 & 0.04 & 0.04 & 0.04 & 0 & 0 & 0 & 0 & 0 & 0 & 0 & 0 & 0 & 0 \\
\hline
&\multicolumn{5}{l}{ $d_1=d_2=80$}\\ \hline
100 & 2.78 & 3.75 & 2.78 & 2.78 & 2.90 & 2.45 & 3.72 & 2.45 & 2.45 & 2.59 & 1.37 & 3.53 & 1.38 & 1.38 & 1.49 \\
200 & 1.90 & 2.83 & 1.90 & 1.90 & 1.94 & 1.47 & 2.52 & 1.47 & 1.46 & 1.51 & 0.78 & 1.43 & 0.78 & 0.78 & 0.76 \\
300 & 0.99 & 0.42 & 0.99 & 0.99 & 0.94 & 0.65 & 0.41 & 0.65 & 0.65 & 0.60 & 0.40 & 0.37 & 0.40 & 0.40 & 0.40 \\
500 & 0.32 & 0.23 & 0.32 & 0.32 & 0.32 & 0.27 & 0.18 & 0.27 & 0.27 & 0.26 & 0.23 & 0.22 & 0.23 & 0.23 & 0.23 \\
1000 & 0.15 & 0.12 & 0.15 & 0.15 & 0.15 & 0.13 & 0.11 & 0.13 & 0.13 & 0.12 & 0.11 & 0.10 & 0.11 & 0.11 & 0.11 \\
\hline
\end{tabular}}
\end{table}

For Model M3 with all very weak factors ($\delta_0=\delta_1=0.6$), Tables \ref{tab:topup.ratio.r5.weak} and \ref{tab:tipup.ratio.r5.weak} report RMSEs of the ER estimators. The results of using IC estimators are not shown, as it is impossible to detect weak factors if we set $\delta_1=0$ in $g_k$.

In summary, the first part of simulation shows that in IC estimators, IC2 and IC4 seem to perform slightly better than IC1, IC3 and IC5 in Table \ref{tab:topup.bic.r5.strong}, while it is reversed in Tables \ref{tab:topup.bic.r5.strongweak} and \ref{tab:tipup.bic.r5.strongweak}. The difference is not significant though. In ER estimators, the choice of the penalty function also seems to have a limited impact on the results. In most cases, ER1 and ER2 are slightly better.

In the third part of simulation, Table \ref{tab:bic.r5.strongweak:delta1} shows the proportion of correct rank identification of the IC estimators in \eqref{penalty:IC} with known $\delta_1$ for Model M2. Moreover, table \ref{tab:bic.r5.weak.delta1} reports the RMSEs of the IC estimators in \eqref{penalty:IC} with known $\delta_1$ for Model M3. Again, given $\delta_1$, the performance of all IC1-IC5 is also very similar. In some cases, IC2 and IC4 are slightly better.

\begin{table}[htbp]
\centering
\caption{Proportion of correct identification of rank $r$ 
using IC estimators with given $\delta_1$ based on TOPUP and TIPUP procedures for Model M2, over 1000 replications}
\label{tab:bic.r5.strongweak:delta1}
\resizebox{\textwidth}{!}
{\begin{tabular}{cccccc|ccccc|cccccc}\hline \hline
 &\multicolumn{5}{c}{initial estimator} & \multicolumn{5}{c}{one step estimator} & \multicolumn{5}{c}{final estimator}\\ \hline
$T$ & IC1 & IC2 & IC3 & IC4 & IC5 & IC1 & IC2 & IC3 & IC4 & IC5 & IC1 & IC2 & IC3 & IC4 & IC5\\ \hline
&\multicolumn{10}{l}{ $d_1=d_2=40$ and TOPUP}\\ \hline
100 & 0 & 0 & 0 & 0 & 0 & 0.002 & 0.010 & 0.004 & 0.012 & 0.001 & 0.650 & 0.650 & 0.650 & 0.651 & 0.650 \\
300 & 0 & 0 & 0 & 0 & 0 & 0.048 & 0.286 & 0.074 & 0.358 & 0 & 0.894 & 0.895 & 0.894 & 0.898 & 0.894 \\
500 & 0 & 0.001 & 0 & 0.005 & 0 & 0.106 & 0.619 & 0.168 & 0.689 & 0.006 & 0.960 & 0.962 & 0.960 & 0.962 & 0.958 \\
1000 & 0 & 0.428 & 0 & 0.676 & 0 & 0.352 & 0.947 & 0.494 & 0.974 & 0.226 & 0.990 & 0.997 & 0.993 & 0.998 & 0.989 \\
\hline
&\multicolumn{10}{l}{ $d_1=d_2=40$ and TIPUP}\\ \hline
100 & 1 & 1 &  1 & 1 & 1 & 1 & 1 & 1 & 1 & 1 & 1 & 1 & 1 &  1 & 1 \\
300 & 1 & 1 &  1 & 1 & 1 & 1 & 1 & 1 & 1 & 1 & 1 & 1 & 1 &  1 & 1 \\
500 & 1 & 1 &  1 & 1 & 1 & 1 & 1 & 1 & 1 & 1 & 1 & 1 & 1 &  1 & 1 \\
1000 & 1 & 1 &  1 & 1 & 1 & 1 & 1 & 1 & 1 & 1 & 1 & 1 & 1 &  1 & 1 \\
\hline
\end{tabular}}
\end{table}

\begin{table}[htbp]
\centering
\caption{Root mean squared errors (RMSEs) of the IC estimators with given $\delta_1$ based on TOPUP and TIPUP procedures for Model M3, averaging over 1000 replications}
\label{tab:bic.r5.weak.delta1}
\resizebox{\textwidth}{!}
{\begin{tabular}{cccccc|ccccc|cccccc}\hline \hline
 &\multicolumn{5}{c}{initial estimator} & \multicolumn{5}{c}{one step estimator} & \multicolumn{5}{c}{final estimator}\\ \hline
$T$ & IC1 & IC2 & IC3 & IC4 & IC5 & IC1 & IC2 & IC3 & IC4 & IC5 & IC1 & IC2 & IC3 & IC4 & IC5\\ \hline
&\multicolumn{10}{l}{ $d_1=d_2=40$ and TOPUP}\\ \hline
100 & 1.02 & 1.01 & 1.01 & 1.01 & 1.39 & 0.95 & 0.91 & 0.94 & 0.90 & 0.99 & 0.97 & 0.98 & 0.97 & 0.99 & 0.92 \\
300 & 1.00 & 1.00 & 1.00 & 1.00 & 1.00 & 1.00 & 0.87 & 0.99 & 0.80 & 1.00 & 0.18 & 0.23 & 0.18 & 0.26 & 0.15 \\
500 & 1.00 & 1.00 & 1.00 & 1.00 & 1.00 & 0.99 & 0.43 & 0.96 & 0.33 & 1.00 & 0 & 0 & 0 & 0 & 0 \\
1000 & 1.00 & 1.00 & 1.00 & 1.00 & 1.00 & 0.92 & 0.05 & 0.76 & 0.03 & 0.98 & 0 & 0 & 0 & 0 & 0 \\
\hline
&\multicolumn{10}{l}{ $d_1=d_2=40$ and TIPUP}\\ \hline
100 & 0.86 & 0.91 & 0.88 & 0.91 & 0.74 & 0.89 & 0.92 & 0.89 & 0.93 & 0.77 & 0.88 & 0.92 & 0.89 & 0.92 & 0.76 \\
300 & 0 & 0.03 & 0 & 0.03 & 0.03 & 0 & 0.03 & 0 & 0.04 & 0 & 0 & 0.03 & 0 & 0.07 & 0 \\
500 & 0 & 0 & 0 & 0 & 0 & 0 & 0 & 0 & 0 & 0 & 0 & 0 & 0 & 0 & 0 \\
1000 & 0 & 0 & 0 & 0 & 0 & 0 & 0 & 0 & 0 & 0 & 0 & 0 & 0 & 0 & 0 \\
\hline
\end{tabular}}
\end{table}

\end{document}